\documentclass{jfm}
\usepackage{amsmath}
\usepackage{graphicx}
\usepackage{etoolbox}
\usepackage{tikz}
\usetikzlibrary{shapes.geometric}
\usepackage{epstopdf, epsfig}
\usepackage{cleveref}
\usepackage[font=small]{caption}
\usepackage[labelformat=empty, position=top]{subcaption}
\usepackage[export]{adjustbox}
\usepackage{color}
\usepackage{xcolor}
\usepackage{mathrsfs}
\usepackage[inline]{enumitem}

\newrobustcmd*{\mytriangle}[1]{\tikz{\filldraw[draw=#1,fill=#1] (0,0) --
		(0.2cm,0) -- (0.1cm,0.2cm);}}
\newrobustcmd*{\mycircle}[1]{\tikz{\filldraw[draw=#1,fill=#1] (0,0) circle [radius=0.1cm];}}
\newrobustcmd*{\myline}[1]{\tikz{\filldraw[draw=#1,fill=#1] (0,0.2cm) -- (0.2cm,0.2cm);}}

\definecolor{mu001}{RGB}{128,128,0}
\definecolor{mu01}{RGB}{255,20,147}
\definecolor{mu1}{RGB}{0,191,191}
\definecolor{mu10}{RGB}{255,215,0}
\definecolor{mu100}{RGB}{0,100,0}

\shorttitle{Turbulence modulations of emulsions in HIT}
\shortauthor{M. Crialesi-Esposito, M. Rosti, S. Chibbaro, L. Brandt}

\title[Modulation of homogeneous and isotropic turbulence in emulsions]{Modulation of homogeneous and isotropic turbulence in emulsions}

\author{Marco Crialesi-Esposito\aff{1}\corresp{\email{marcoce@kth.se}}, 
	   Marco Edoardo Rosti\aff{2},
	   Sergio Chibbaro\aff{3}, 
	   Luca Brandt\aff{1,4}}

\affiliation{\aff{1}FLOW Centre, KTH Royal Institute of Technology, Stockholm, Sweden
\aff{2}Complex Fluids and Flows Unit, Okinawa Institute of Science and Technology Graduate University, 1919-1 Tancha, Onna-son, Okinawa 904-0495, Japan
\aff{3} Institute Jean le Rond $\partial$'Alembert, Sorbonne Universite, Paris, France
\aff{4} Department of Energy and Process Engineering, Norwegian University of Science and Technology (NTNU), Trondheim, Norway
}

\begin{document}

\maketitle

\begin{abstract}
We present a numerical study of emulsions in homogeneous and isotropic turbulence at $Re_\lambda=137$. The problem is addressed via Direct Numerical Simulations (DNS), where the  Volume of Fluid (VOF) is used to represent the complex features of the liquid-liquid interface.
We consider a mixture of two iso-density fluids, where fluid properties are varied with the goal of understanding their role in turbulence modulation, in particular the volume fraction ($0.03<\alpha<0.5$), viscosity ratio ($0.01<\mu_d/\mu_c<100$) and large scale Weber number ($10.6<We_\mathcal{L}<106.5$).  The analysis, performed by studying integral quantities and spectral scale-by-scale analysis, reveals that energy is consistently transported from large to small scales by the interface, and no inverse cascade is observed. Furthermore, the total surface is found to be directly proportional to the amount of energy transported, while viscosity and surface tension alter the dynamic that regulates energy transport. We also observe the $-10/3$ and $-3/2$ scaling on droplet size distributions, suggesting that the dimensional arguments which led to their derivation are verified in HIT conditions. 
\end{abstract}

\begin{keywords}
\end{keywords}

\section{Introduction}
\label{sec:intro}
Emulsions are multiphase flows 
of two immiscible (totally or partially) liquid phases with similar densities. Such flows are extremely common in industrial applications such as pharmaceutical \citep{nielloud2000pharmaceutical,spernath2006microemulsions}, food processing \citep{mcclements2015food}, oil production \citep{kokal2005crude,mandal2010characterization,kilpatrick2012water} and waste treatment. Emulsions are also relevant for environmental flows such as oil spilling in oceans, when the oil droplets distribution becomes fundamental for quantifying environmental damages \citep{Li1998,french2004oil,Gopalan2010}. Many studies have been performed on the rheological behavior of emulsions in the past \citep{einstein1906neue,einstein1911berichtigung,Pal2000,Pal2001,Jansen2001,DeVita2019}, while the current knowledge on their behavior in turbulent flows is limited \citep{Yi2021}.

The two fluids are usually referred to as continuous phase (or carrier phase in case of strong advection) and dispersed phase (or droplet-phase) depending on whether the volume fraction $\alpha$ is respectively greater or lower than $0.5$; the system is denoted as binary flow when $\alpha = 0.5$. 
As the density ratio is usually considered to be close to 1, gravity effects are negligible with respect to the stirring and advection needed to sustain turbulence in the flow. For this reason, four dimensionless numbers can be used to describe these flows, namely the Reynolds number $Re$, 
 the Weber number $We$, the volume fraction of the dispersed phase and the viscosity contrast. 
Depending on the specific configuration under investigation, the definition of these numbers can change, yet they completely define the case studied provided the two fluid have the same density.

Several aspects of fundamental importance in emulsions, such as turbulence modulation, droplet size distributions and inter-phase energy fluxes, are not fully understood. We therefore aim to partially fill this gap by means of numerical simulations. In the following we provide an overview of the main results available in literature. Results for bubble/droplet laden flows are also discussed when relevant to the present work.

\subsection{Observations on droplet size distribution}
The Droplet Size Distribution (DSD) is a key aspect of emulsions, as its prediction becomes fundamental in most applications.  In his early seminal  work, \citet{Kolmogorov1949} discussed the criteria under which a droplet undergoes breakup when subject to surrounding turbulence. Kolmogorov first proposed a dimensional argument according to which surface tension forces need to be locally balanced by turbulent energy fluctuations. This idea was later addressed in \cite{Hinze1955} and translated into a critical Weber number $We_c$ of order 1 at which breakup occurs, leading to the definition of the Hinze scale $d_H$ as the minimum droplet diameter at which breakup may occur due to pressure fluctuations. A general definition for this scale is:
\begin{equation}
	d_H = \left(\frac{We_c}{2}\right)^{3/5} \left(\frac{\sigma}{\rho_c}\right)^{3/5}\varepsilon^{-2/5},
	\label{eq:hinzeGen}
\end{equation}
where $\sigma$ is the surface tension coefficient, $\rho_c$ is the carrier phase density and $\varepsilon$ is the energy dissipation rate. 
This estimate proved valid for bubbles \citep{Masuk2021, Chan2020a} and emulsions \citep{Perlekar2012,Mukherjee2019,Rosti2020,Yi2021}. 
Different $\mathcal{O}(1)$ values have been reported for $We_c$ in numerical \citep{Riviere2021} and experimental works \citep{Deane2002,Lemenand2017}, from $0.5$ up to $5$;  for dilute emulsions in turbulence $We_c\approx 1.17$,  according to the values from both numerical \citep{Perlekar2012}  and experimental \citep{Yi2021} data. 

For bubbles larger than the Hinze scale, \cite{Garrett2000} found that, in isotropic turbulent conditions, droplets break with a cascade process, 
and the diameter distribution follows a $d^{-10/3}$ power-law. This deterministic process can accurately describe bubble size distributions in breaking waves obtained
in experiments \citep{Garrett2000,Deane2002,Qi2020} and 
numerical simulations \citep{Deike:2016ir,Chan2020a}. 
The same power law has also been proposed for emulsions, based on diffuse-interface numerical simulations \citep{Skartlien2013,Mukherjee2019,Soligo2019}. 
For bubbles smaller than the Hinze scale,  \cite{Deane2002} suggested the existence of a fragmentation process; in this case, a $d^{-3/2}$ power-law is used to accurately fit experimental data. Agreement with this empirical power-law has been observed in Homogeneous and Isotropic Turbulence (HIT) both for bubbles \citep{Riviere2021} and  emulsions \citep{Mukherjee2019}.
The transition between the two power-laws is defined  by the Hinze scale. A consequence of this transition 
 is that droplets with $d\gg d_H$ generate both local and  non-local bubble/droplet production, as  they can fragment in both droplets larger or smaller than the Hinze scale \citep{Riviere2021}. Although both power-laws have been derived under the hypothesis of dilute conditions ($\alpha \lesssim 0.05$) they have been recently observed  in HIT studies of dense emulsions \citep{Mukherjee2019}, rising the question on the effective role of coalescence in the process. 

The connection between bubbles and emulsions is non-trivial and deserves special attention. In his work, \cite{Hinze1955} discussed how $We_c$ depends on the fluid properties of the dispersed phase. He assumed that $We_c=C[1-f(N_{Vi})]$, with $f$ a generic function of the viscosity group 
$N_{Vi}=\mu_d/\sqrt{\rho_d\sigma d}$, where  $\mu_d$ is the dispersed phase viscosity. On the other hand, 

$d_H$ was derived under the assumption of a dilute emulsion, hence the density in \Cref{eq:hinzeGen} refers to the carrier phase, as the phase where the energy dissipation rate $\varepsilon$ could be measured in experiments. This allows the direct application of the Hinze criteria in flows where density/viscosity ratios are significant as in air-water flows.  
However,  significant uncertainties are discussed in literature about the properties of the function $f$ and the role of the dispersed phase properties remains mostly unknown \citep{Masuk2021}. Also unknown is the role of turbulence inhomogeneity and anisotropy, which, according to \cite{Hinze1955}, may be a further source of non-linear effects in the determination of $We_c$. In fact, in flows where the energy dissipation rate shows strong spatial variations, $We_c$ varies for each bubble/droplet and it assumes meaning only on an average sense, making it difficult to disentangle  the effects of turbulence anisotropy and property contrast. 
Despite all these uncertainties, correlations from \cite{Hinze1955,Garrett2000,Deane2002}, derived for isotropic turbulent conditions, applies in most studies with strong property contrasts and large-scale anisotropy. This is likely due to the underlying assumption that the breakup process is purely inertial, as it only depends on $\varepsilon$ \citep{Garrett2000}. 
Thus, bubble breakup studies become relevant also for the present study.

It is finally worth noticing that the flow configuration appears to have a significant impact on DSD and 
experimental observations in shear flows can depart quite substantially from the discussed power-law behaviors. The recent work of \cite{Yi2021} presents strong experimental evidences of gamma/log-normal DSD in Taylor-Couette flow, confirming the previous findings of \cite{Pacek1998}. These configurations are characterized by strong anisotropy, making the comparison with data obtained for emulsions and bubbles in HIT difficult. 
On the other hand, \cite{Soligo2019} studied breakup and coalescence of emulsions dynamic in a turbulent channel flow. These authors observed the appearance of the $-10/3$ power-law for the droplet size distribution in presence of surfactants. It is interesting to observe that, in this numerical study, the scaling from \cite{Garrett2000} seems to apply in anisotropic configurations.  Fortunately, there has been a significant effort in recreating local HIT conditions in experiments in the latest years \citep{Debue2018,Dubrulle2019,Knutsen2020} and new studies are expected to provide new insights on these aspects.

\subsection{Studies of two-fluid  turbulence}
With the advent of more powerful computational resources, a significant number of studies have considered droplets in turbulent flows, yet almost only through diffuse-interface methods which 
may display significant mass loss.
In their study of emulsions in HIT turbulence, \cite{Perlekar2012} show that a statistical stationary state can be reached for the droplet size distributions. In the study, the authors used the Pseudo-Potential Lattice Boltzmann method \citep{Biferale2011}, which compensates  mass losses (due to droplets dissolution) by artificially re-inflating existing ones.
Simulations of the Cahn-Hilliard-Navier-Stokes formulation are presented in \cite{Perlekar2014} for binary fluids.
These authors found that enforcing large-scale HIT arrests coarsening. This result is particularly significant for emulsions (of which binary fluids represent a special case) as it shows that turbulence is the main factor to determine the droplets size. Furthermore, these authors report modified energy spectra for the mixtures, with a crossover  in correspondence to the Hinze scale.  

\cite{Komrakova2015} used a free-energy lattice Boltzmann method to numerically simulate emulsion breakup in HIT, induced by an external large-scale linear forcing. Their findings show that energy spectra present deviations with respect to the single-phase configuration and that the numerical method employed may alter the small-scale dynamics of the flow. Finally, increased coalescence is found for volume fractions $\alpha>0.05$ also owing to the nature of the diffused interface method. 

Droplet interactions with turbulence have been studied by \cite{Dodd2016} in decaying isotropic turbulence. Amongst several observations, these authors discuss the effects of droplet breakup and coalescence on the turbulent kinetic energy budget. Droplet coalescence lowers the total amount of area, hence decreases the surface energy and consequently increases the kinetic energy locally, while the opposite occurs in the case of breakup.    
More recently, \cite{Mukherjee2019} have studied emulsions in HIT conditions using a pseudo-potential Lattice Boltzmann method, discussing droplet statistics and their correlation with the surrounding turbulence. They confirm the findings of \cite{Perlekar2014} for energy spectra pivoting at the Hinze scale, demonstrating that energy is subtracted from large scales and injected at small scales, while no direct observation of the underlying mechanism is presented. These authors also show that the droplet generation  can be described through the Weber number spectra. In the same work, Mukherjee and co-workers discuss and demonstrate the need of using a forcing scale smaller than the turbulent-box size in order to achieve a polydisperse droplet distribution. 
It is yet important to note that \cite{Mukherjee2019} used a pseudopotential lattice-Boltzmann method, which leads to a significant loss of the dispersed mass
 during the simulation, as fairly discussed by the authors.

As concerns binary fluids, \cite{Perlekar2019} shows how the presence of interfaces leads to a different energy transfer mechanism, confirming the conclusions in
 \cite{Dodd2016}.
The author uses  the scale-by-scale (SBS) energy balance to show that  the energy absorption at larger scales 
is mainly given by the interface source term in the Cahn-Hilliard equation used by the author to describe the multiphase nature of the flow. Furthermore,   \cite{Perlekar2019} shows that small-scale statistics are almost unchanged when changing $We$ while they  are  affected by the Reynolds number.  This study complements the previous  findings in binary fluids \citep{Perlekar2014,Perlekar2017}, where coarsening was analyzed in 3D and 2D turbulence by means of a spinoidal decomposition.
\cite{Rosti2020} study droplets in Homogeneous Shear Turbulence (HST), focusing on the  effect of the droplet initial diameter and the shear-rate magnitude; the results show that a statistically stationary regime (i.e.\ balance of coalescence and breakup events and energy balance convergence) can be reached, while the Taylor-scale Reynolds number $Re_\lambda$ decreases with increasing surface tension.

Despite the growing literature on the subject, many issues remain to be fully understood. 
In particular, most of the studies have been carried out using diffuse-interface approaches, which cannot exactly represent the surface-terms effects yet key in many occasions.
In this sense, our work complements the very recent one by \cite{Riviere2021} focused on the bubble break-up dynamics.

\subsection{Objectives of the present study}

In the present  work, we use Direct Numerical Simulations (DNS) to study the effects of viscosity ratio, volume fraction and surface tension on the emulsion turbulent behavior. The chosen setup is tri-periodic HIT, with turbulence sustained throughout the simulation time. The analysis is performed at  $Re_\lambda\approx 137$, large enough to  represent realistic turbulent flows, while volume fraction, viscosity ratio and surface tension are varied to cover most relevant applications  \citep{Jansen2001}. We analyze the  turbulence through global and phase-averaged energy balance, energy spectra, SBS energy budget, and Probability Density Functions (PDF) for the intermittency analysis. Furthermore, we discuss droplet size distributions for all cases.
In summary, we will show that
\begin{enumerate*}[label=(\roman*)]
	\item the energy balance is significantly altered by the properties of the dispersed phase;
	\item surface tension forces induce an additional mechanism for energy transfer from larger scale towards the energy dissipation range;
	\item the modified energy transport mechanism alters  the energy spectra;
	\item the presence of the interface increases intermittency and alters the small scale statistics;
	\item the droplet size distribution displays both the $d^{-3/2}$ and $d^{-10/3}$ power-laws, with remarkable accuracy also for $d<d_H$.
\end{enumerate*}

\section{Methodology}
\label{sec:method}
\subsection{Governing equations and numerical method}
\label{subsec:met:govEq}

We consider an incompressible flow obeying the continuity and Navier-Stokes equations:
\begin{subequations}
\label{eqgroup:govenringEq}
\begin{equation}
 \partial_i u_i = 0
\label{eq:DNSmass}
\end{equation}
\begin{equation}
\rho(\partial_t u_i+u_j\partial_j u_i) = -\partial_i p + \partial_i \left[ \mu (\partial_i u_j + \partial_j u_i) \right] +f^\sigma_i +f^T_i
\label{eq:DNSmom}
\end{equation}
\end{subequations}
where $u_i$ is the velocity in the $i$-th direction, $p$ is the pressure, $\rho$ and $\mu$ the local density and viscosity. 
The forcing term  $f^\sigma_i = \sigma \xi \delta_s n_i$ represents the surface tension force, where $\sigma$ is the surface tension, $\xi$ the local interface curvature, $n_i$ the $i$-th component of the surface normal vector and $\delta_s$ the Dirac delta function that ensures the surface force is applied only at the interface \citep{tryggvason2011direct}. The last term in equation \Cref{eq:DNSmom} is the forcing needed to sustain  turbulence by injecting energy at the large scales; among the several  algorithms available to force sustained homogeneous and isotropic turbulence \cite[e.g.][]{Eswaran1988,Rosales2005,Mallouppas2013,Bassenne2016}, we use here the Arnold-Beltrami-Childress (ABC) forcing \citep{Mininni2006},
\begin{equation}
\label{eq:ABC}
\begin{aligned}
f_x &= A\: sin \:\kappa_0 z + C\: cos \:\kappa_0 y \\
f_y &= B\: sin \:\kappa_0 x + A\: cos \:\kappa_0 z \\
f_z &= C\: sin \:\kappa_0 y + B\: cos \:\kappa_0 x.
\end{aligned}
\end{equation}
with $x$, $y$ and $z\in[0,2\pi]$. As reported by \cite{Podvigina1994}, the ABC forcing creates an unstable single-phase flow  for $1/\nu>20$, with $\nu$ the kinematic viscosity and $\kappa_0$ the forcing wavelength. 

The description of the code and the algorithm used can be found in \cite{Rosti2019,Rosti2020}, together with several validations. The method is therefore only shortly described here, see also \cite{Costa2018} for references to the code structure. 

The equations are discretized on a staggered uniform Cartesian mesh: the spatial derivatives are computed using second-order centered finite differences and a second-order Adam-Bashford scheme is used for the time integration. The pressure splitting method presented in \cite{Dodd2014} is used to obtain a constant-coefficient Poisson equation, which we then solve with the direct FFT-based pressure solver presented in \cite{Costa2018}.
 
The interface between the two fluids is described with the Volume of Fluid (VOF) method, in particular  the Multi-dimensional Tangent Hyperbola INterface Capturing (MTHINC) algorithm developed by \cite{Ii2012}.  The advection equation for the VOF can be written in divergence form as
\begin{equation}
\label{eq:VOF}
\partial_t \phi +  \partial_i u_i \mathcal{H} = \phi \partial_i u_i.
\end{equation}
where $\mathcal{H}$ is the color function assuming the value of 0 and 1 in each of the fluids, and
$\phi$ the cell-averaged value of $\mathcal{H}$. 
In the MTHINC method, the function $\mathcal{H}$ is locally approximated using the hyperbolic tangent,
\begin{equation}
    \mathcal{H}(x',y',z') \approx \tfrac{1}{2}(1+tanh(\beta(P(x',y',z')+d))),
    \label{eq:tanh}
\end{equation}
where  $(x',y',z')\in[0,1]$ is the cell-cented local coordinate system, $\beta$ is a sharpness parameter (equal to 1 in the current work), $d$ a normalization factor and $P$ is the three-dimensional surface function, assumed here to be quadratic \citep{Ii2012}. The advantage of the method is that \Cref{eq:tanh} allows to solve the fluxes in \Cref{eq:VOF}  by semi-analytical integration.
Once the VOF function $\phi$ is known, we evaluate the local fluid properties as
\begin{equation}
\begin{aligned}
\rho &= \rho_d \phi + \rho_c(1-\phi)    \\
\mu  &= \mu_d \phi + \mu_c(1-\phi).    
\end{aligned}
\label{eq:prop}
\end{equation}
where the subscripts $c$ and $d$ indicates carrier and dispersed phase. 
Finally, the Continuum Surface Force (CSF) model is used to compute the surface tension force \citep{Brackbill1992}, with the normal evaluated with Youngs' method and the curvature as in \cite{Ii2012}. 

\subsection{Flow configuration}
\label{subsec:met:setup}

All the simulations are performed using the same ABC forcing, injecting energy at wavenumber $\kappa_0=2\pi/\mathcal{L}=2$, with $A=B=C=1$, corresponding to $Re_\lambda\approx137$ for the single phase flow (see \Cref{subsec:met:sp-case} for the characteristics of the reference single-phase flow and definition of the meaningful observables). As reported in literature \citep{Komrakova2015, Mukherjee2019} forcing the second wavelength is recommended in order to avoid coalescence induced by large turbulent structures in periodic domains.

In addition to the Reynolds number, the emulsion flows are characterised by 4 non-dimensional parameters.  The volume fraction, $\alpha=\mathcal{V}_d/\mathcal{V}$, defined as the ratio between the volume occupied by the dispersed phase $\mathcal{V}_d$ and the total volume $\mathcal{V}=(2\pi)^3$, the viscosity ratio $\mu_d/\mu_c$, where the subscripts $d$ and $c$ indicate the dispersed and carrier phase, and the Weber number, $We_\mathcal{L}=\rho_c\mathcal{L} u_{rms} ^2/\sigma$, where  $u_{rms}$ is the space-time average of the root-mean-square velocity of the single-phase case (which can be related to the forcing amplitude $A=B=C$) and
$\mathcal{L}$ the scale of the ABC forcing.
Finally, the density ratio,  $\rho= \rho_c/\rho_d$, is kept constant equal to 1 in this study.  

Here, we will vary the dispersed phase volume fraction, the viscosity ratio and the Weber number;
the parameters pertaining the different simulations discussed below are presented in \Cref{tab:testMat}. 
Note, finally, that the table also indicates  the integration time $N_\mathcal{T}$ 
required to reach statistical convergence of the turbulent quantities and droplet-size-distribution (DSD) 
in units of large eddies turnover times, $\mathcal{T}=\mathcal{L} u_{rms}$ \citep{Mininni2006}. The simulations are considered at convergence when global energy production balances dissipation (see \Cref{subsec:met:obs} and \Cref{eq:enBalMP} for their definition) with an error of less than $4\%$, also implying that the area derivative over time is negligible (see  \Cref{subsec:met:obs} for further details).
Interestingly, $N_\mathcal{T}$ varies significantly with the physical configuration. 
In particular, starting with the reference cases BE1 and BE2, we observe that increasing  $\mu_d/\mu_c$ longer times are needed to reach a statistically stationary state, which we will attribute to a decrease of the breakup rate.  A similar behavior is observed when decreasing $We_\mathcal{L}$, when higher surface tension forces decrease the probability of breakup events. Finally, large structures become unavoidable when increasing the volume fraction $\alpha$ \citep{Komrakova2015,Mukherjee2019}, which implies longer simulation times.

\begin{table}
    \centering
    \begin{tabular}{c c c c c c c}
         & $N$ & $\mu_d/\mu_c$ &$We_\mathcal{L}$ &$\sigma$  & $\alpha$ & $N_\mathcal{T}$ \\ 
\rule{0pt}{4ex}SP1 & 256 & - & - & - & - & 136 \\
               SP2 & 512 & - & - & - & - & 136 \\
\rule{0pt}{4ex}%
        BE1 & 512 & 1    & 42.6 & 0.46 & 0.03 & 115\\
        BE2 & 512 & 1    & 42.6 & 0.46 & 0.1  & 100\\
\rule{0pt}{4ex}%
        V11 & 512 & 0.01 & 42.6 & 0.46 & 0.03 & 115\\
        V12 & 512 & 0.1  & 42.6 & 0.46 & 0.03 & 100\\
        V13 & 512 & 10   & 42.6 & 0.46 & 0.03 & 64\\
        V14 & 512 & 100  & 42.6 & 0.46 & 0.03 & 60\\
\rule{0pt}{4ex}%
		V21 & 512 & 0.01 & 42.6 & 0.46 & 0.1 & 115\\
		V22 & 512 & 0.1  & 42.6 & 0.46 & 0.1 & 100\\
		V23 & 512 & 10   & 42.6 & 0.46 & 0.1 & 64\\
		V24 & 512 & 100  & 42.6 & 0.46 & 0.1 & 60\\
\rule{0pt}{4ex}
        C12 & 512 & 1    & 42.6 & 0.46 & 0.06 & 100\\
        C13 & 512 & 1    & 42.6 & 0.46 & 0.2  & 100\\
        C14 & 512 & 1    & 42.6 & 0.46 & 0.5  & 100\\
        C24 & 1024 & 1    & 42.6 & 0.46 & 0.5  & 100\\
        C34 & 256 & 1    & 42.6 & 0.46 & 0.5  & 100\\
\rule{0pt}{4ex}%
        W11 & 512 & 1 & 10.6 & 0.046 &0.03 & 160\\
        W12 & 512 & 1 & 106.5 & 0.046 &0.03 & 100\\

    \end{tabular}
    \caption{Parameter settings for the simulations considered in this study: number of grid points in each direction $N$, viscosity ratio $\mu_d/\mu_c$, Weber number  $We_\mathcal{L}$ with surface tension $\sigma$, volume fraction $\alpha$ and integration time to reach statistical convergence $N_\mathcal{T}$. All simulations are performed with $\mu_c=0.006$ and same ABC forcing. 
    Each case is denoted by a letter indicating the parameter which is varied: V for viscosity ratio, C volume fraction and W Weber number. SP are the single-phase flows and BE are configurations which recur in different parameterizations (base emulsions). }
    \label{tab:testMat}
\end{table}

\begin{figure}
    \centering
    \includegraphics[width=0.3\textwidth]{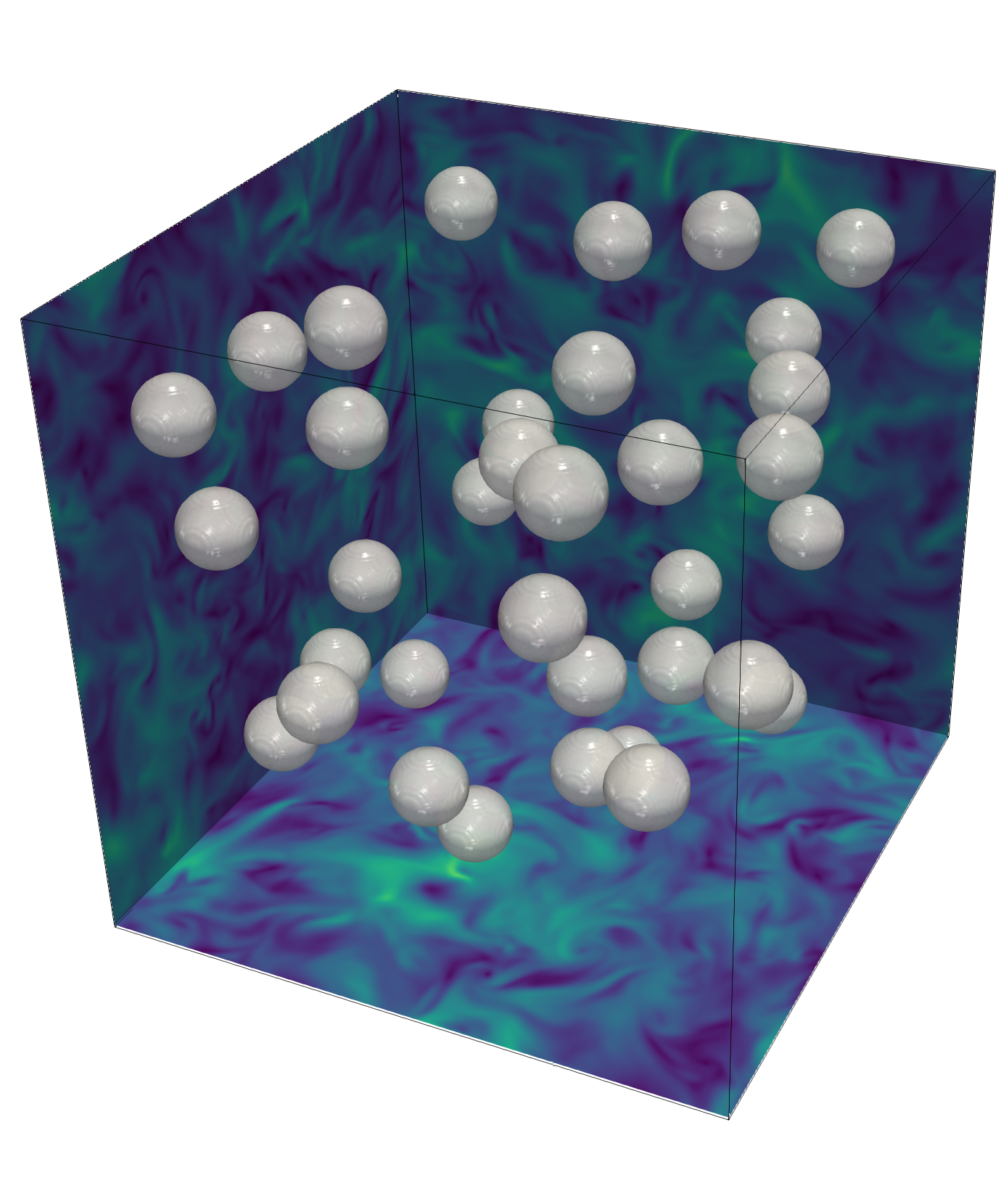}
    \put(-110,130){{$t_0$}}
    \includegraphics[width=0.3\textwidth]{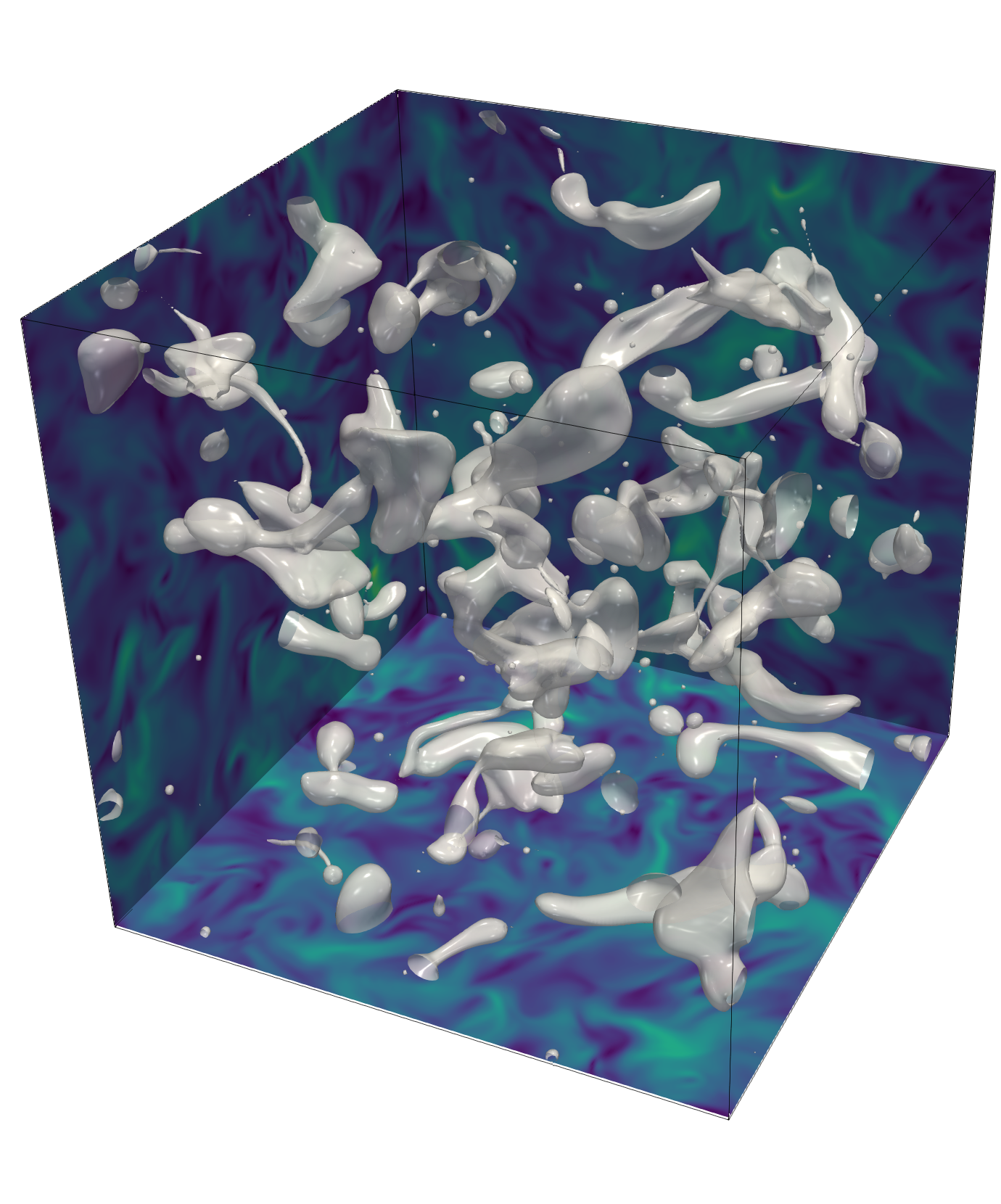}
    \put(-110,130){{$t_1=\mathcal{T}/4$}}
    \includegraphics[width=0.3\textwidth]{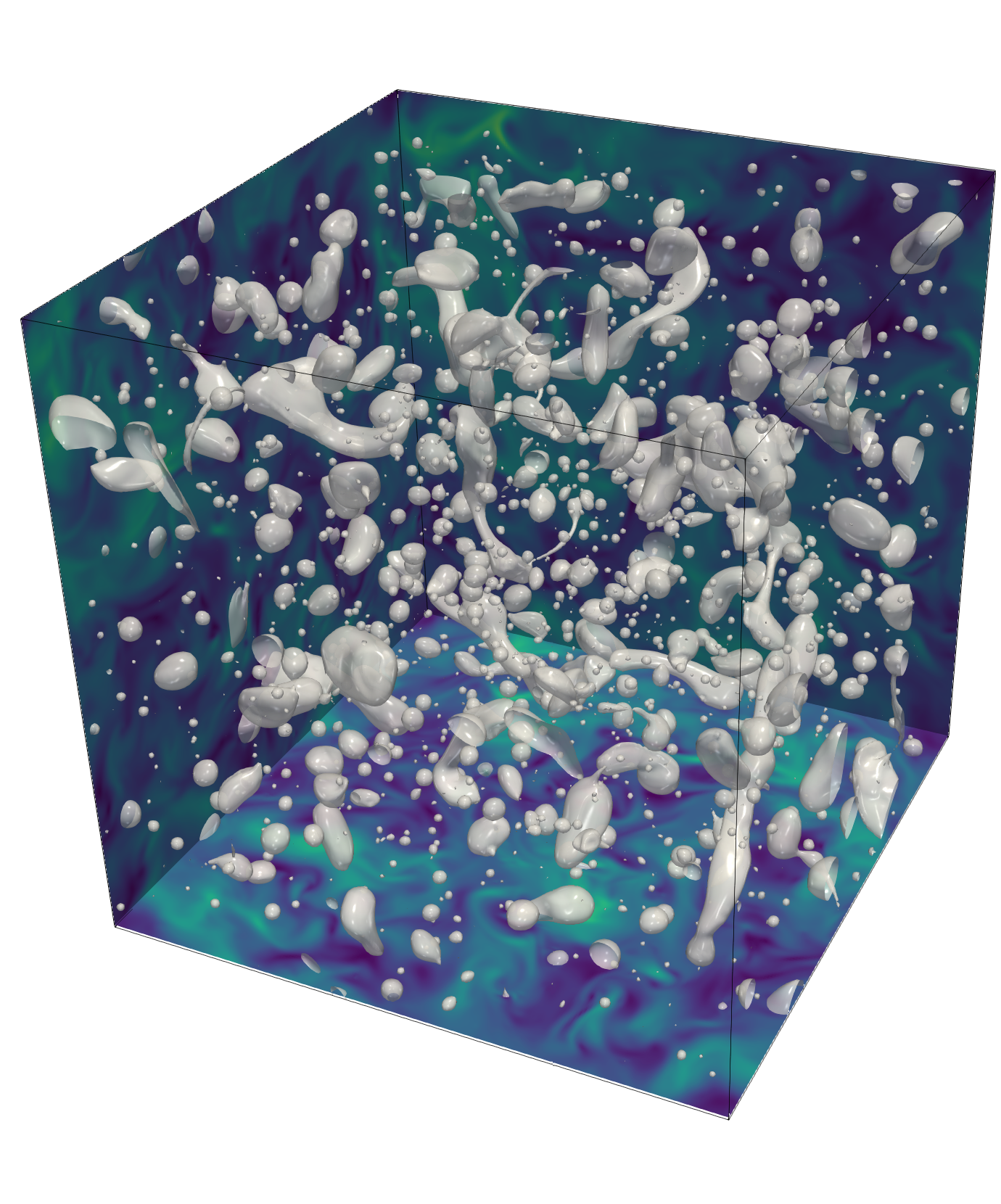}
    \put(-110,130){{$t_{2}=10\mathcal{T}$}}
    \caption{Initial evolution of the emulsion flow (example reported for case BE1). The droplet are initialized at $t_0$ in a developed turbulent field. As turbulence is maintained, breakup and coalescence start occurring ($t_1$), and statistical convergence in the DSD is achieved after a few turnover times ($t_2$)}
    \label{fig:time_seq}
\end{figure}

Visualisations of the transient phase to reach the final steady state are reported in \Cref{fig:time_seq} for the reference case BE1 with $\alpha=0.03$.  The simulation starts at $t_0$ using the fully developed single-phase HIT field from case SP2. 
The dispersed phase is initialised as an ensemble of spheres, which soon deform in the flow as shown in panel b) pertaining time $t_1=\mathcal{T}/4$. At statistical convergence, $t \approx10\mathcal{T}$, when statistics are collected, we observe a poly-dispersed distribution of asymmetric droplets.
Note finally that for $\alpha \leq 10\%$ the simulations are initialised using spherical droplets of size $d_0\approx 0.12L$, while a single spherical droplet of initial size $d_0=(6\alpha L^3/\pi)^{1/3}$ was used for larger values of $\alpha$. We have checked that the initial distribution has no effect on the final droplet size distribution, as  
also reported  in \cite{Mukherjee2019} for a similar configuration.

\subsection{Observables,  phase-averaged energy balance and scale-by-scale budget}
\label{subsec:met:obs}

In this section, we discuss the theoretical tools and the physical observable that will be discussed throughout the study.
The objective of this study is to understand the turbulence modulations induced by a second phase, focusing on comparisons of the 
energy spectra and the SBS analysis. 
In particular, we will consider 
the Taylor scale Reynolds number, the energy spectra, the phase-averaged energy balance, Probability Density Functions (PDF) of velocity fluctuations and vorticity, and the SBS budgets. The Taylor scale Reynolds number is defined as
\begin{equation}
    Re_\lambda =  \left(\frac{u_i u_i}{3}\right)^{1/2} \frac{\lambda}{\nu},
    \label{eq:relam}
\end{equation}
where  $\lambda=(5\nu u_i u_i/\varepsilon)^{1/2}$ is the Taylor scale, with the energy dissipation rate computed as $\varepsilon=\nu\partial_i u_j \partial_j u_i$. For the reference single-phase flow $Re_\lambda=137$.
Here, we compute  $\varepsilon$ and all the relevant observables at each computational grid point and then average in space and time. This procedure is required due to material properties discontinuities when $\mu_d/\mu_c$ is varied. Note that, from now on, $\varepsilon$ will denote the space-time averaged value. 

For the a multiphase flow, the energy balance is obtained by multiplying \Cref{eq:DNSmom} by the velocity $u_i$
\begin{equation}
\rho\left(\frac{\partial_t u_i u_i}{2} + \frac{\partial_j u_i u_i u_j}{2}\right) = 
-\partial_i u_i p + \mu\partial_j u_i \partial_i u_j  + \partial_j \mu u_i \left(\partial_j u_i + \partial_i u_j\right)
+u_i f_\sigma + u_i f_\sigma.
\label{eq:enBal1}
\end{equation}
We define the volume average as
\begin{equation}
\langle \cdot \rangle_m = \frac{1}{\mathcal{V}_m}\int_{\mathcal{V}_m} \cdot \, d\mathcal{V},
\label{eq:volAvg}
\end{equation}
where the subscript $m$ represents an integral over the dispersed phase $d$, the carrier phase $c$ or the total volume, if omitted. Applying the operator $\langle\cdot\rangle$ to \Cref{eq:enBal1} leads to
\begin{subequations}
	\begin{eqnarray}
	&\rho\partial_t  k = \mathcal{P}-\varepsilon + \Psi_\sigma  \\
	& k=\langle u_i u_i/2 \rangle, \quad \mathcal{P} = \rho\langle u_i f_i\rangle, \quad \varepsilon =  \langle \nu \partial_j u_i \partial_i u_j\rangle, \quad \Psi_\sigma = \langle u_i f_i\rangle;
	\end{eqnarray}
	\label{eq:enBalSP}
\end{subequations}
where $k$ is the turbulent kinetic energy.

Due to the homogeneity of the HIT configuration, the transport term arising from the nonlinear transport in \Cref{eq:enBal1} vanishes. Further details on its derivation for the case of emulsions can be found in \cite{Dodd2016,Rosti2020}. It can be proven that $\Psi_\sigma\propto \partial_t\mathcal{A}$ \citep{Dodd2016} (with $\mathcal{A}$ the total interface area) and that the contribution of the surface tension to the total energy variation also vanishes,  since the time derivative is zero at the statistically stationary state.  Hence, we obtain that the production term $\mathcal{P}$ is perfectly balanced by the energy dissipation $\varepsilon$.

Next, we consider the phase-averaged energy balance, following the approach described in \cite{Dodd2016,Rosti2020}. 
Averaging  \Cref{eq:enBal1} on each phase (i.e. enforcing eq.~\ref{eq:volAvg}), we obtain the phase-average energy balance:
\begin{subequations}
	\begin{eqnarray}
		&\rho\partial_t k_m = \mathcal{P}_m - \varepsilon_m + \mathcal{T}^\nu_m +  \mathcal{T}^p_m\\
		\label{eq:enBalMP}
		&k_m=\langle u_i u_i/2 \rangle_m, \quad \mathcal{P}_m = \rho\langle u_i f_i\rangle_m, \quad \varepsilon_m = \langle \nu \partial_j u_i \partial_i u_j\rangle_m, \\
		&\mathcal{T}^\nu_m = \langle \partial_j \mu u_i \left(\partial_j u_i + \partial_i u_j\right)\rangle_m \quad \mathcal{T}^p_m = -\langle\partial_i u_i p\rangle_m.
		\label{eq:enBalMP_terms}
	\end{eqnarray}
\end{subequations} 
Here, $\mathcal{P}_m$ and $\varepsilon_m$ indicate production rate and viscous dissipation rate per unit volume
in each phase. 
The terms $\mathcal{T}^\nu_m$ and  $\mathcal{T}^p_m$ are the viscous and pressure transport densities and represent the coupling between the two phases; when the sum of these two, $\mathcal{T}_m = \mathcal{T}^\nu_m +\mathcal{T}^p_m$ is positive, energy is absorbed from phase $m$, when negative energy is transferred to the other phase. Again, in statistical stationary conditions, $\partial_t k_m\approx 0$.

We now move to spectral space and present the SBS balance. This is derived for the two-fluid flows following the formulation in \cite{Olivieri2020a,Olivieri2020}; for more details the reader is refereed to  \cite{Frisch1995,Alexakis2018}. 
Taking the Fourier transform of the momentum equations (eq.~\ref{eq:DNSmom}), we obtain 
\begin{equation}
    \partial_t \Tilde{u_i} + \Tilde{G}_i = -\mathrm{i}\kappa\widetilde{p/\rho} - \Tilde{V}_i + \Tilde{f_\sigma} +\Tilde{f_i}.
    \label{eq:NSspec}
\end{equation}
Denoting the  Fourier transform of a quantity $J(x_i,t)$ as $\Tilde{J}(\kappa_i,t)=\mathscr{F}\{J(x_i,t)\}$, with $\kappa_i$  the $i$th component of the wavelength vector,
in the expression above $\Tilde{G}_i = \mathscr{F}\{u_j\partial_j u_i\}$ and  $\Tilde{V}_i = \mathscr{F}\{\partial_i(\nu[\partial_i u_j + \partial_j u_i]\} $.
 Note that as the viscosity $\mu$ is a function of space and time, we actually compute the dissipation term in physical space
to avoid a convolution in the spectral space. 
We next multiply \Cref{eq:NSspec} with the complex conjugate of the velocity $\Tilde{u}_i^*$ and drop the pressure term by imposing the incompressibility condition $\kappa_i\Tilde{u}_i=0$ as in this work $\rho_c=\rho_d=1$.  Multiplying the complex conjugate of \Cref{eq:NSspec} by $\Tilde{u}_i$, summing the equations obtained for $\Tilde{u}$ and $\Tilde{u}^*$ and averaging in time, we finally obtain
\begin{equation}
    \partial_t E(\kappa_i) = T(\kappa_i) + \mathcal{D}(\kappa_i) +\mathcal{S}_\sigma(\kappa_i) + \mathcal{F}(\kappa_i)
    \label{eq:spectrBal}
\end{equation}
where
\begin{itemize}
    \item  $E = \langle\Tilde{u}_i\Tilde{u}_i^*\rangle_t$ is the time-averaged kinetic energy in the spectral domain, whose time derivative is zero at statistical steady state;
    \item  $T = -\langle\Tilde{G}_i\Tilde{u}_i^* + \Tilde{G}_i^*\Tilde{u}_i\rangle_t$ is the time-averaged energy transfer due to the non-linear term;
    \item  $\mathcal{D} = -\langle\Tilde{V}_i\Tilde{u}_i^* + \Tilde{V}_i^*\Tilde{u}_i\rangle_t$ is the time-averaged viscous dissipation;
    \item  $\mathcal{S}_\sigma = -\langle\Tilde{f_\sigma}_i\Tilde{u}_i^* + \Tilde{f_\sigma}_i^*\Tilde{u}_i\rangle_t$ is the time-averaged work of the surface tension force at the different scales;
    \item  $\mathcal{F} = \langle\Tilde{f}_i\Tilde{u}_i^* + \Tilde{f}_i^*\Tilde{u}_i\rangle_t$ is the time-averaged energy input due to the large-scale forcing.
\end{itemize}
All of the above are three-dimensional fields in spectral space.  Note that, at steady state when the total interfacial area is constant,  $\mathcal{S}_\sigma$ integrates to zero \citep{Dodd2016}  so that this term can be effectively seen as an energy transport due to the surface tension.
Finally, we perform a spherical-shell integral in spectral space and express each term in the budget as a function of the magnitude of the wavevector  $\kappa$. This operation results in, e.g.,\
\begin{equation}
	T(\kappa) = \sum_{\kappa<|\kappa_i|<\kappa+1} T(\kappa_i).
	\label{eq:shellInt}
\end{equation}
This term represents the shell-to-shell energy transfer function for the non-linear term of the momentum equation and similarly for the other terms above. If we  instead 
perform the integration over a sphere (i.e. for all $|\kappa_i|<\kappa$) we obtain the cumulated SBS budget:
\begin{equation}
	\partial_t\sum_{|\kappa_i|<\kappa}E(\kappa_i) = \Pi(\kappa) + \sum_{|\kappa_i|<\kappa} \mathcal{D}(\kappa_i) + \Pi_\sigma (\kappa) + \sum_{|\kappa_i|<\kappa} \mathcal{F} (\kappa_i).
	\label{eq:spectrBalCum}
\end{equation}
In the expression above, the fluxes $\Pi(\kappa) = \sum_{|\kappa_i|<\kappa} T(\kappa_i)$ and $\Pi_\sigma(\kappa) = \sum_{|\kappa_i|<\kappa} \mathcal{S}_\sigma(\kappa_i)$,
indicate the energy flux from all the largest scale to $\kappa_i$, 
and are typically used  to study scalings in the inertial range (where $\Pi(\kappa)=\varepsilon$) and the direction of the energy cascade \citep{Alexakis2018}. 
The remaining terms represent the energy injected and the dissipation at all scales below $\kappa_i$. 
The cumulative SBS budget can easily be related to the energy balance in the physical domain: \Cref{eq:spectrBalCum} and \Cref{eq:enBalSP} are equivalent for $\kappa=\kappa_{max}$, hence it can be easily demonstrated that $\Pi(\kappa_{max})=\Pi_\sigma(\kappa_{max})=0$ and $\varepsilon = \mathcal{D}(\kappa_{max})=  \mathcal{P} = \mathcal{F} (\kappa_{max})$.
In this work, we will mostly show the shell-by-shell energy budget (eq.~\ref{eq:spectrBal}, integrated using \ref{eq:shellInt}, referred to SBS if not differently specified), as more suited for detailed comparisons at each scale, while the cumulative energy budget is used for the single phase flow only.

\subsection{Analysis of the reference single-phase flow and grid convergence}
\label{subsec:met:sp-case}
\begin{figure}
    \centering
    \includegraphics[width=0.5\textwidth]{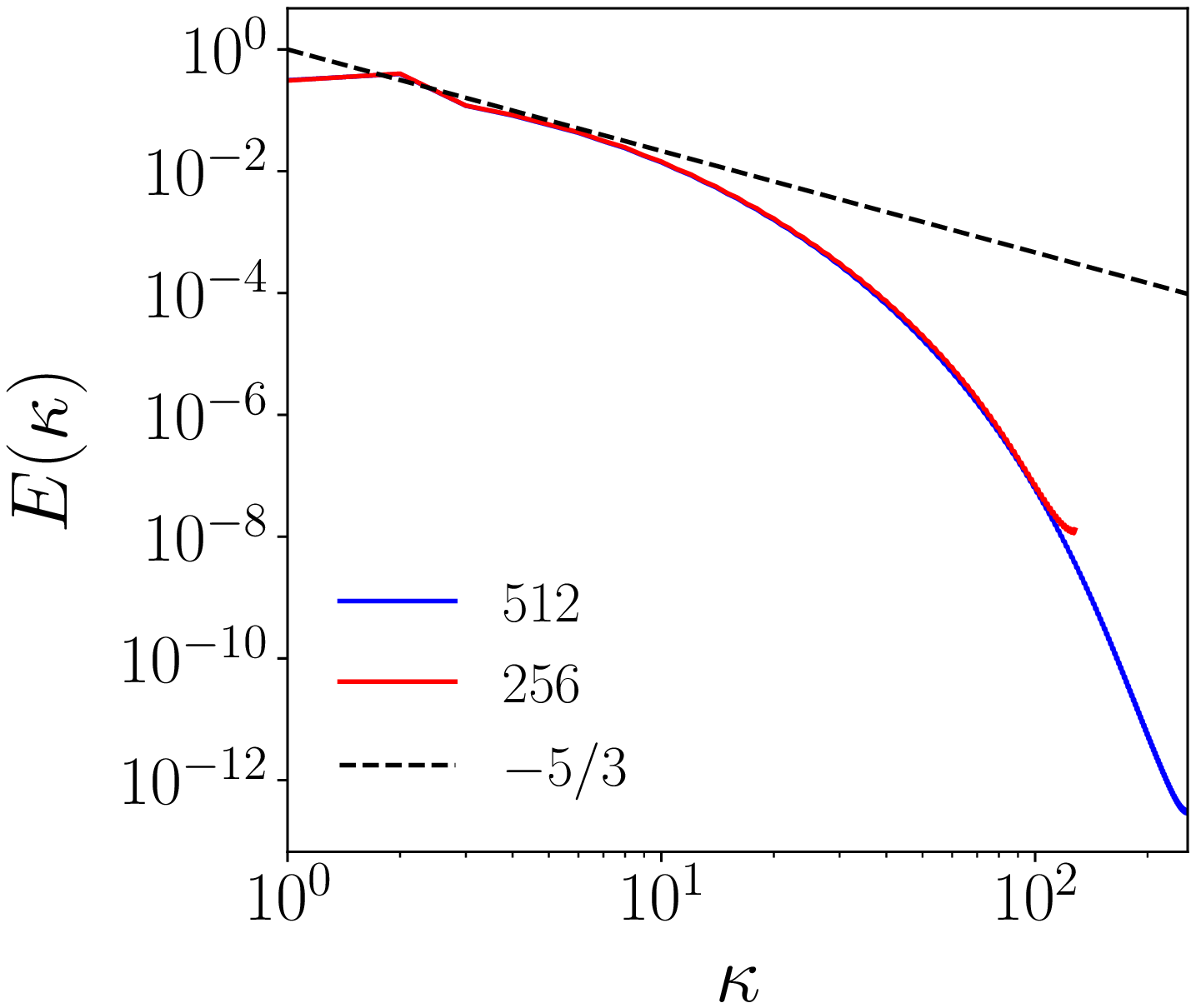}
    \put(-190,140){(\textit{a})}
    \includegraphics[width=0.5\textwidth]{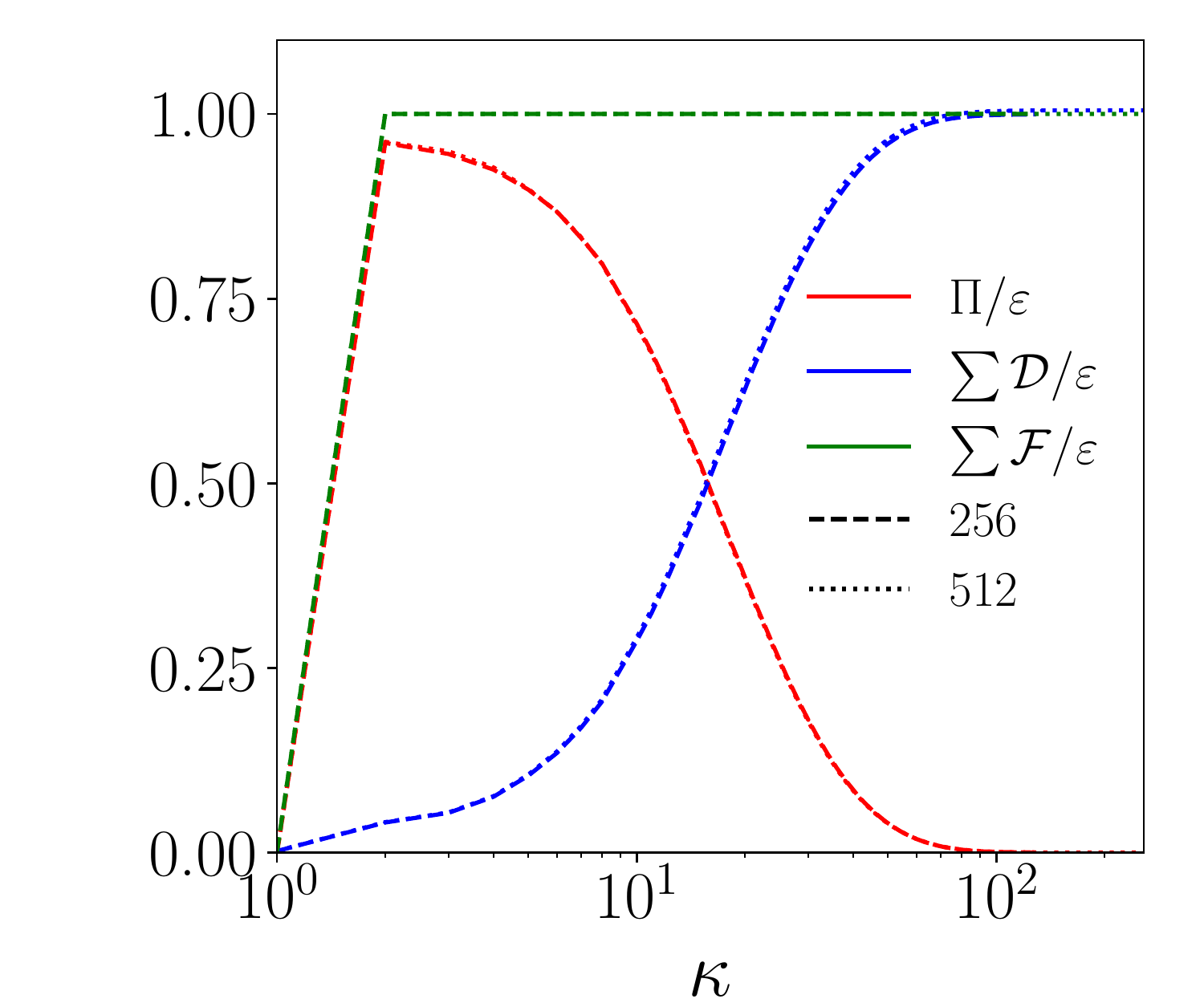}
    \put(-190,140){(\textit{b})}
    \caption{Spectral analysis of single-phase HIT (cases SP1 and SP2) and grid-resolution analysis. (\textit{a}) Spectra for single-phase simulation with $N=256$ (red continuous line) and $N=512$ (blue continuous line) against the -5/3 law for the inertial subrange (dashed black line). (\textit{b}) Energy scale-by-scale cumulative balance for the single-phase simulations with $N=256$  and $N=512$ grid points.}
    \label{fig:spectBal_sp}
\end{figure}

We will now motivate the choice of the resolution adopted for the emulsion simulations, i.e.\,  $N=512$ points in each direction. To this aim, we will first analyze the behavior observed in single phase turbulence.
The energy spectra and the cumulative SBS balance pertaining the single-phase flow are shown in \Cref{fig:spectBal_sp}.
A good agreement between the cases SP1 and SP2 is evident in panel (a).
The $\kappa^{-5/3}$ law for the inertial range extends over almost a decade, showing a fully developed turbulent flow at a moderately high Reynolds number. 
The 2 cases yield the same result in terms of $ \varepsilon $,
with a relative error of less than $5\%$. 
The cumulative SBS balance shows that, due to the moderate $Re_\lambda$, the viscous term $\mathcal{D}$ is not negligible already at large scales, where it dissipates approximately $3\%$ of the injected energy. This is in agreement with the observation that a fully developed inertial range is only partially observable for $Re_\lambda\lesssim200$ in single-phase turbulence \citep{Ishihara2009}. A substantial dissipative range is present for $\kappa\ge 10^2$, indicating an accurate computation of  the smallest scales. In this region we can observe that $\Pi(\kappa_{max})=0$ and $\sum_{|\kappa_i|<\kappa} \mathcal{D}(\kappa_i)=\varepsilon$.
As a consequence of the imposed ABC forcing, energy injection is clearly observed for $\kappa=2$. 
The SBS budget shows almost no difference between the results from the two grid resolutions, and that all relevant processes are already accurately captured at the lower resolution, $N=256$, down to the smallest scales.

\begin{figure}
	\centering
	\includegraphics[width=0.5\textwidth]{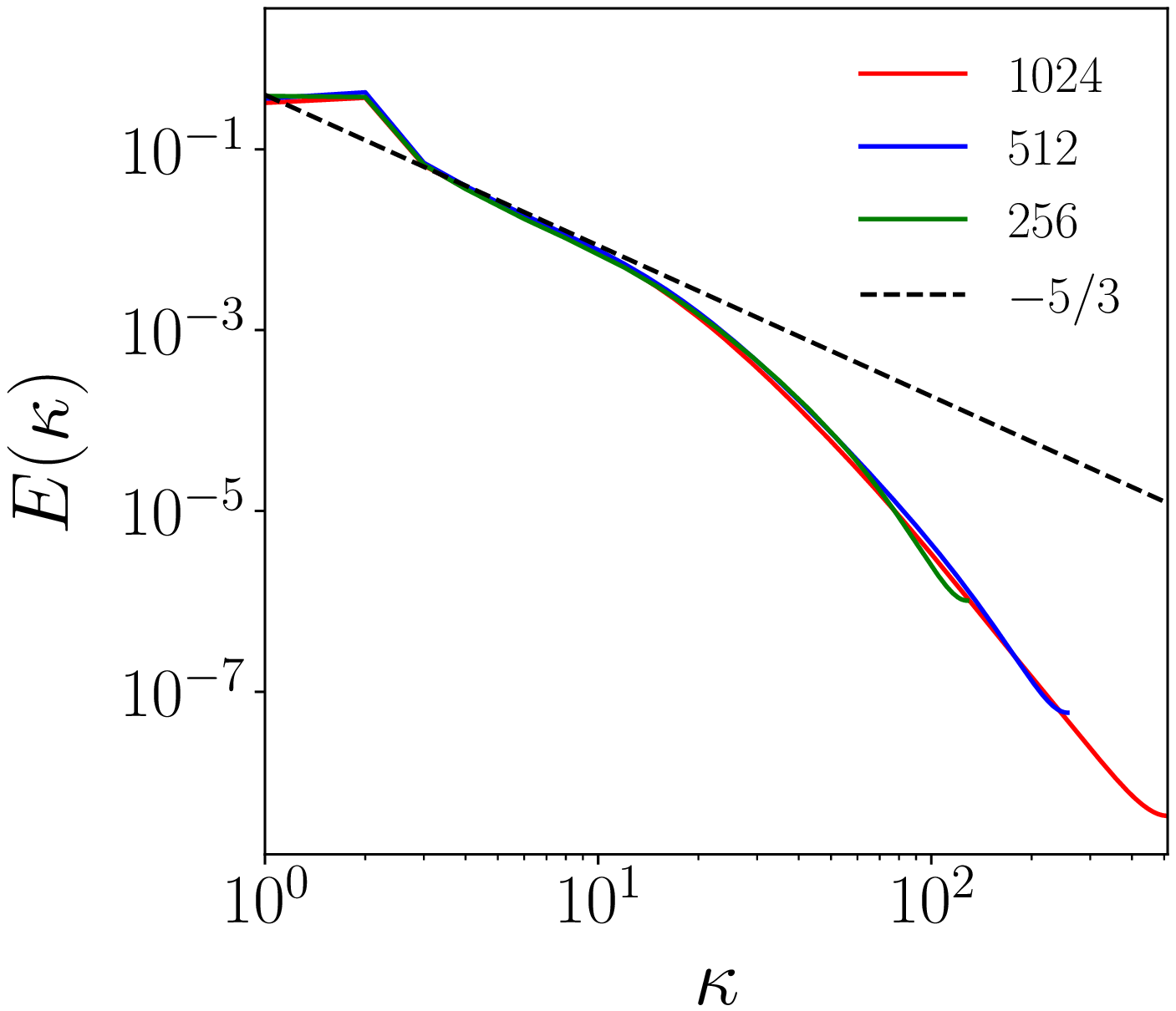}
	\put(-190,140){(\textit{a})}
	\includegraphics[width=0.5\textwidth]{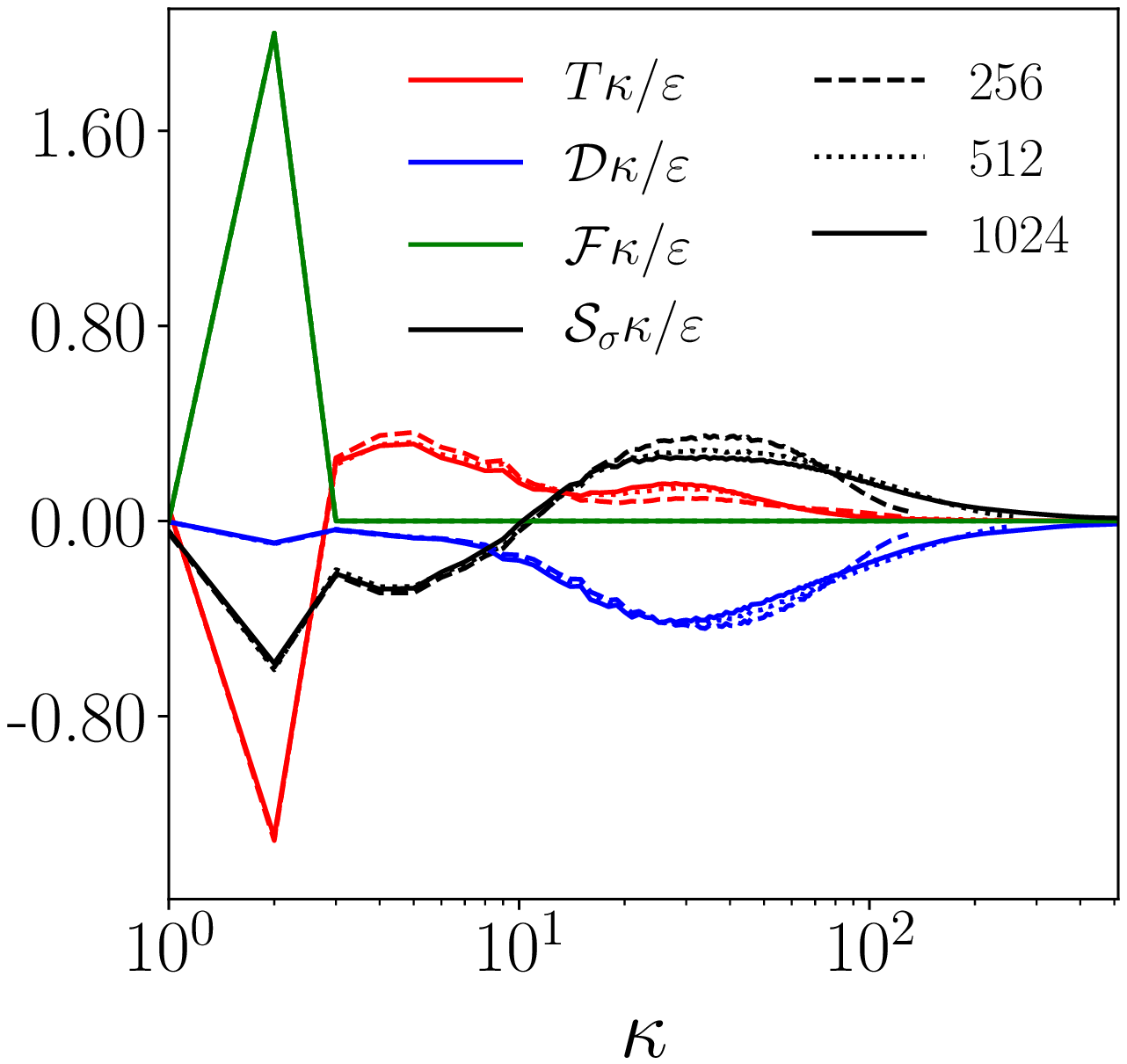}
	\put(-190,140){(\textit{b})}
	\caption{Grid resolution study on spectral analysis of multiphase simulations  (cases C14, C24 and C34 at $\alpha=0.5$, $We_\mathcal{L}=42.6$ and $\mu_d/\mu_c=1$). (\textit{a}) Energy spectra for simulation with $256^3$ (green continuous line), $512^3$ (blue dashed line) and $1024^3$ (continuous red line) compared against the -5/3 law for the inertial subrange (dashed black line). (\textit{b}) Energy scale-by-scale balance for multiphase simulations with grids $256^2$, $512^3$ and $1024^3$. }
	\label{fig:spectBal_mp}
\end{figure}

To investigate convergence of the multiphase flows, we consider the case $We_\mathcal{L}=42.6$, $\mu_d/\mu_c=1$ and $\alpha=0.5$ for three different resolutions, from $N=256$ to $N=1024$ corresponding to cases C14, C24, and C34, see \Cref{fig:spectBal_mp}. This configuration was selected because it is the one with the largest interfacial area and largest fluctuations, $d\mathcal{A}/dt$, and hence where larger errors in the averaged energy budget  are expected.
 Nevertheless
the spectra in panel (\textit{a}) do not seem to be significantly affected by the grid resolution and the dissipative range is observed also at the lowest resolution, $N=256$. 

A more stringent test  is the convergence of the SBS budget, here (and hereafter)  shown in its shell-by-shell form, see panel (\textit{b}) 
 where all terms are normalized by $\varepsilon$ and pre-multiplied by the wavelength $\kappa$ to improve readability. 
 Comparing the data at different resolution, the energy injection at large scales $\mathcal{F}$  and the energy transfer by the non-linear term $T$ are almost identical in the inertial range. The energy dissipation $\mathcal{D}$ and the transfer due to interfacial forces $\mathcal{S}_\sigma$ display some differences 
 for $\kappa\gtrsim 10 $. 
 If we consider the integral contribution from the surface tension, which should theoretically be zero, 
 $\Pi_\sigma(\kappa_{max})/\varepsilon\approx 0.08$ for the lowest resolution, $N=256$; this value decreases for $N=512$ and remains almost constant at $N=1024$,  $\Pi_\sigma(\kappa_{max})/\varepsilon\approx 0.04$. 
 It is worth underlying that this is the largest error encountered among all cases,  since $\Pi_\sigma(\kappa_{max})/\varepsilon\approx 0.01$ for most of the other cases discussed in this study. Overall, the energy not resolved by the the simulation with a grid of $N=512$ and the differences in the SBS transfer functions can be considered as negligible for the scope of the present study, where we wish to primarily examine the energy transfer towards the smallest scales.

The data in figure \ref{fig:spectBal_mp} highlight already important features of emulsions in HIT, which will be consistently observed in all cases studied. 
First, the energy at the smallest scales increases with respect to the single-phase case (panel a). Secondly,  the presence of the interface alters the behavior of the turbulent flow,  with the surface tension forces $\mathcal{S}_\sigma$ transferring energy from large scales towards the dissipative range. Because of this, 
the total energy transported by the non-linear term $T$ reduces and the dissipative range extends towards smaller scales, where we observe a balance between the work of surface tension and viscous dissipation.  These modified flow features are similar to what found by \citet{Olivieri2020a,Olivieri2020} for fiber suspensions in turbulent flows. 

\section{Results}
\label{sec:results}

\subsection{Emulsions at different  volume fractions}
\label{sec:res:alpha}
\begin{figure}
    \centering
    \includegraphics[width=0.3\textwidth]{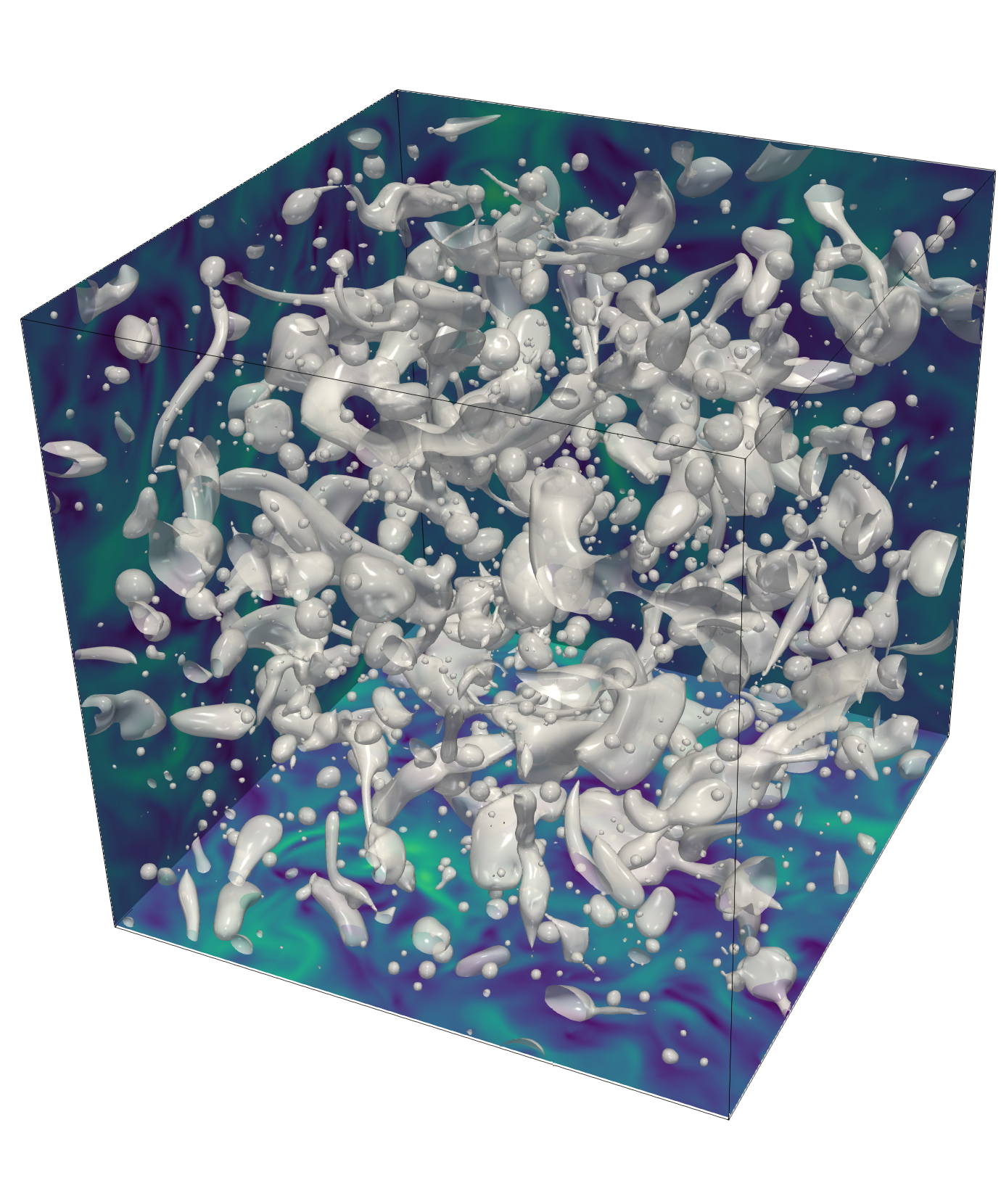}
    \put(-100,140){{$\alpha=0.06$}}
    \includegraphics[width=0.3\textwidth]{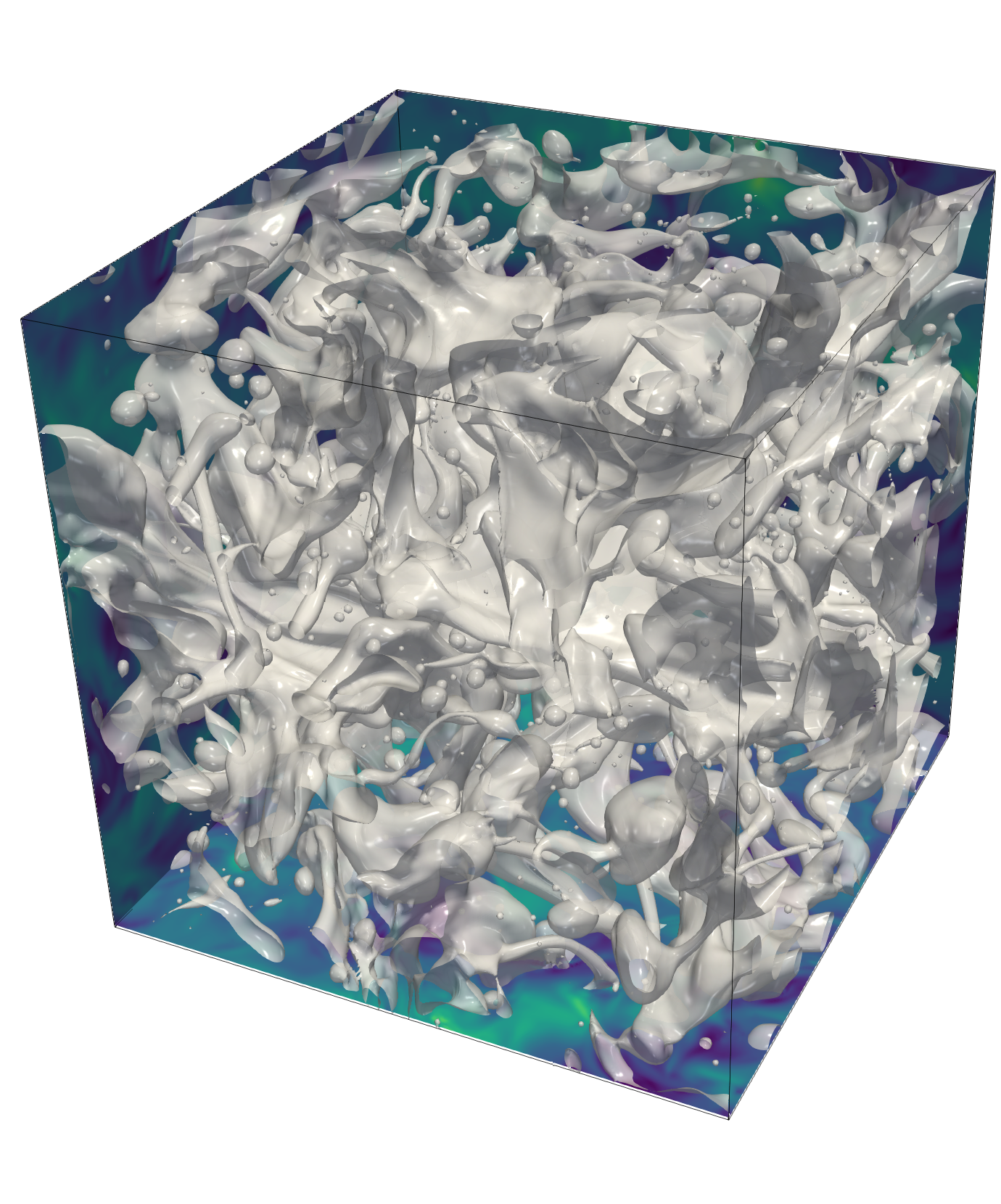}
    \put(-100,140){{$\alpha=0.2$}}
    \includegraphics[width=0.3\textwidth]{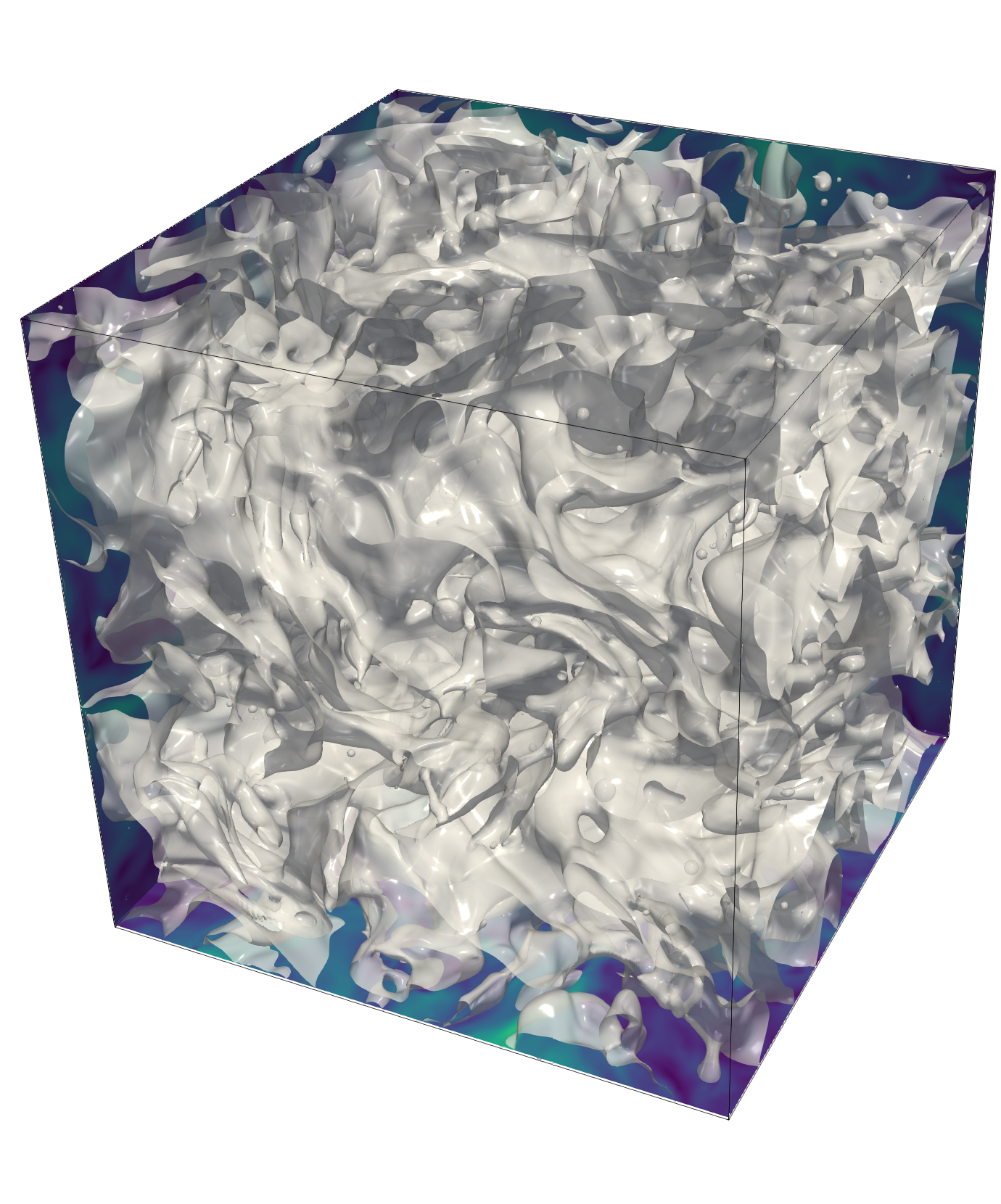}
    \put(-100,140){{$\alpha=0.5$}}
    \caption{ Render of the two-fluid interface (corresponding to the value of the VOF function $\phi=0.5$) for different values of the volume fractions $\alpha$ (left to right, 0.06, 0.2 and 0.5).  The vorticity fields are shown on the box faces on a planar view. }
    \label{fig:alphaRender}
\end{figure}

We first examine  the influence of the dispersed-phase volume fraction on the turbulent flow, cases BE1 and Cxx in \Cref{tab:testMat}, corresponding to increasing values of $\alpha$ from 3 to 50\%. A render of the cases discussed here is shown in \Cref{fig:alphaRender}, where the isocontour of VOF fields are shown for volume fractions $0.06$, $0.2$ and $0.5$. 

\begin{figure}
	\centering
	\includegraphics[width=1\textwidth]{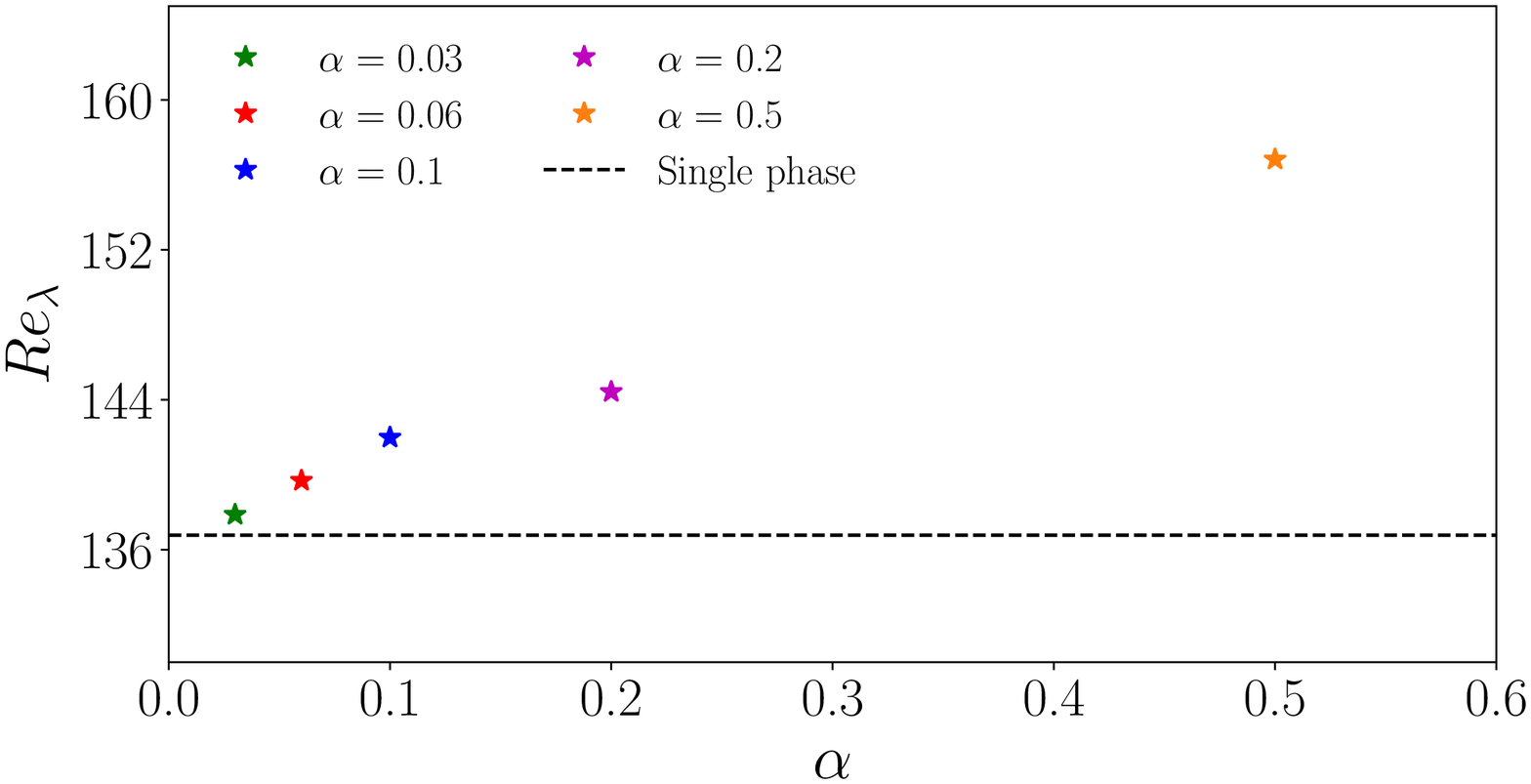}
	\put(-180,65){\includegraphics[width=0.4\textwidth]{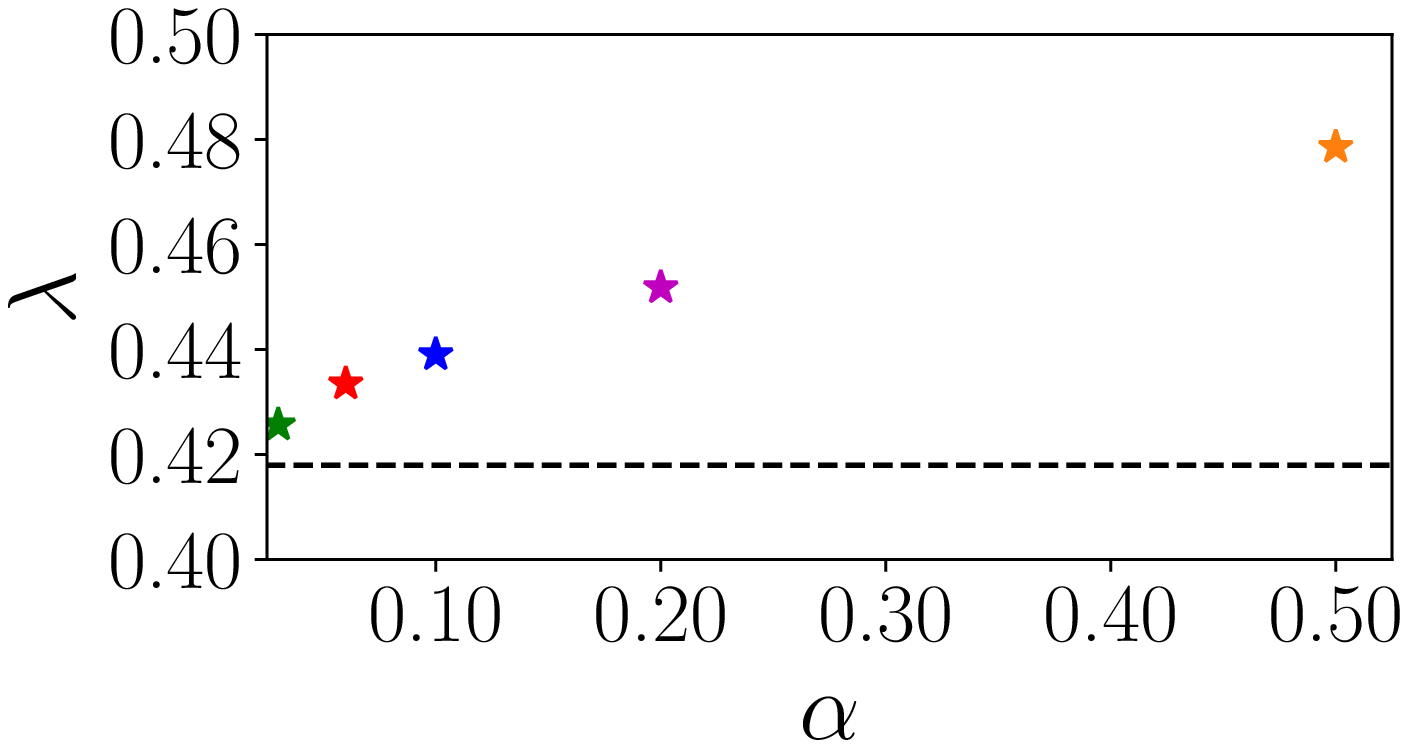}}
	
	\caption{Taylor Reynolds number, $Re_\lambda$,  versus the dispersed-phase volume fraction $\alpha$, for viscosity ratio $\mu=1$ and density raio $\rho=1$.
	The inset shows the Taylor scale, $\lambda$ versus the different values of $\alpha$ under investigation.}
	\label{fig:relam_alpha}
\end{figure}

The modulation of the turbulence is first quantified in terms of integral quantities.
\Cref{fig:relam_alpha} shows  $Re_\lambda$, computed according to \Cref{eq:relam}, versus the volume fraction $\alpha$. 
$Re_\lambda$ increases  almost linearly with $\alpha$, by approximately 15\% for $\alpha=0.5$.  
A similar trend is found for $\lambda$, as shown in the inset of \Cref{fig:relam_alpha}. Considering that the average of $\varepsilon$ and $k$ is approximately 
constant in all cases (variations of $\pm 3\%$), the increase of $Re_\lambda$ and $\lambda$ is therefore due to  the local variations of the ratio $k/\varepsilon$.

\begin{figure}
	\centering
	\includegraphics[width=0.9\textwidth]{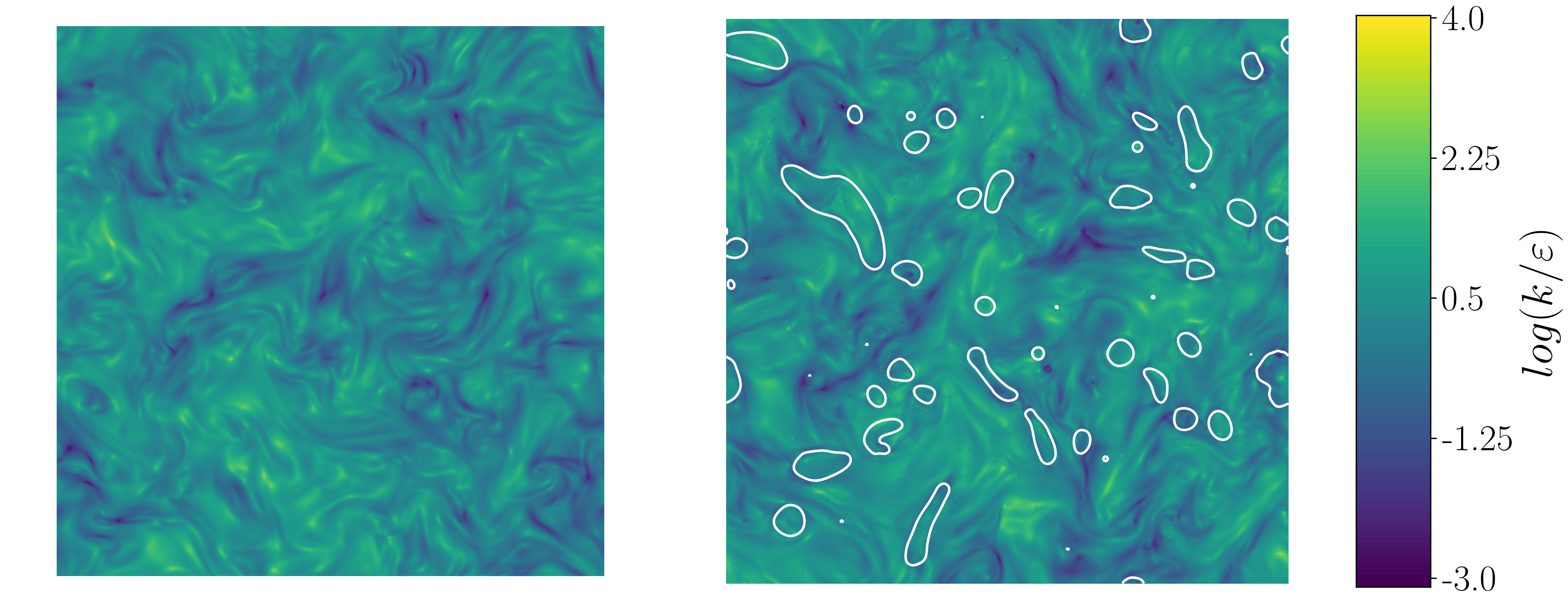}
	\put(-300,130){Single phase}
	\put(-150,130){$\alpha=0.1$}
	\put(-350,120){(\textit{a})}
	\put(-205,120){(\textit{b})}
	
	\includegraphics[width=0.9\textwidth]{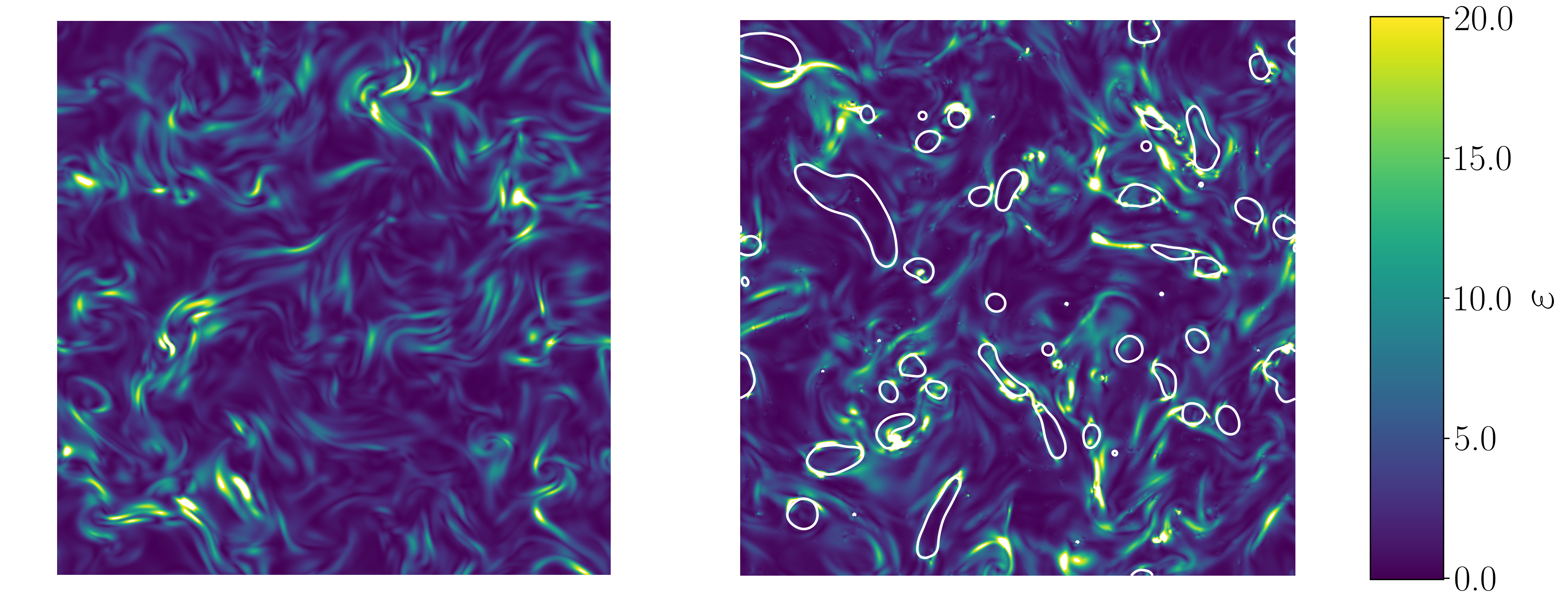}
	\put(-350,120){(\textit{c})}
	\put(-205,120){(\textit{d})}
	\caption{(\textit{a,b}) Contours of the ratio $k/\varepsilon$ with logarithmic scale in two planes. (\textit{c,d}) energy dissipation rate $\varepsilon$. The left panels present results for the single phase case SP2, while the right panels results for case BE2 ($\alpha=0.1$). The white lines represent the VOF iso-contours for $\phi=0.5$.}  
	\label{fig:planes_alpha}
\end{figure}
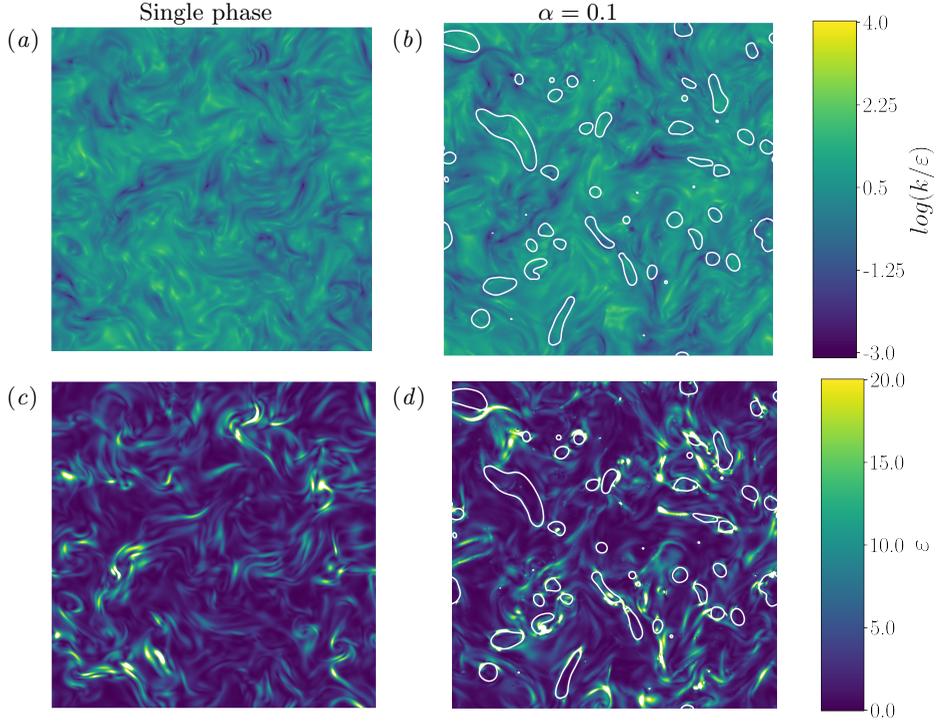

In particular, the increased values $(k/\epsilon)$ for similar averaged values of the two quantities is attributed to the increased correlation between regions of strong turbulent kinetic energy and low dissipation.
A graphical evidence is presented in \Cref{fig:planes_alpha}, where we show the instantaneous ratio $k/\epsilon$ for the single-phase (case SP2 in panel \textit{a}) and multiphase flow (case BE2 with $\alpha=0.1$ in panel \textit{b}) in logarithmic scale. The figure shows that when the dispersed phase is present,  large regions of fluid with higher $k/\epsilon$ are observed far from the droplet interface (denoted with white line).
This can be explained as follows: as the total dissipation is constant, the local increase of   $\epsilon$ near the interface, as also observed in \cite{Dodd2016}, correspond to a decrease of the dissipation rate in large portions of the fluid, those far from an interface.
Considering that the turbulent kinetic energy is less affected by the presence of the interface, the ratio $(k/\epsilon)$ increases in average.
To support this statement, \Cref{fig:planes_alpha}(\textit{c},\textit{d}) depicts the instantaneous energy dissipation rate for the same planes. In the emulsion
(panel \textit{d}), higher values of $\varepsilon$ are found close to the droplet interface and to the clustering regions, while for the single-phase flow (panel \textit{c}) no specific pattern is observed.

\begin{figure}
	\centering
	\includegraphics[width=1\textwidth]{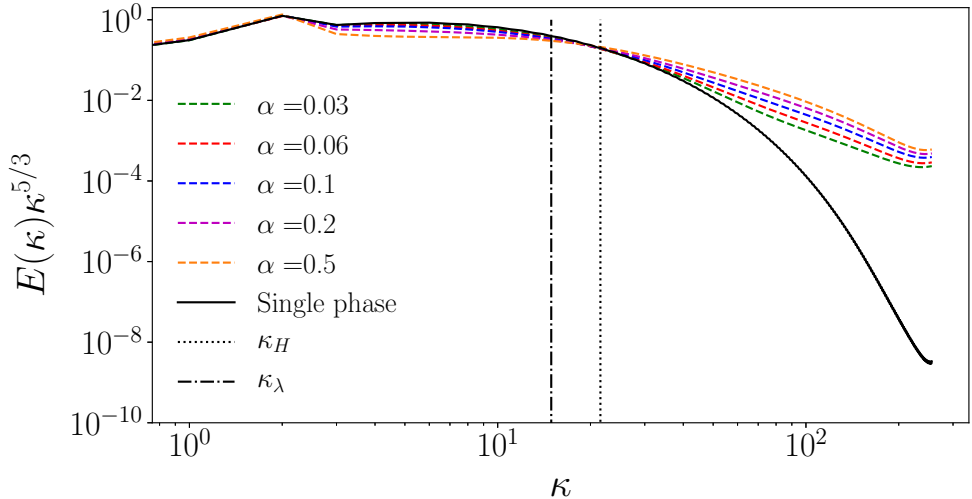}
	\caption{Compensated energy spectra for simulations at different volume fraction $\alpha$; the dot-dashed line represent Taylor scale $\lambda$, while the dotted line the Hinze scale $d_H$.}
	\label{fig:spectra_alpha}
\end{figure}

The one-dimensional  energy spectra $E(\kappa)$ multiplied by $\kappa^{5/3}$, i.e.\ the so-called compensated spectra, are displayed in \Cref{fig:spectra_alpha} for the different $\alpha$  considered.
The Taylor scale of the single-phase flow is indicated by the dot-dashed line, while the vertical dotted black line is used for the wavenumber $\kappa_H$ corresponding to the Hinze scale, defined as: 
\begin{equation}
    d_H = 0.725\varepsilon^{-2/5}(\rho_c/\sigma)^{-3/5}.
    \label{eq:hinze}
\end{equation}
Note that,  the prefactor $0.725$ in \Cref{eq:hinze} is set accordingly to the original work of \cite{Hinze1955} for emulsions in HIT conditions,  corresponding to $We_c=1.17$.

The data in \Cref{fig:spectra_alpha} reveal that  the presence of the dispersed phase reduces the energy with respect to the single-phase case (SP1)  for $\kappa<\kappa_H$. At the same time, the energy content increases at the smaller scales, $\kappa>\kappa_H$, in the dissipative range of the spectra. 
As noted in previous studies \citep{Mukherjee2019}, the amount of energy subtracted to the large scales is proportional to the volume fraction $\alpha$. Interestingly, 
the wavenumber at which the curves cross over from reduced to increased energy content corresponds to the Hinze scale. For brevity, we will 
denote as \textit{pivoting point} the wavelength where the spectra of the multiphase cases intersect the one from the single-phase reference case.

Pivoting points were not clearly observed in some previous studies on emulsions \citep{Mukherjee2019, Perlekar2019}, while they are clearly visible in others \citep{Perlekar2014,Dodd2016,Rosti2020};
this is possibly due to the
different methods used to simulate the dispersed phase: the ability of the VOF to accurately resolve the interface reduces the energy dissipation by the surface tension term in the dissipative range.  Such energy dissipation is indeed clearly observed by \cite{Perlekar2019} who present results obtained by solving the Cahn-Hilliard equation
in a diffuse-interface formulation. 
As mentioned in \Cref{subsec:met:sp-case}, these numerical artefacts do not have significant effects on the dynamics at the inertial range, while they affect the dissipative range. This aspect will be discussed also later in this section.

\begin{figure}
\centering
	\includegraphics[width=0.5\textwidth]{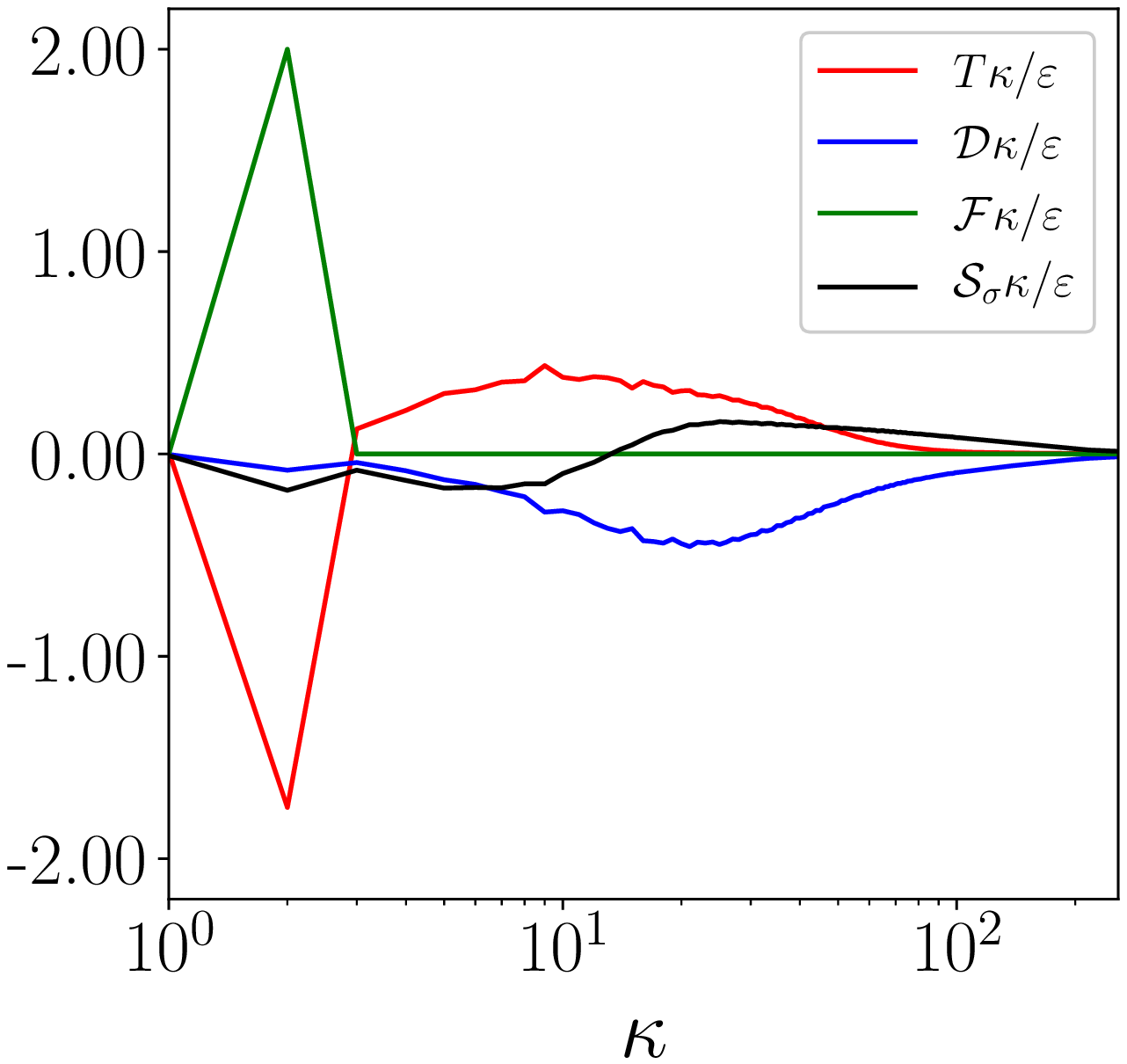}
	\put(-200,140){(\textit{a})}
	\includegraphics[width=0.5\textwidth]{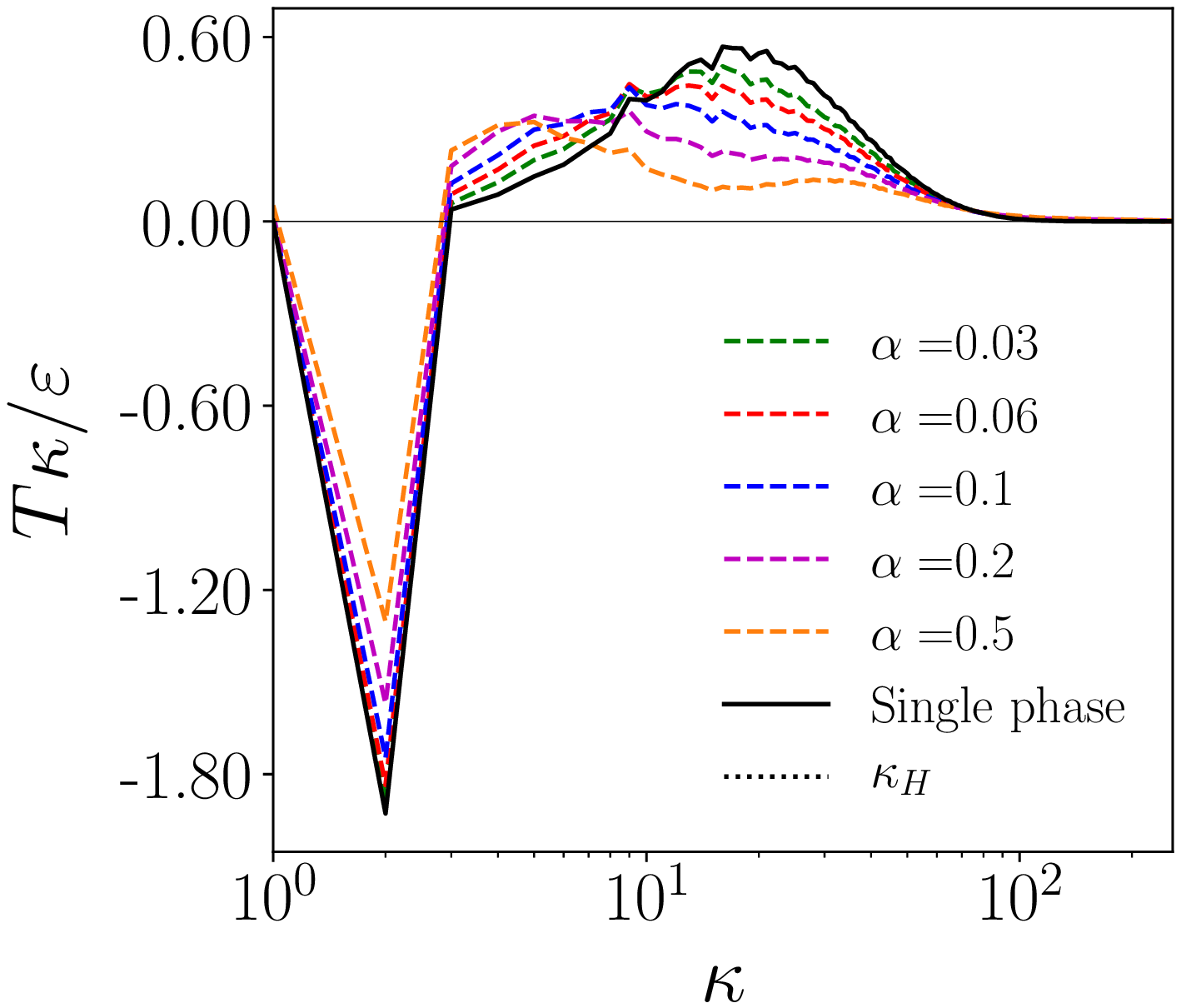}
	\put(-190,140){(\textit{b})}
	
	\includegraphics[width=0.5\textwidth]{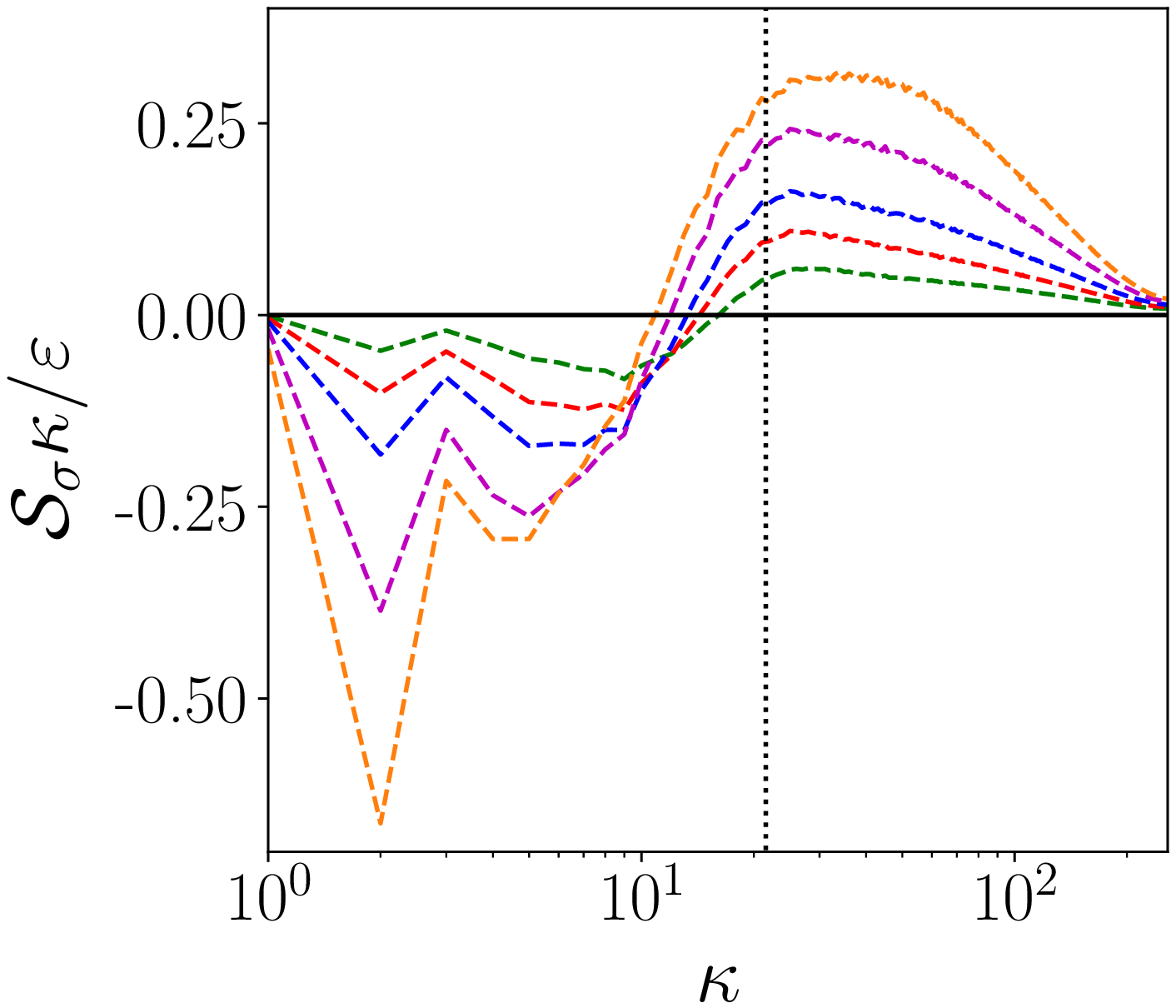}
	\put(-200,140){(\textit{c})}
	\includegraphics[width=0.5\textwidth]{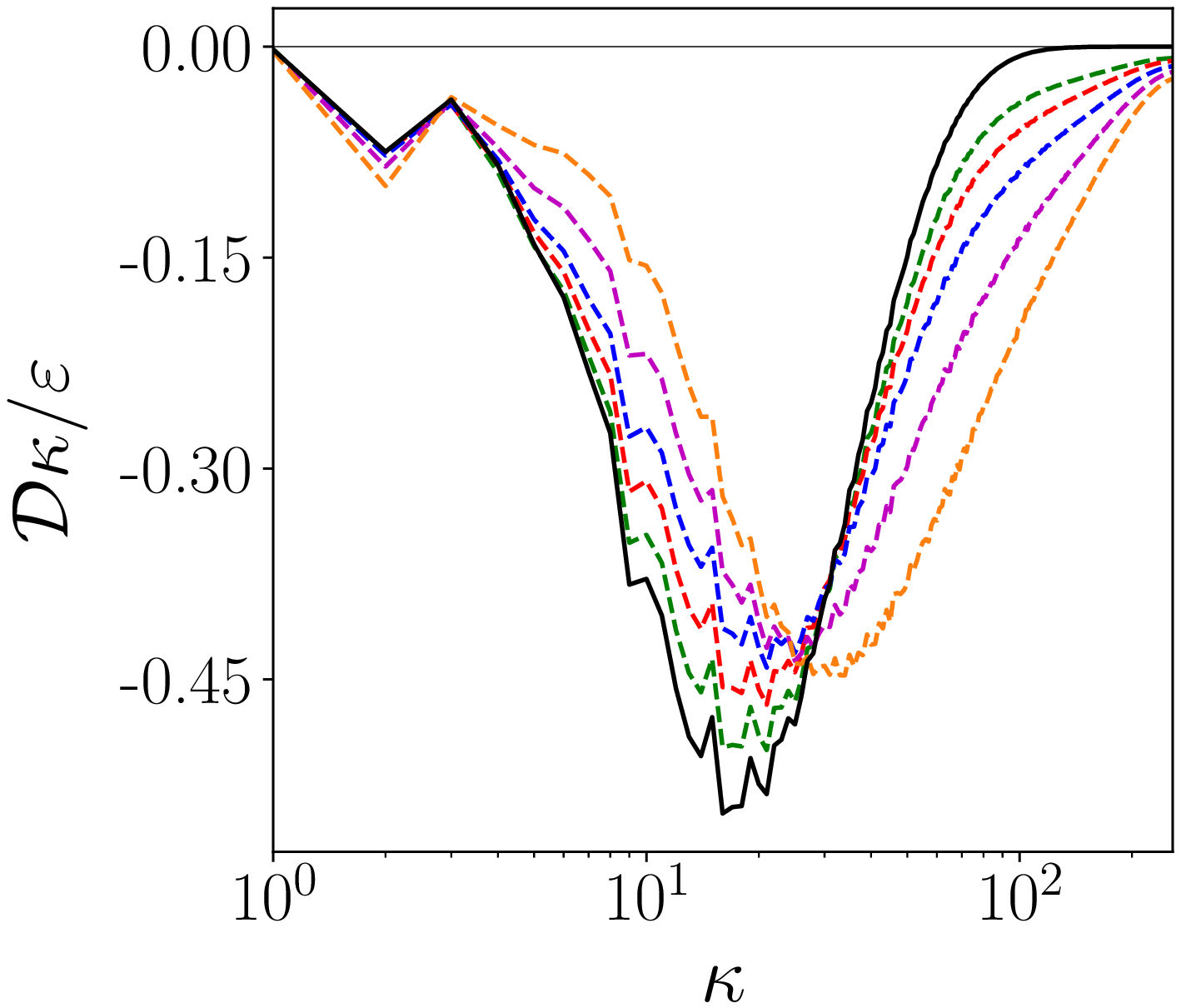}
	\put(-190,140){(\textit{d})}
\caption{Scale-by-scale energy budget for different volume fractions $\alpha$. (\textit{a}) Full energy balance for the case BE2 with $\alpha=0.1$; (\textit{b}) the energy transfer $T$ due to the non-linear terms; (\textit{c}) the energy transfer $\mathcal{S}_\sigma$ associated with the surface tension term; (\textit{d}) energy dissipation rate  $\mathcal{D}$.
}
\label{fig:balance_alpha}
\end{figure}

Insight on the energy transfer among the different scales is gained be using the SBS analysis. 
The full SBS energy budegt, i.e.\ the contributions from the different terms in \Cref{eq:spectrBal}, is displayed in 
 \Cref{fig:balance_alpha}(\textit{a}) for case BE2,  chosen as
 an illustrative example with an intermediate value of $\alpha=0.1$. 
The external forcing is injecting energy  at $\kappa=2$, which is absorbed by the non-linear transfer term  $T$, for a large majority, and by the surface tension term $\mathcal{S}_\sigma$, for a small part. The non-linear term transfers energy towards smaller scales, larger values of $\kappa$.
The surface tension term, $\mathcal{S}_\sigma$, acts as a dissipative process at large scales, where it absorbs approximately the same energy as the dissipative term $\mathcal{D}$; however for  $10<\kappa<20$, we observe a significant change in the energy transport mechanism: $\mathcal{S}_\sigma$ becomes positive, hence contributing to transferring energy towards the small scales, similarly to $T$, a process active until $\kappa_{max}$. 
It is important to note that the surface tension transport remains active also at small scales in the dissipative range, consequently extending the range of wavelengths where the dissipation term remains active. These observations confirm the previous findings   obtained in \cite{Perlekar2019} for binary mixtures.

The details of the effect of the volume fraction $\alpha$ on each term of the SBS balance are displayed in \Cref{fig:balance_alpha}(\textit{b-d}). 
We first analyze the non-linear transfer term $T$ in panel (\textit{b}). As $\alpha$ increases, $T$ absorbs progressively less energy at the injection frequency $\kappa=2$. Consequently, less energy is transferred towards smaller scales by nonlinear advection. The energy flux $\Pi$ (not shown) do not display an inverse cascade for any $\alpha$. 
Furthermore, we notice that no energy  is transferred to the far end of the dissipative range, which is resolved over a large range of scales in all cases.    

The contribution from the surface tension $\mathcal{S}_\sigma$, see  \Cref{fig:balance_alpha}(\textit{c}), confirms that interfacial stresses  absorb part of the energy injected into the domain at $\kappa=2$. The energy absorbed by the surface tension term at large scales is approximately proportional to $\alpha$. The surface tension term becomes positive at smaller scales, where energy is released. The positive peak is reached at approximately the Hinze scale for all cases. As for the energy absorption, the magnitude of the peak scales proportionally to $\alpha$. We also notice that, for any $\alpha$, the surface tension terms act also in the dissipative range, where the non-linear term $T$ is zero. 

The behavior observed so far for $T$ and $\mathcal{S}_\sigma$ provide a clear explanation for the previous observations on the energy spectra. At small wavenumber, the energy cascade produced by the non-linear energy transfer is partially inhibited by the presence of the interfacial forces. For high wavenumbers, $T$ progressively reaches zero, but the energy previously subtracted by the interfacial stresses at large scale is redistributed at small scales, which can be seen in \Cref{fig:spectra_alpha} as an energy increase at high wavenumbers.

To close the SBS balance, we examine the viscous dissipation term $\mathcal{D}$, see figure \Cref{fig:balance_alpha}(\textit{d}). 
First, we note that only a small amount of the injected energy  (less than $5\%$ for all cases) is absorbed by the dissipation term at the scale of the forcing, $\kappa= 2$.  The overall effect of the dispersed phase is to shift the energy dissipation towards smaller scales. This constitutes the natural reaction of the system to the increased activity in the dissipative range caused by the surface tension term. This behavior becomes more evident as $\alpha$ increases and progressively enhances dissipation at those small scales where the single-phase dissipation  is negligible.

\begin{figure}
\centering
	\centering
	\includegraphics[width=0.49\textwidth]{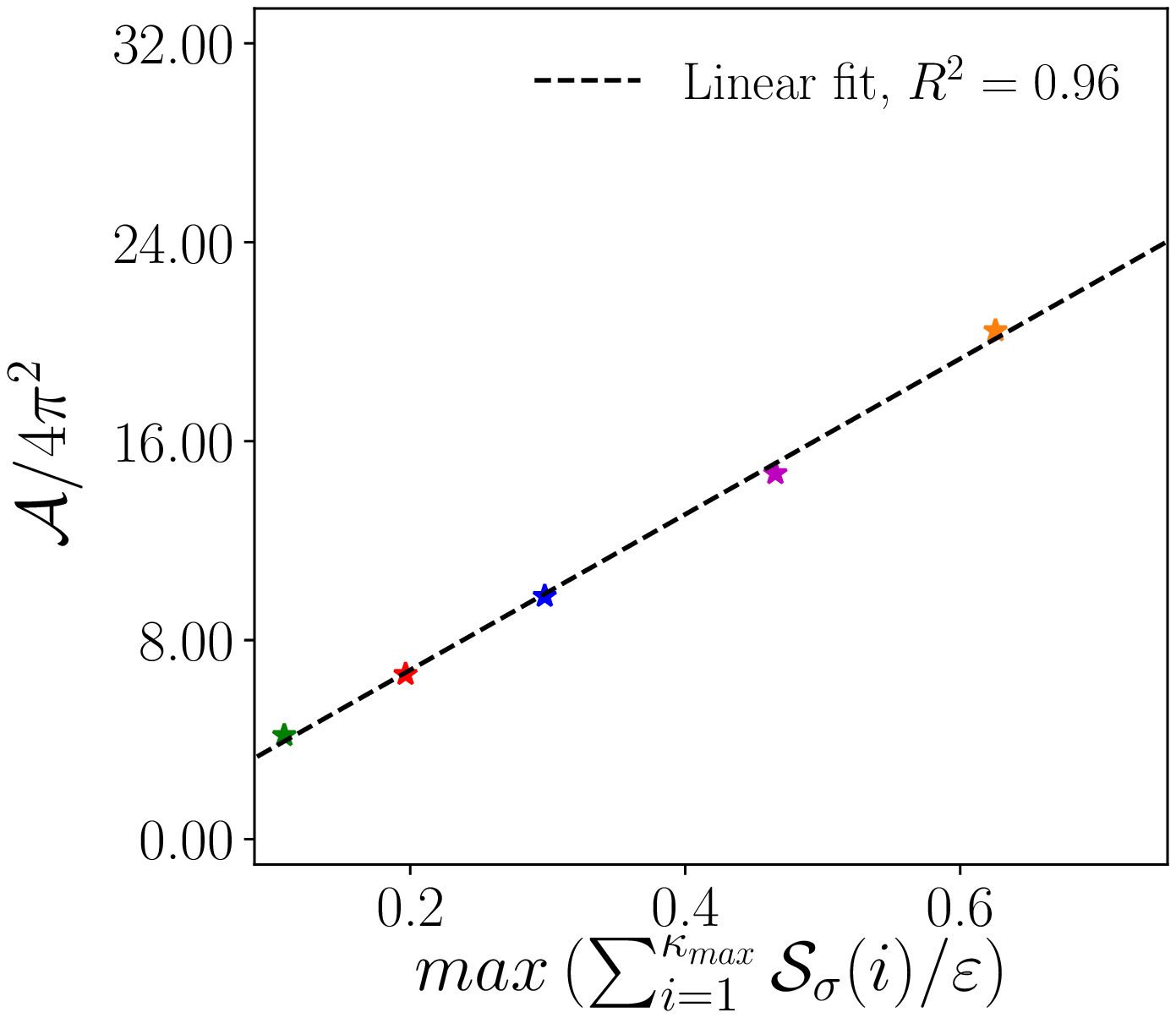}
	\put(-185,140){(\textit{a})}
	\includegraphics[width=0.49\textwidth]{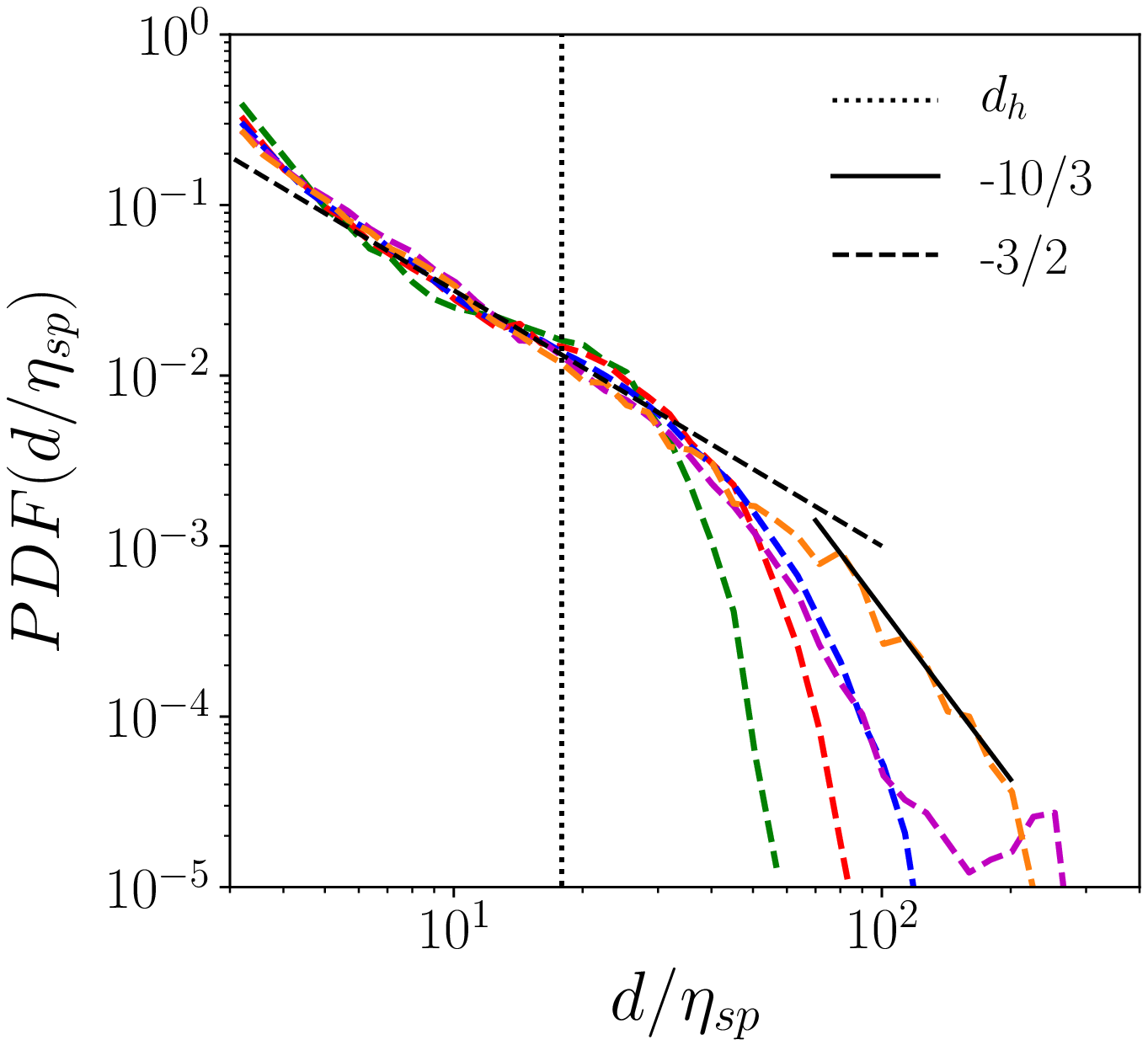}
    \put(-190,140){(\textit{b})}
    
    \includegraphics[width=0.95\textwidth]{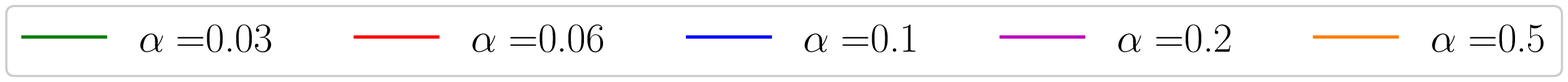}
\caption{(\textit{a}) Correlation between the maximum surface tension term $max\left(\sum_{i=1}^{\kappa_{max}}\mathcal{S}_\sigma(i)\right)$ and the total surface area $\mathcal{A}$ for the different volume fractions $\alpha$. The dashed black line is the linear fit to the data.  (\textit{b}) PDF of the droplet size distribution for different values of $\alpha$. The dashed black line indicates the $d^{-3/2}$ law from \cite{Deane2002}; continuous black line the $d^{-10/3}$ law from \cite{Garrett2000}; dotted black line the Hinze scale $d_H$. The droplet size is normalized by the Kolmogorov scale of simulation SP2 
	, 
$\eta_{sp}$ .}
\label{fig:dsd_alpha}
\end{figure}

Summarising, the surface tension introduces an alternative path for energy transmission from large towards small scales, as discussed  for binary flows in \cite{Perlekar2019}.
The amount of energy transferred by the surface tension is directly proportional to the total droplets surface area $\mathcal{A}$, as shown in \Cref{fig:dsd_alpha}(\textit{a}) where we display the maximum energy transferred via surface tension $max\left(\sum_{i=1}^{\kappa_{max}}\mathcal{S}_\sigma(i)\right)$ and the total area of the dispersed phase $\mathcal{A}$ for the 
different volume fractions under consideration with a linear fit to the data. This observation 
reinforces our previous conclusion that
 the interface transfers energy among different scales by disrupting larger turbulent structures and creating smaller ones, hence affecting the canonical -5/3 slope of the turbulence spectra.
Note also that, while in mono-dispersed flows this results in a deviation at a specific spectral frequency  \citep{Dodd2016}, in poly-dispersed flows this behaviour is seen at all scales.

We next consider the dynamics of the dispersed phase.
We first examine  the DSD for all the values of volume fraction studied, see \Cref{fig:dsd_alpha}(\textit{b}),  where we display the droplet diameters normalized by the single-phase (SP1) Kolmogorov scale $\eta_{sp}$

. The dashed black line depicts the $d^{-3/2}$ law by \cite{Deane2002} and the solid line the $d^{-10/3}$ law by \cite{Garrett2000} valid for larger droplets. 
For small droplets, the -3/2 law is well captured also for marginally resolved droplets (with $d/\eta_{sp}<6$). For droplets larger than the Hinze scale, the -10/3 law is also a very good fit, with increasing accuracy for increasing values of $\alpha$. Our data are in agreement with the findings by \cite{Mukherjee2019} and explained by
higher coalescence probability 
 at higher volume fractions,
leading to a bigger population of larger droplets.
Interestingly, the Hinze scale turns out to approximately define the transition between the -3/2 and the -10/3 scalings as proposed in  \cite{Deane2002}, although for higher values of $\alpha$, the onset of the $d^{-10/3}$ power-law occurs at larger diameters.  As the droplet distributions can be, to a good approximation, represented by these 2 laws, it follows that $\mathcal{A}\propto\alpha$, explaining why $\mathcal{A}$, $\mathcal{S}_\sigma$ and $\alpha$ are  linearly correlated (see panel \textit{a} of the same figure).

\begin{figure}
	\centering
	\includegraphics[width=0.5\textwidth]{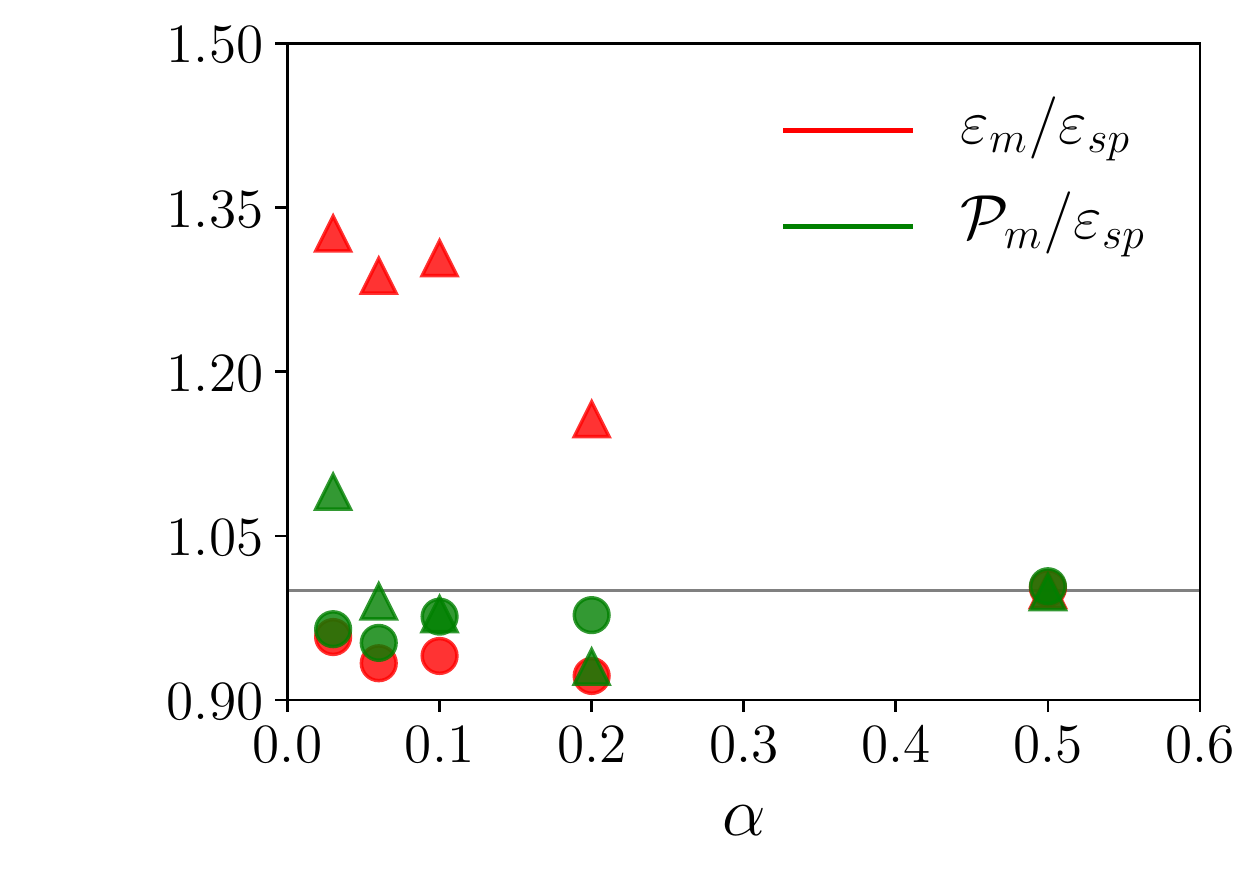}
	\put(-180,130){(\textit{a})}
	\includegraphics[width=0.5\textwidth]{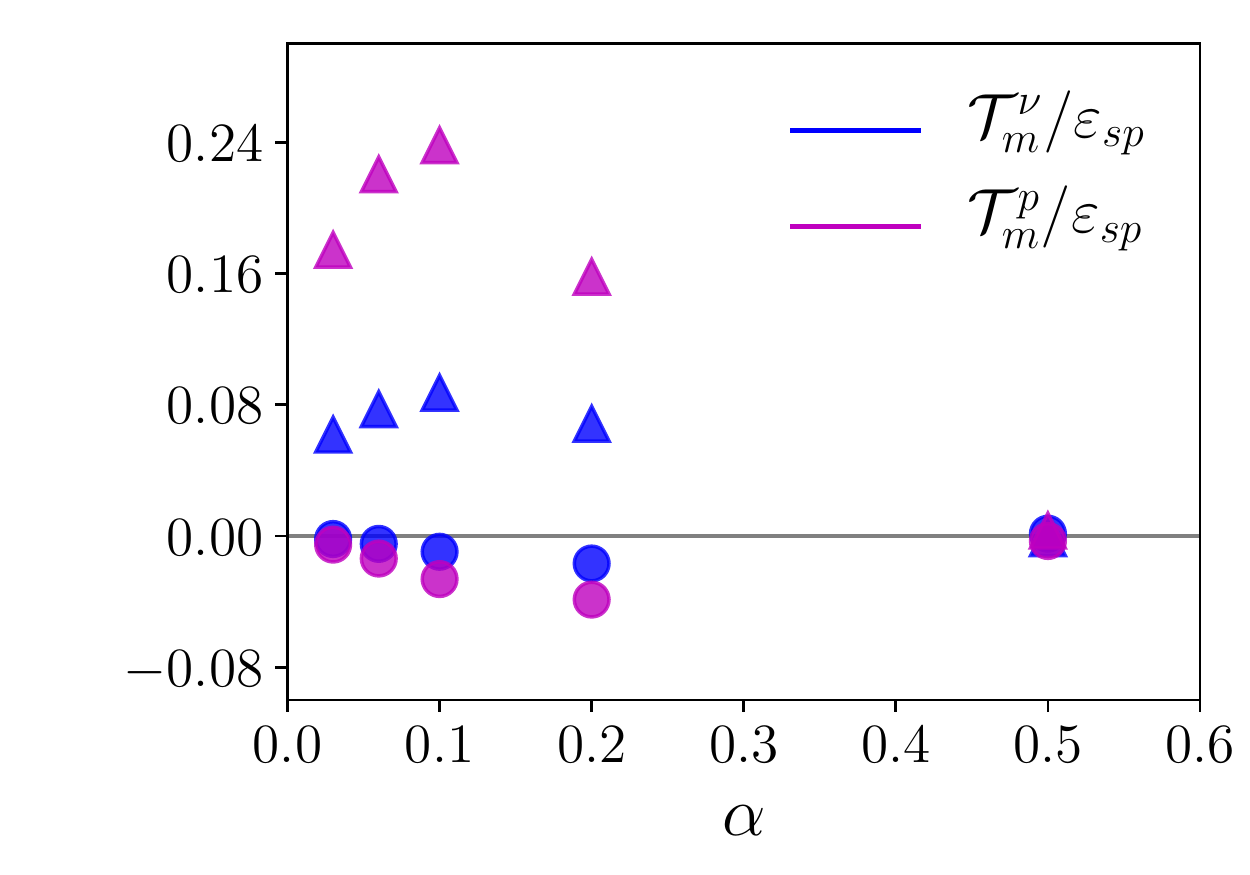}
	\put(-180,130){(\textit{b})}
	\caption{Phase-averaged energy balance versus the dispersed phase volume fraction $\alpha$, see  \Cref{eq:enBalMP}. In each plot, colored triangles (\mytriangle{black}) represent the dispersed phase ($m=d$) while circles (\mycircle{black}) the carrier phase ($m=c$). Each term is normalized by the single phase energy dissipation $\varepsilon_{sp}$ computed for case SP2. The energy production $\mathcal{P}_m$ and energy dissipation $\varepsilon_m$  are reported in panel (\textit{a}), while viscous energy transport $\mathcal{T}^\nu_m$  and the pressure energy transport $\mathcal{T}^p_m$ in panel (\textit{b}).
	}
	\label{fig:balPhase_alpha}
\end{figure}

We now consider the phase-averaged energy budget, introduced in  \Cref{subsec:met:sp-case}. The different terms of \Cref{eq:enBalMP}, production, dissipation and transport by pressure and viscous forces, are shown in \Cref{fig:balPhase_alpha}, normalized by the single-phase dissipation.
We first observe that the total production and dissipation 
\[ \mathcal{P}= \alpha \mathcal{P}_d + (1- \alpha) \mathcal{P}_c \approx\varepsilon\approx 0.95\varepsilon_{sp} \] 
for $\alpha<0.5$.  
The energy production density $\mathcal{P}_m$ (green markers in panel \textit{a}), is higher in the dispersed phase for low volume fractions, while it is comparable to that of the carrier phase for $\alpha>0.1$. 
The energy dissipation rate per unit volume in the dispersed  phase $\varepsilon_d$  (green markers in panel \textit{b}) is also larger at low volume fractions and monotonically decreases with increasing  $\alpha$. 
The dissipation in the carrier phase, $\varepsilon_c$, also decreases, as it compensates for the energy transport $\mathcal{T}_m$ from the carrier flow towards the dispersed phase. 
The viscous transport (see blue markers in panel \textit{b} of the same figure) is significantly lower than its pressure-induced counterpart, $\mathcal{T}^p_m$, although they exhibit a similar behavior: they first increase until  $\alpha=0.1$ and then decrease to reach zero for a binary  mixture.  Note again, that the total transport term is zero, i.e.\ the sum of  $\mathcal{T}^p$ and $\mathcal{T}^{\nu}$ from both phases.
The case $\alpha=0.5$ deserves a specific mention. In this case, production and dissipation in the two phases are equal, hence the transport term $\mathcal{T}_m=0$. Intuitively,  it is not possible to define unambiguously a carrier and dispersed phase in binary mixtures; while the  energy is locally transported from one phase to the other, the global average is zero for both pressure and viscous transfer. 

\begin{figure}
	\centering
	\includegraphics[width=0.33\textwidth]{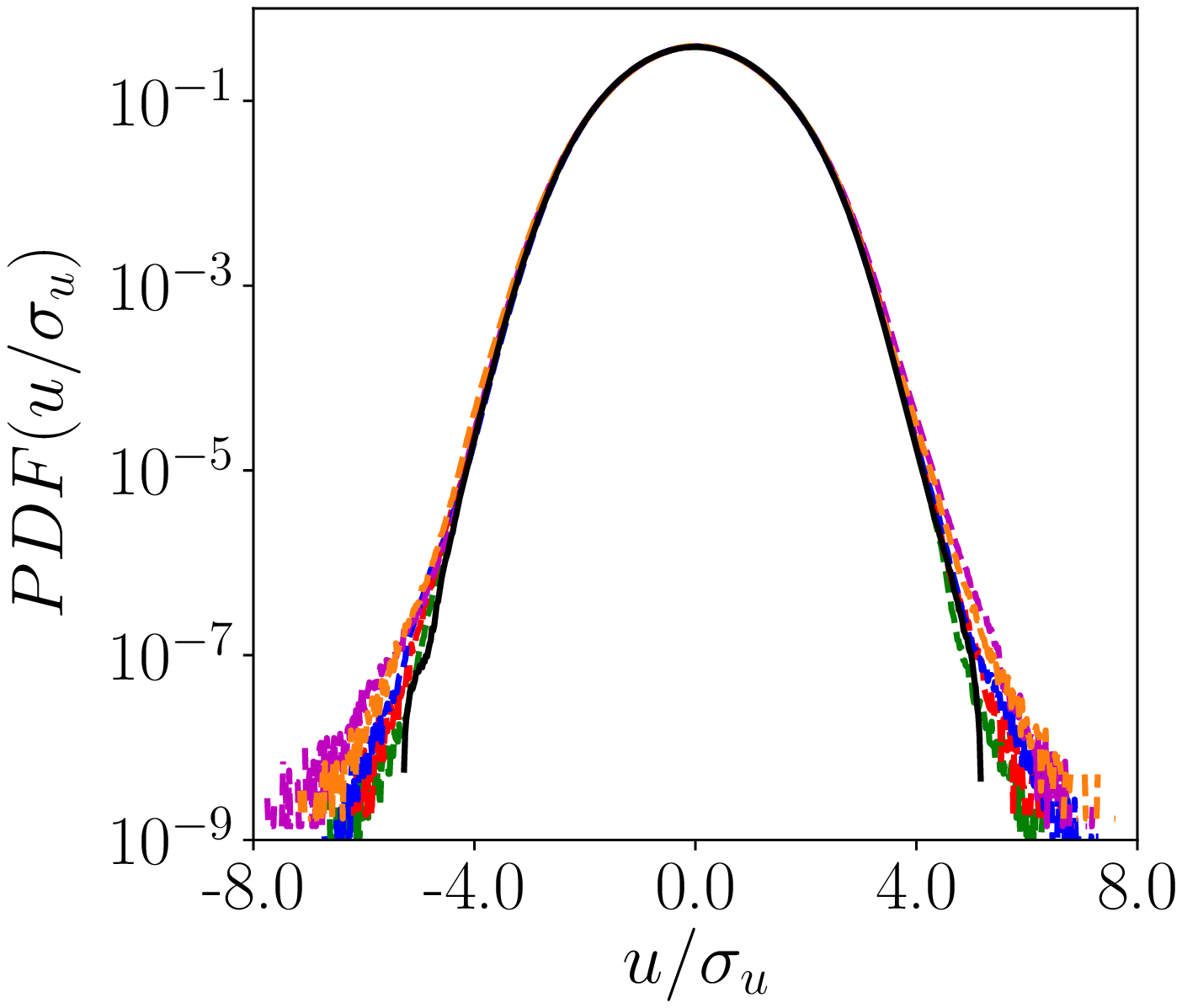}
	\put(-120,110){(\textit{a})}
	\includegraphics[width=0.33\textwidth]{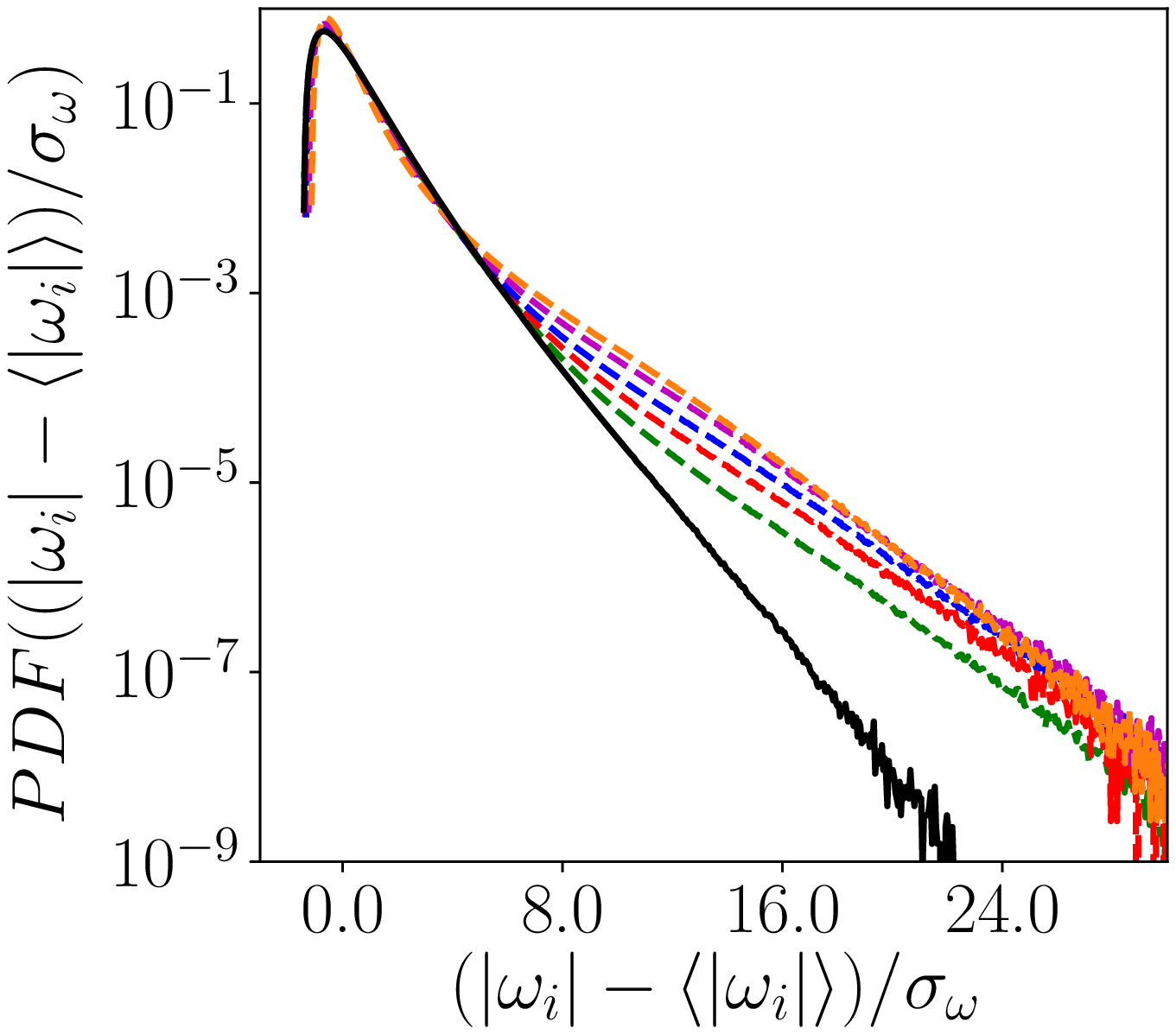}
	\put(-120,110){(\textit{b})}
	\includegraphics[width=0.33\textwidth]{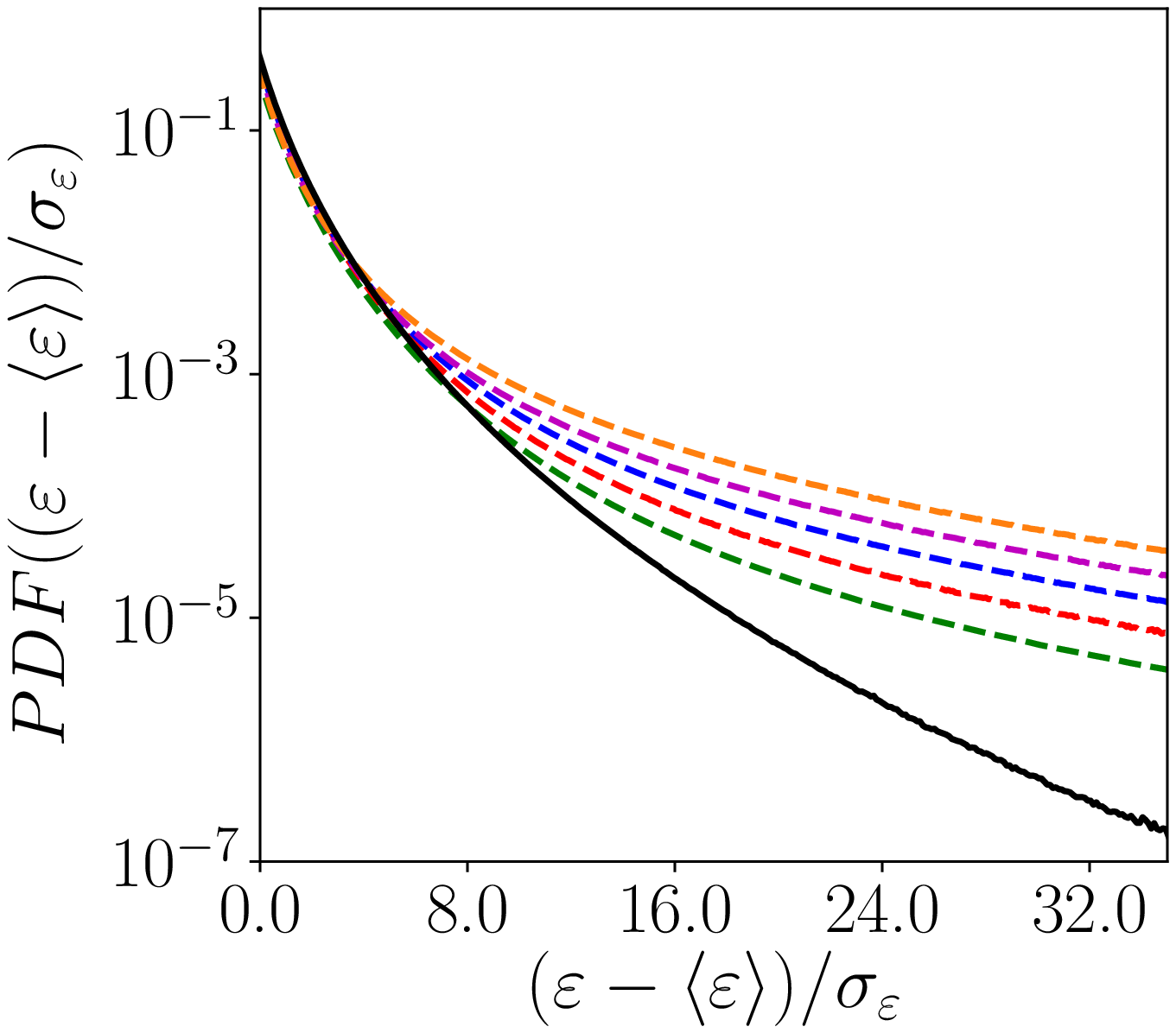}
	\put(-120,110){(\textit{c})}
	
	\includegraphics[width=0.95\textwidth]{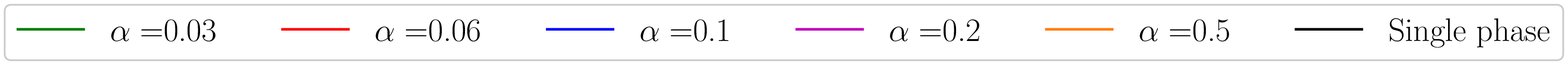}
	\caption{PDF of velocity fluctuations $u$, vorticity $\omega$ and dissipation $\epsilon$. All quantities are normalized as standard score.}
	\label{fig:PDF_alpha}
\end{figure}

We finally analyze the PDF of velocity fluctuations $u_n=u/\sigma_u$, vorticity fluctuations $\omega_n=(|\omega_i|-\langle |\omega_i| \rangle)/\sigma_\omega$ and energy dissipation $\varepsilon_n=(\varepsilon-\langle \varepsilon \rangle)/\sigma_\varepsilon$, normalized by their standard deviation. In \Cref{fig:PDF_alpha}(\textit{a}), we observe that, while the PDF remains symmetric, the tails of the PDF of the velocity fluctuation strongly deviate from the typical pseudo-Gaussian behavior of single-phase turbulence 
\citep{Sreenivasan1997,Jimenez2000a,Ishihara2009}. 
As concerns the vorticity in panel (\textit{b}), no deviation is observed in the Gaussian core \cite[as defined in][]{Sreenivasan1997}. 
However, the distributions of the multiphase flows strongly depart from the single-phase case in the tails. 
In particular, the exponentially decaying tails have a higher exponent in the case of emulsions, indicating more events with strong vorticity.
Interestingly, while increasing the volume fraction does not influence the value of the exponent, increasing $\alpha$ induces deviations in the distributions already at lower values of $\omega_n$. 
We observe  a similar behavior for  $\varepsilon_n$
in panel (\textit{c}):
 the intermittency of the single-phase flow
is amplified by the presence of the interface. As for the vorticity,  departures from the single-phase distributions are observed at lower values of $\varepsilon$  when increasing the volume fraction $\alpha$. 
As a final general remark, the analysis of the PDFs reveals that strong deviations are induced by the presence of the interface, already at low volume fractions,  
overall increasing the intermittent behavior of the flow. As no collapse is observed for the normalized variables, it can be inferred that the small-scale statistics are affected by the presence of the interface.

\subsection{Influence of viscosity ratio}
\label{sec:res:visc}
\begin{figure}
    \centering
    \includegraphics[width=0.3\textwidth]{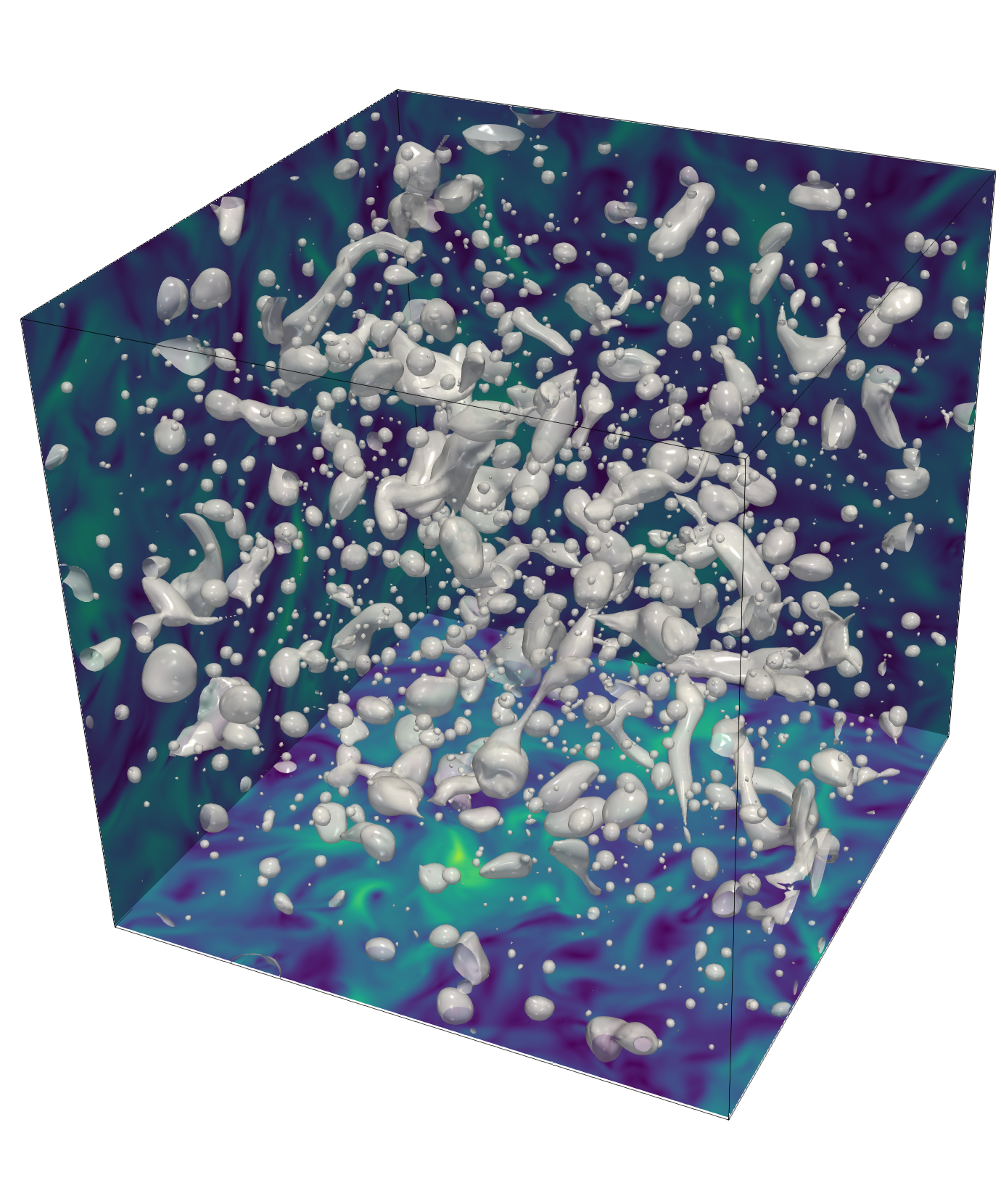}
    \put(-100,140){{$\mu_d/\mu_c=0.01$}}
    \includegraphics[width=0.3\textwidth]{figures/tinf_c.png}
    \put(-100,140){{$\mu_d/\mu_c=1$}}
    \includegraphics[width=0.3\textwidth]{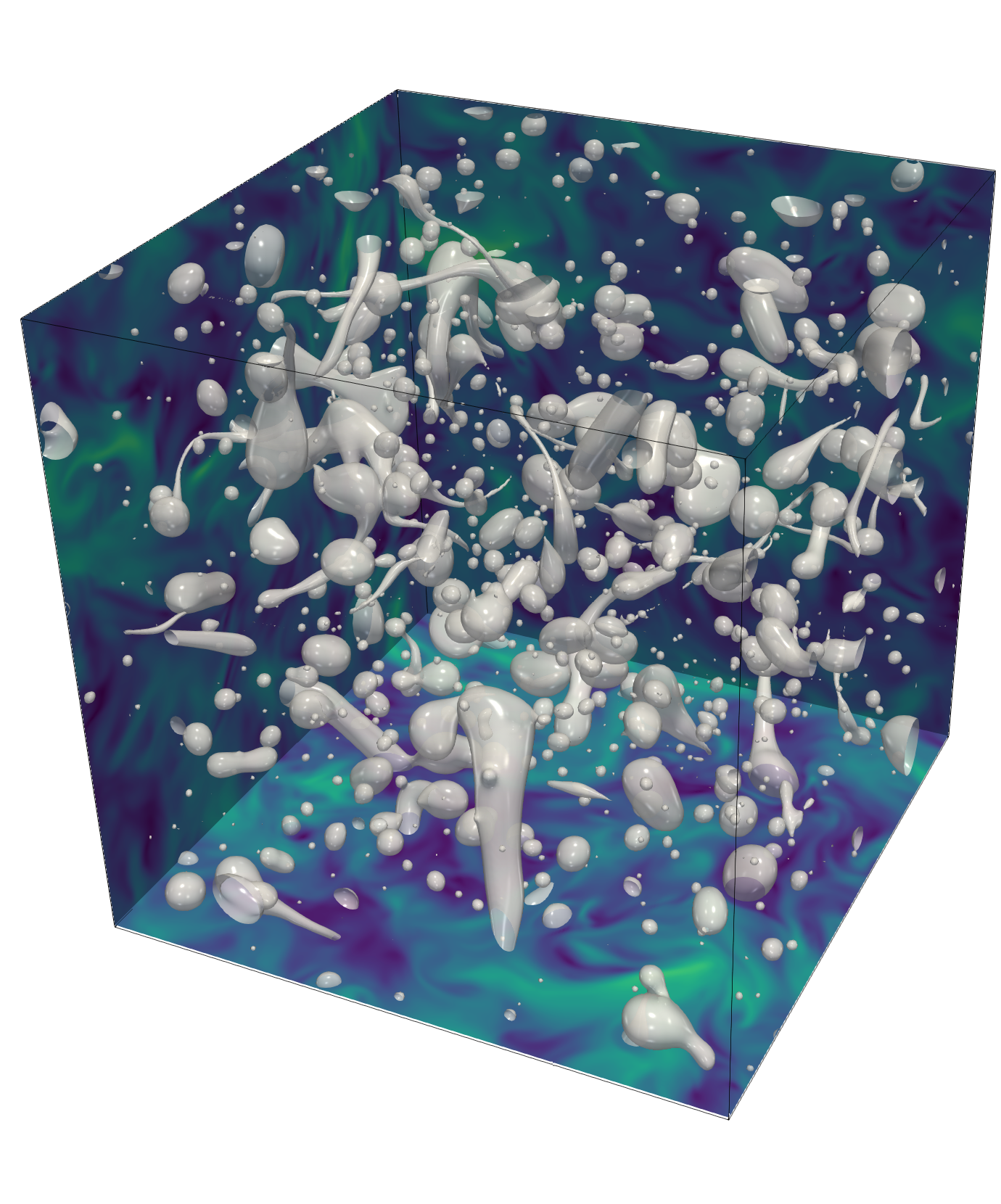}
    \put(-100,140){{$\mu_d/\mu_c=100$}}
    \caption{Render of the two-fluid interface (corresponding to the value of the VOF function $\phi=0.5$) for different values of the viscosity ratio $\mu_d/\mu_c$ (left to right, 0.06, 0.2 and 0.5). 
    The vorticity fields are shown on the box faces on a planar view. 
    All simulations are performed at $\alpha=0.03$ and $We_\mathcal{L}=42.6$.}
    \label{fig:muRender}
\end{figure}

We consider now the influence of the viscosity ratio on the flow turbulence, i.e. cases  BEx, V1x and V2x in \Cref{tab:testMat}. The viscosity ratios analyzed span the range
 $0.01<\mu_d/\mu_c<100$, while $We_\mathcal{L}=42.7$ for all cases. Two values of the volume fractions are considered, $\alpha=0.03$ (series V1x) and $\alpha=0.1$ (series V2x).
A render 
of the two-fluid interface (corresponding to the value of the VOF function $\phi=0.5$) is shown in \Cref{fig:muRender} for cases V11, BE1 and V14 (from left to right). As $\mu_d/\mu_c$ increases, larger droplets appear;  at low viscosity ratios we find a significantly higher number of droplets.

\begin{figure}
	\centering
		\includegraphics[width=0.49\textwidth]{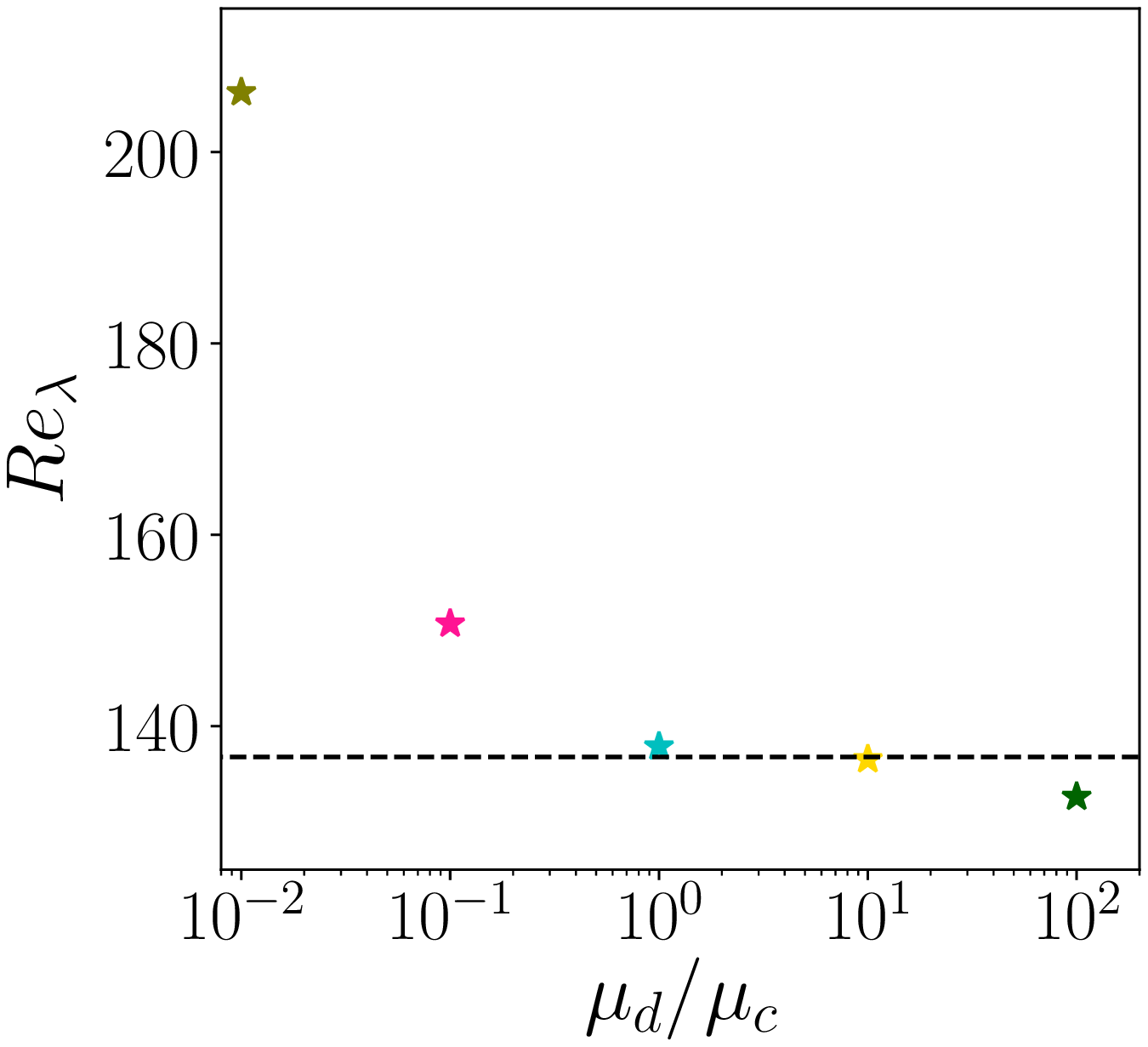}
		\put(-108,60){\includegraphics[width=0.25\textwidth]{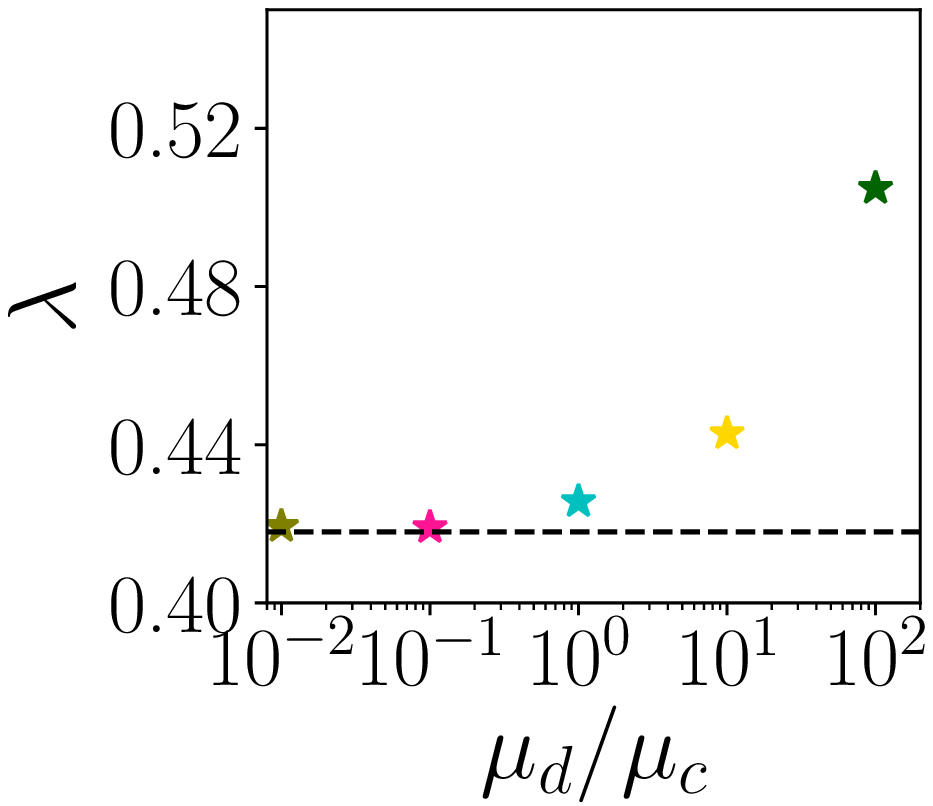}}
		\put(-185,140){(\textit{a})}
		\put(-55,140){$\alpha=0.03$}
		\includegraphics[width=0.49\textwidth]{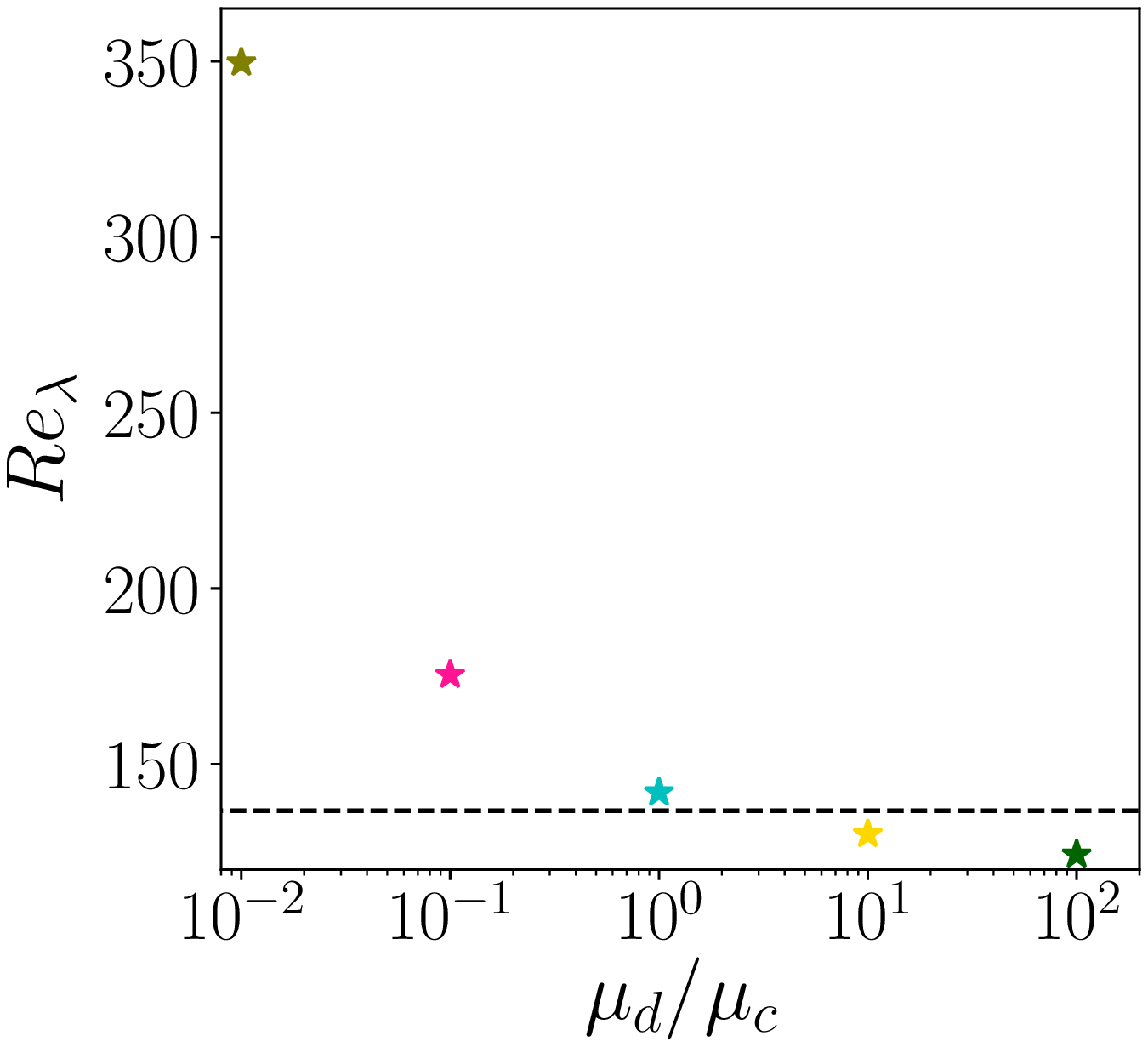}
		\put(-108,60){\includegraphics[width=0.25\textwidth]{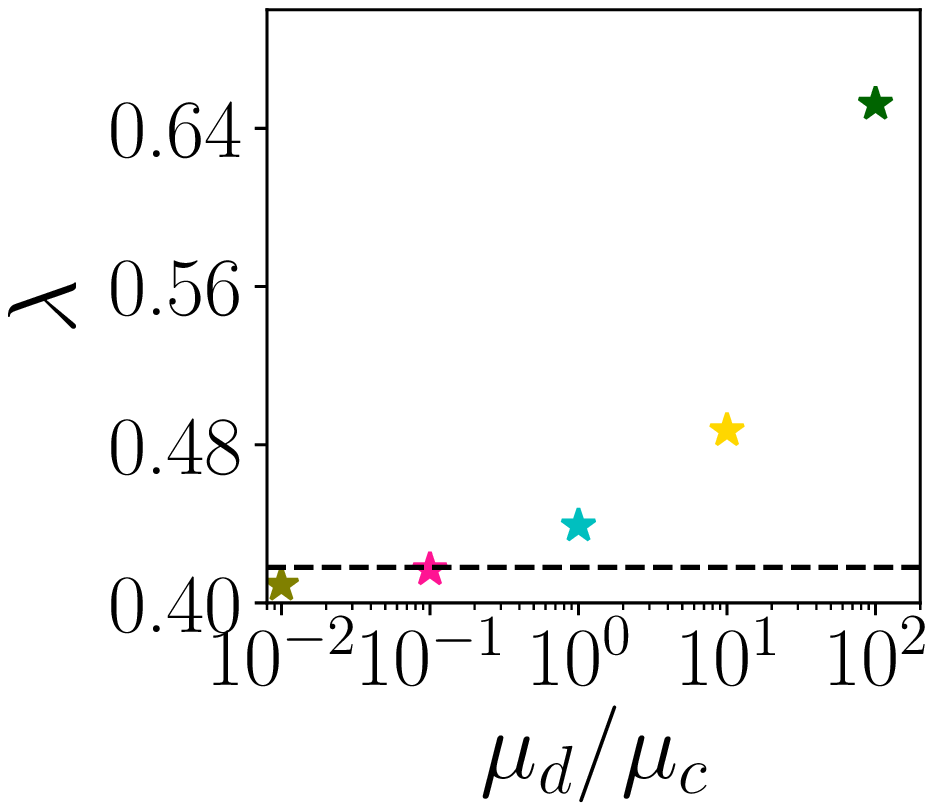}}
		\put(-190,140){(\textit{b})}
		\put(-50,140){$\alpha=0.1$}
		
		\includegraphics[width=0.8\textwidth]{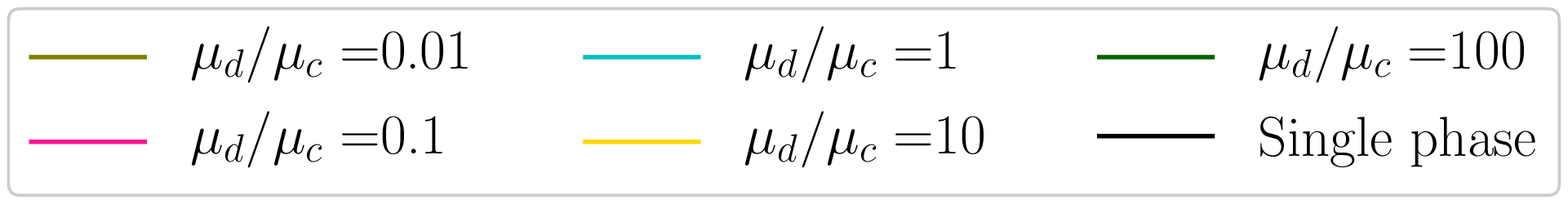}
	\caption{Taylor Reynolds number of the emulsion flows for the different viscosity ratios examined. $Re_\lambda$ is shown versus $\mu_d/\mu_c$. The panel on the left shows cases BE1 and V1x with volume fraction $\alpha=0.03$ whereas the panels on the right show cases BE2 and V2x with $\alpha=0.1$. The inset shows the evolution of $\lambda$ with the viscosity ratio.}
	\label{fig:ReLam_mu}
\end{figure}

We start by examining the Taylor Reynolds number of the emulsion flows for the different viscosity ratios under investigation.
Panels (\textit{a}) and (\textit{b}) of \Cref{fig:ReLam_mu} show the variation of $Re_\lambda$ versus the viscosity ratio for the two volume fractions considered, $\alpha=0.03$ and $\alpha=0.1$.   
As expected, $Re_\lambda$ decreases with the viscosity ratio.  Significant variations in $Re_\lambda$ are observed already for small volume fractions, the effects being amplified for $\alpha=0.1$. In the insets of the same figure, we can observe that $\lambda$ (i.e.\ the local variations of $k/\varepsilon$) does not increase linearly with $\mu_d/\mu_c$, indicating increased velocity fluctuations for the dispersed phase at lower viscosity.

\begin{figure}
	\centering
	\includegraphics[width=0.49\textwidth]{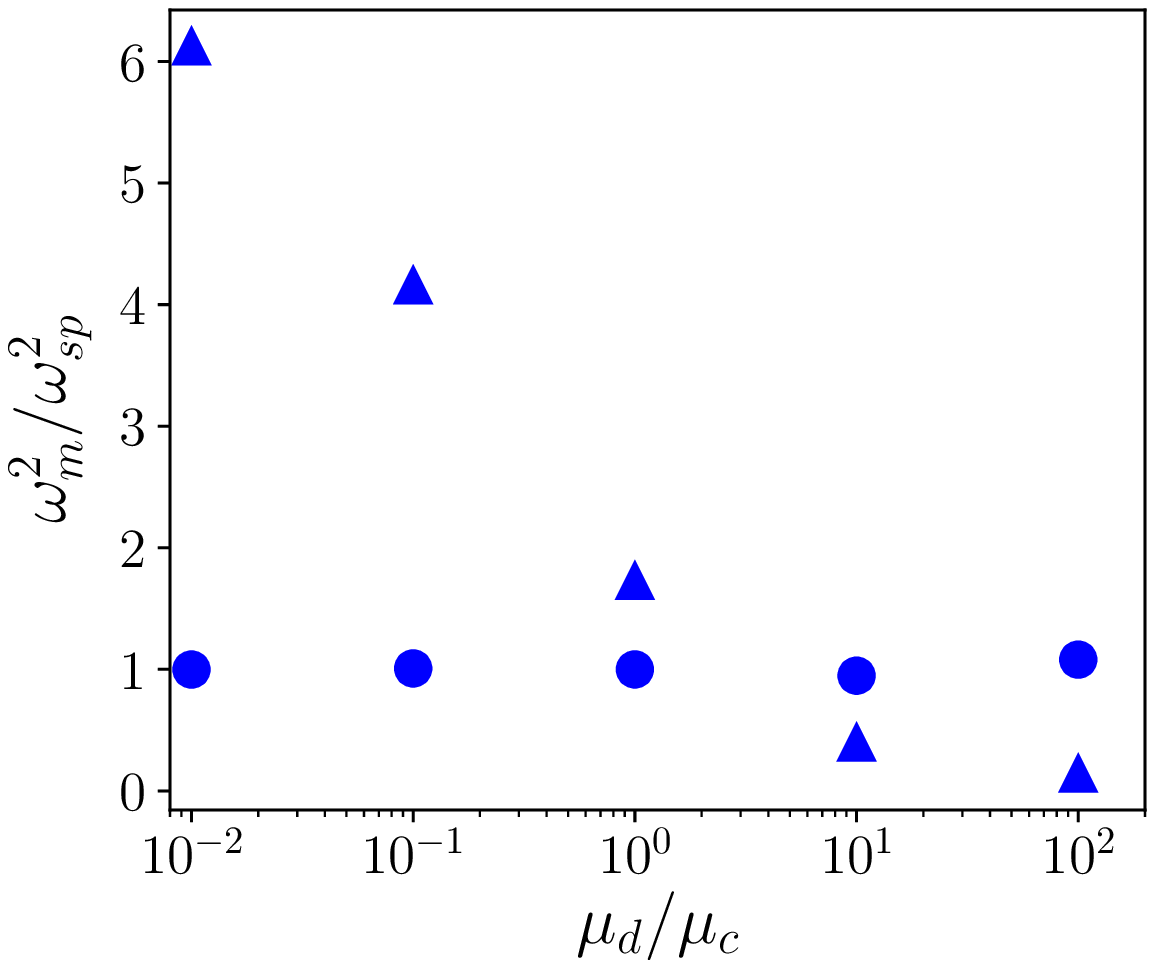}
	\put(-185,140){(\textit{a})}
	\put(-55,120){$\alpha=0.03$}
	\includegraphics[width=0.49\textwidth]{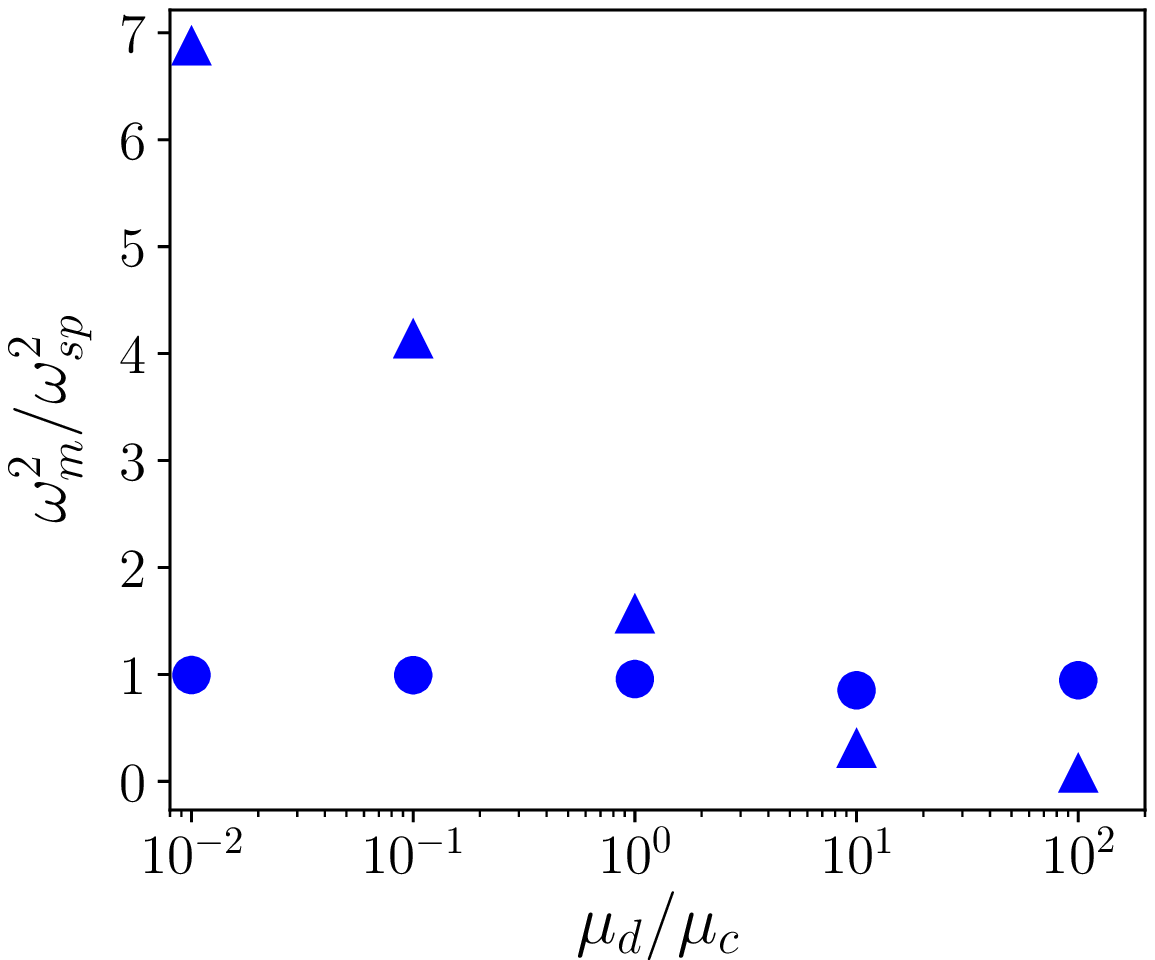}
	\put(-190,140){(\textit{b})}
	\put(-50,120){$\alpha=0.1$}
	\caption{Phase-averaged enstrophy $\omega^2_m$ (normalized by its value in the single-phase case SP1) for different viscosity ratios $\mu_d/\mu_d$. Triangles (\mytriangle{black}) indicates the dispersed phase ($m=d$) while circles (\mycircle{black}) the carrier phase ($m=c$). Panel (\textit{a}) shows results for $\alpha=0.03$, and panel (\textit{b}) for $\alpha=0.1$.}
	\label{fig:enske_mu}
\end{figure}

To better quantify the variations of the flow gradients, we show the phase-averaged (see eq.\ \ref{eq:volAvg})  enstrophy  $\omega^2_m$ 
in \Cref{fig:enske_mu}, 
normalized by the single-phase values from SP1.  The viscosity ratio 
strongly affects enstrophy in the dispersed phase, while the magnitude in the carrier phase is almost constant. Further, smaller variations can be observed when changing the volume fraction from 0.03 to 0.1. 
For $\mu_d/\mu_c\le 1$, the enstrophy in the dispersed phase goes approximately  as $\omega^2_c\propto -log(\mu_d/\mu_c)$.  
As the viscosity of the dispersed phase becomes larger, $\mu_d>\mu_c$, $\omega_d$ decreases below the average value of the single-phase flow and tends towards zero, as high viscosity dampens velocity fluctuations in the dispersed phase. 
It is worth noting that, for incompressible flows, the energy dissipation rate can be defined as $ \varepsilon \equiv\nu |\omega_i|^2$; however, when phase averaging, the two formulations differ for by a term proportional to $\partial_{ii}p$.

\begin{figure}
	\centering
		\includegraphics[width=.49\textwidth]{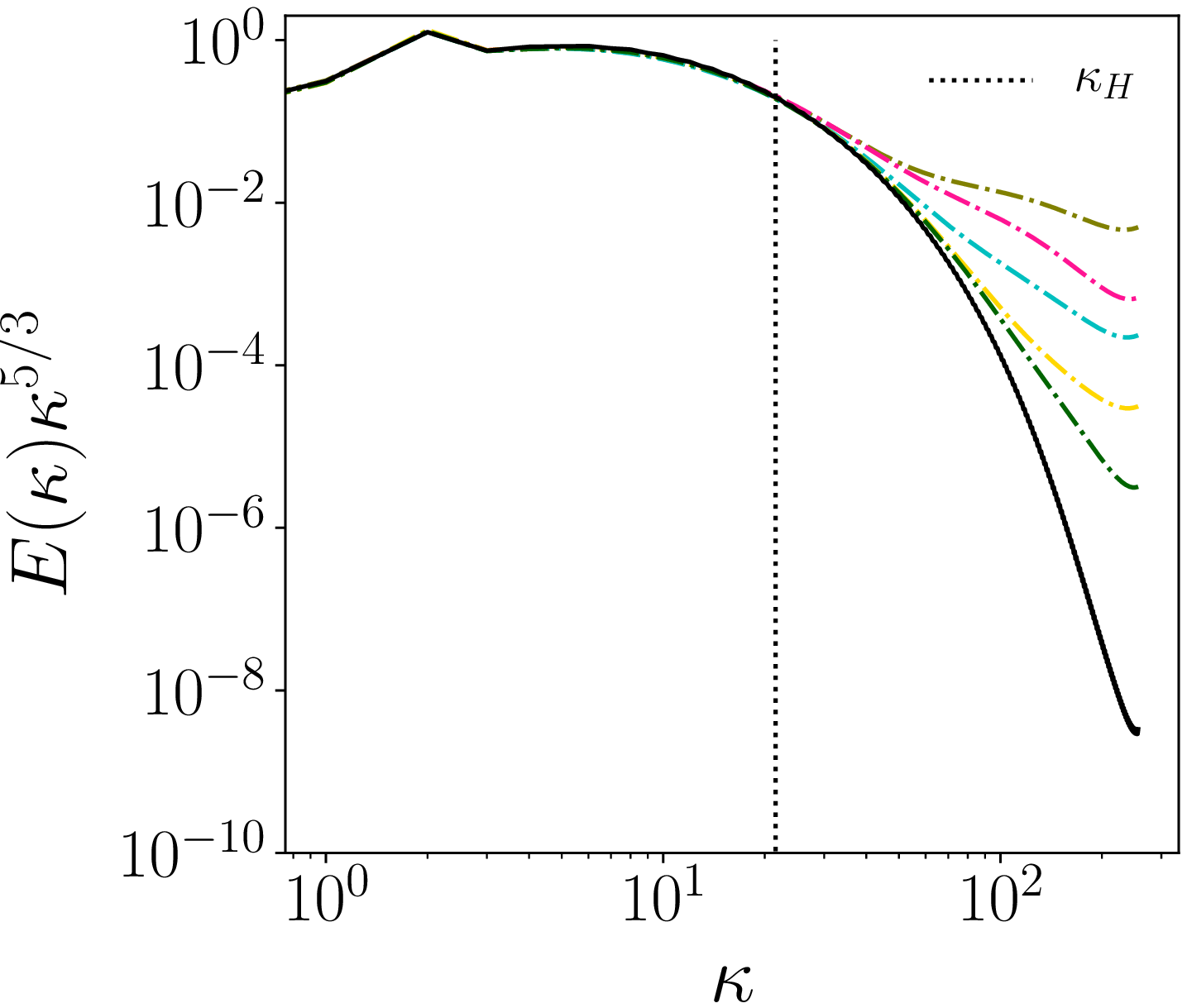}
		\put(-185,140){(\textit{a})}
		\put(-135,100){$\alpha=0.03$}
		\includegraphics[width=.49\textwidth]{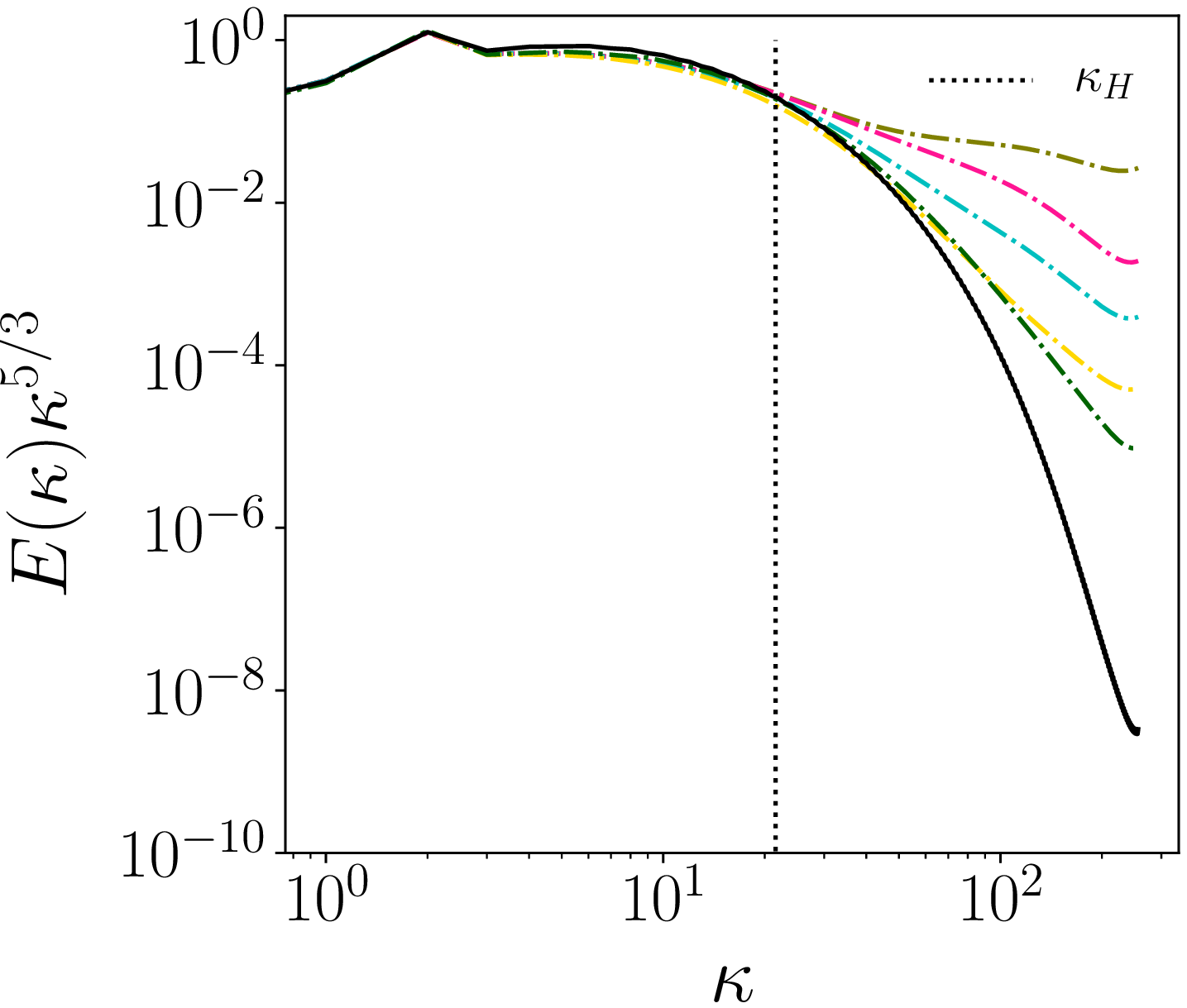}
		\put(-190,140){(\textit{b})}
		\put(-135,100){$\alpha=0.1$}
		
		\includegraphics[width=0.8\textwidth]{figures/legend_mu.eps}
	\caption{ One-dimensional compensated energy spectra for  (\textit{a}): $\alpha=0.03$ and (\textit{b}) $\alpha=0.1$ and different values of the viscosity ratio $\mu_b/\mu_c$. The vertical dotted line indicates the Hinze scale wavelength, $\kappa_H$.}
	\label{fig:spectra_mu}
\end{figure}

We now discuss the influence of viscosity ratio on the compensated energy spectra, shown in \Cref{fig:spectra_mu}(\textit{a}) for $\alpha=0.03$ and in panel (\textit{b}) for $\alpha=0.1$. Similarly to previous observations for \Cref{fig:spectra_alpha}, the  Hinze scale shows, to a good approximation, the pivoting point, below which energy  increases with respect to the single-phase spectra. Differences in the inertial subrange are hardly observable for $\alpha=0.03$, while they become more prominent when the volume fraction is increased, see panel (\textit{b}). Analysis of the data for $\kappa<\kappa_H$ , reveals that the simulations with a dispersed phase present less energy than the single phase case. In the dissipative range the trend emerges more clearly. As $\kappa>\kappa_H$, the less viscous the dispersed phase, the more energy is injected in the smaller scales.  As discussed in the previous section, energy reduces at large scales and increases at small scales when increasing the volume fraction $\alpha$. 

\begin{figure}
	\centering
	\includegraphics[width=.49\textwidth]{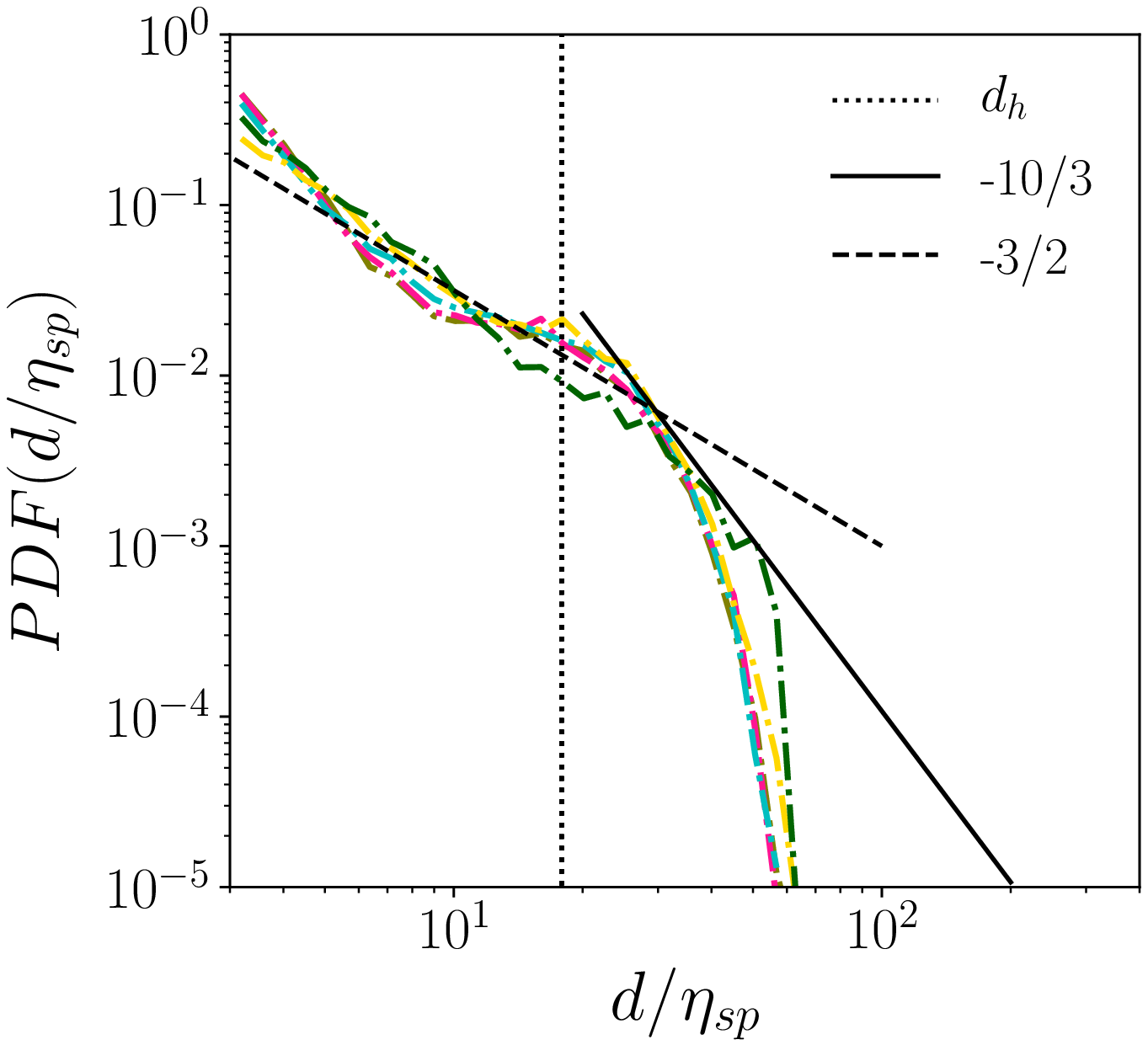}
	\put(-185,140){(\textit{a})}
	\put(-135,80){$\alpha=0.03$}
	\includegraphics[width=.49\textwidth]{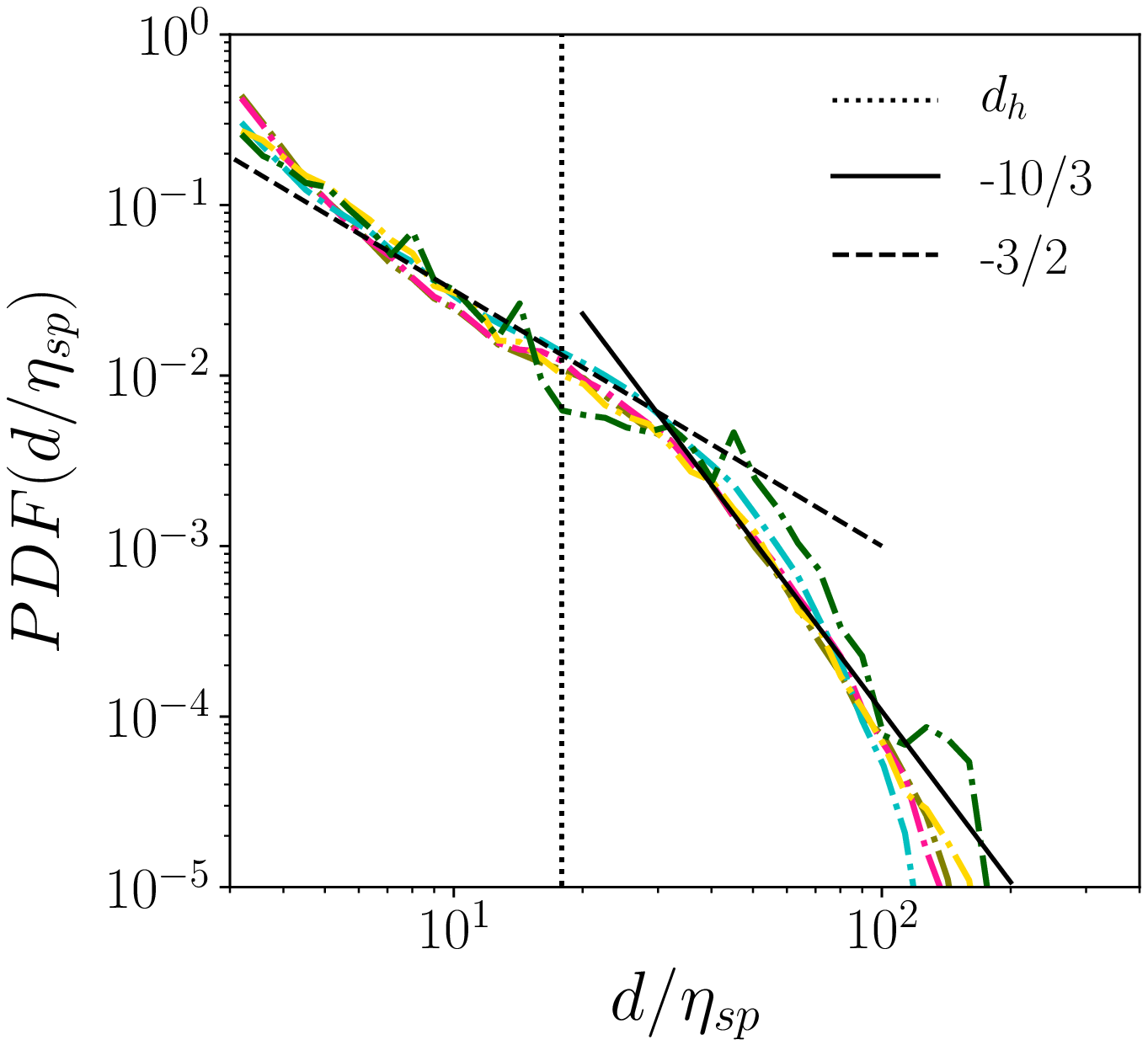}
	\put(-190,140){(\textit{b})}
	\put(-135,80){$\alpha=0.1$}
	
	\includegraphics[width=.95\textwidth]{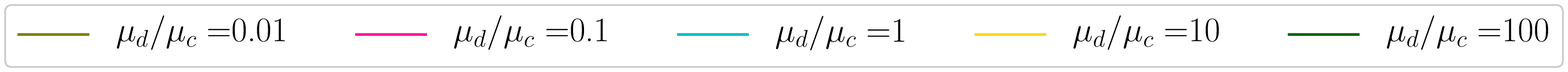}
	\caption{ PDF of the DSD for different values of $\mu$, at $\alpha=0.03$ (panel \textit{a}) and $\alpha=0.1$ (panel \textit{b}). The dashed line represents the $d^{-3/2}$ scaling from \cite{Deane2002}, while the continuous black line shows the $d^{-10/3}$ law from \cite{Garrett2000}.}
	\label{fig:dsd_mu}
\end{figure}

\Cref{fig:dsd_mu} shows the DSD for all configurations with different viscosity ratios. As for the data in \Cref{sec:res:alpha}, we also display the -3/2 power-law, which well describes the distribution of small droplets, and the $d^{-10/3}$ law from \cite{Garrett2000} for larger droplets. In this range, $d>d_H$, the -10/3 law is observed only in a limited region of the spectrum. As noted previously, this is most likely due to the low volume fraction considered. 
The variation of $\mu_d$ has an influence on large droplets, as  higher viscosity in the dispersed phase increases the probability of formation of these large droplets. This was also observed qualitatively in \Cref{fig:muRender} and confirms previous findings \citep{Roccon2017}.

\begin{figure}
	\centering
	\includegraphics[width=0.5\textwidth]{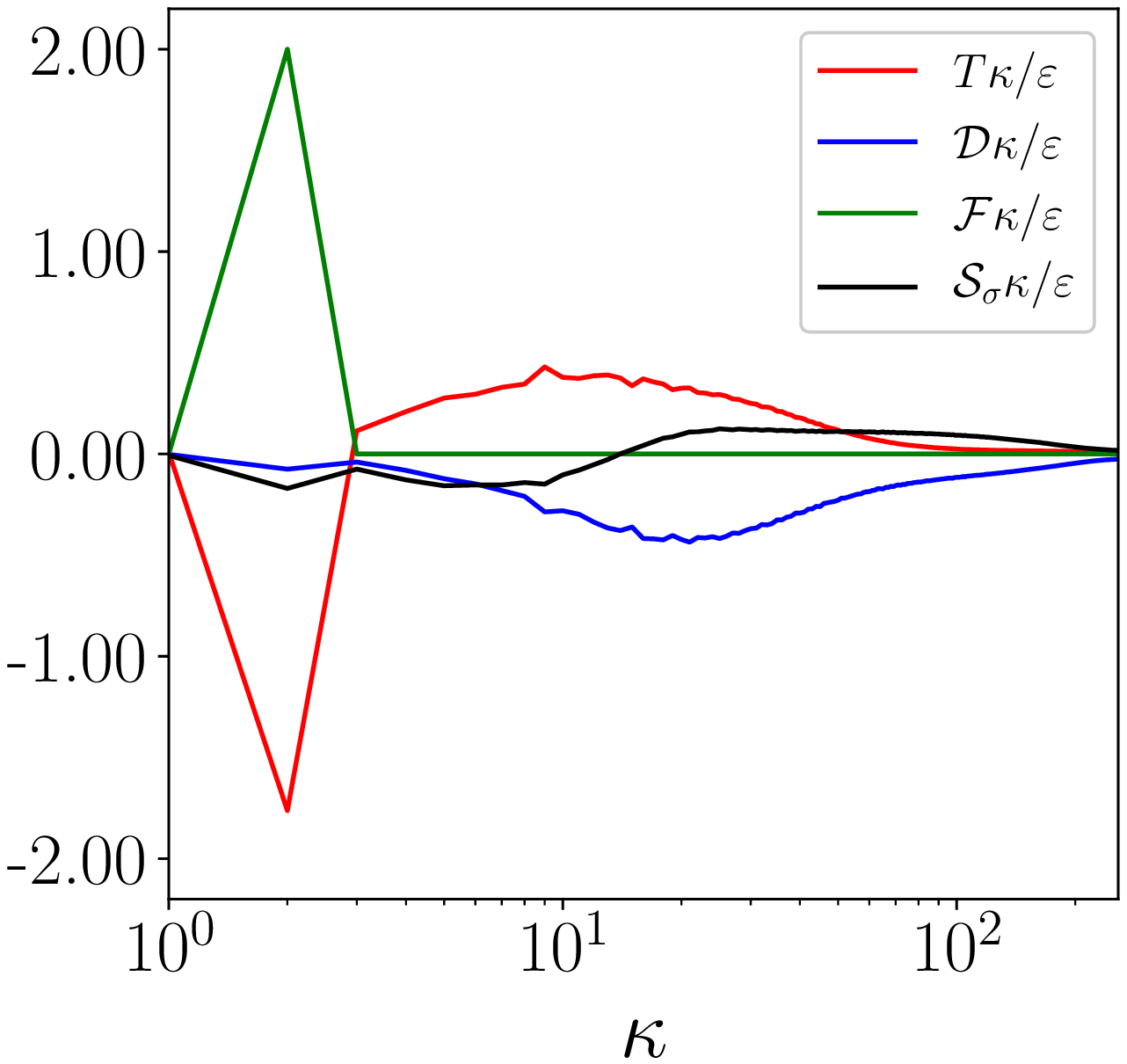}
	\put(-200,140){(\textit{a})}
	\includegraphics[width=0.5\textwidth]{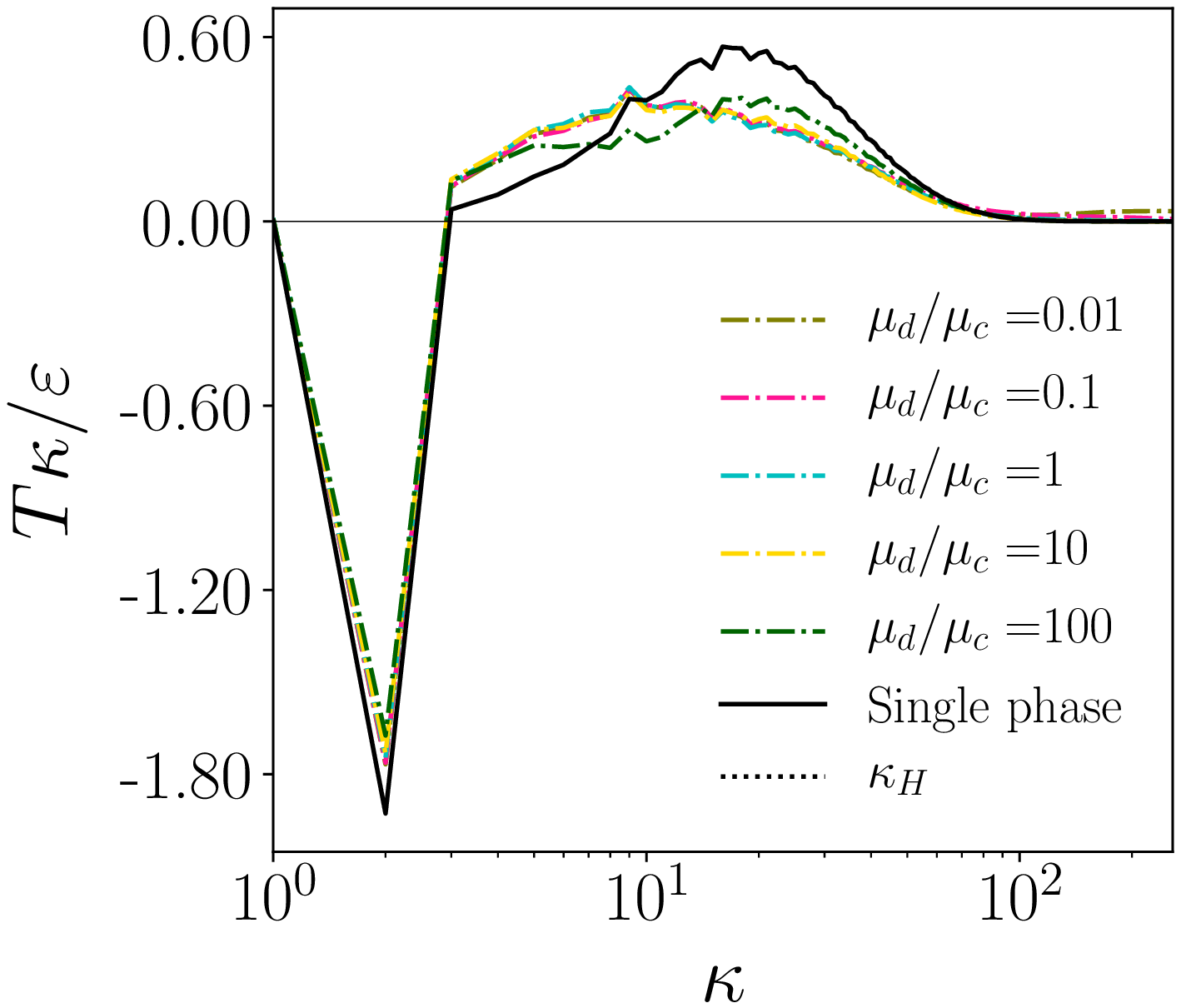}
	\put(-190,140){(\textit{b})}\\
	\includegraphics[width=0.5\textwidth]{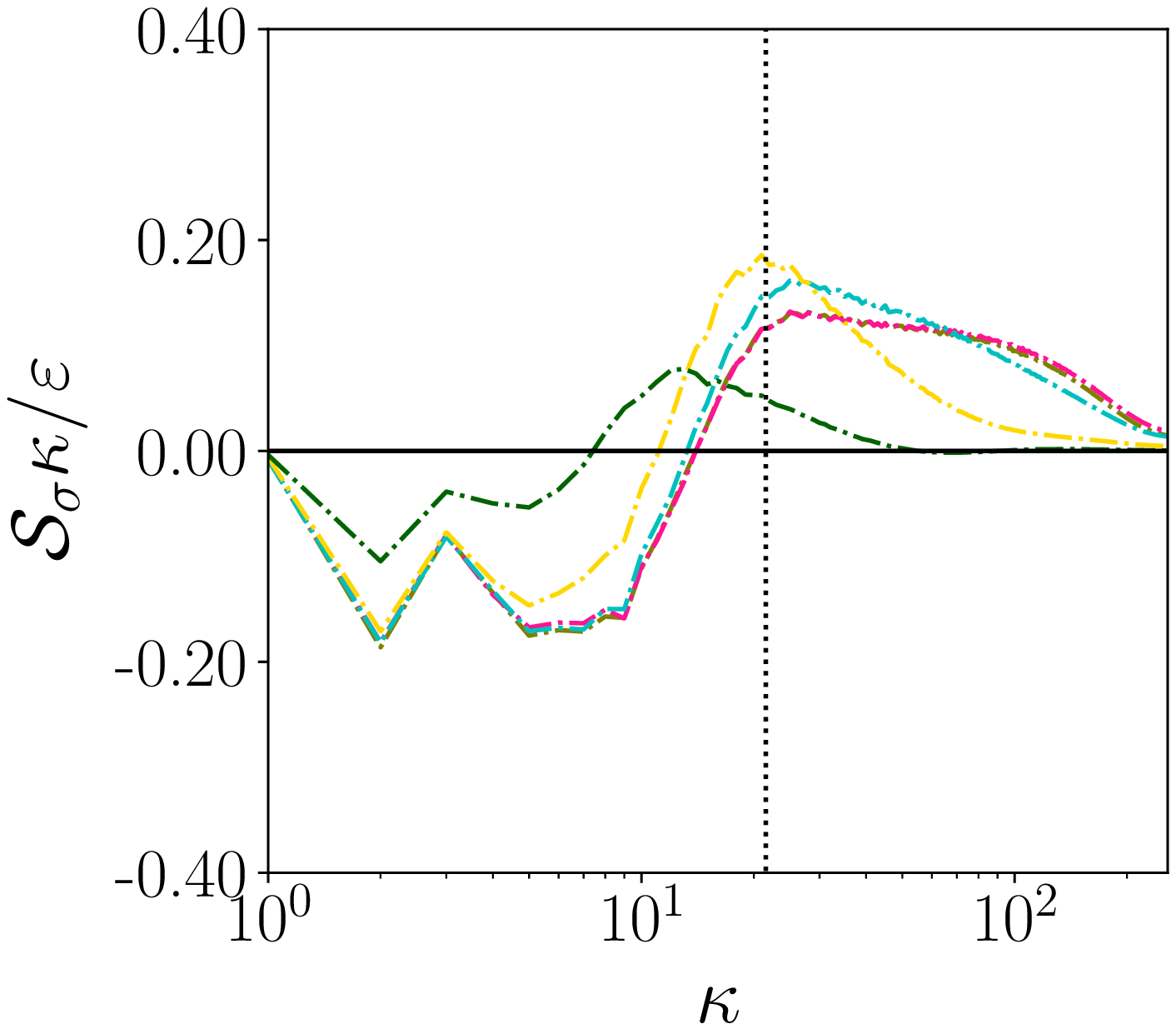}
	\put(-200,140){(\textit{c})}
	\includegraphics[width=0.5\textwidth]{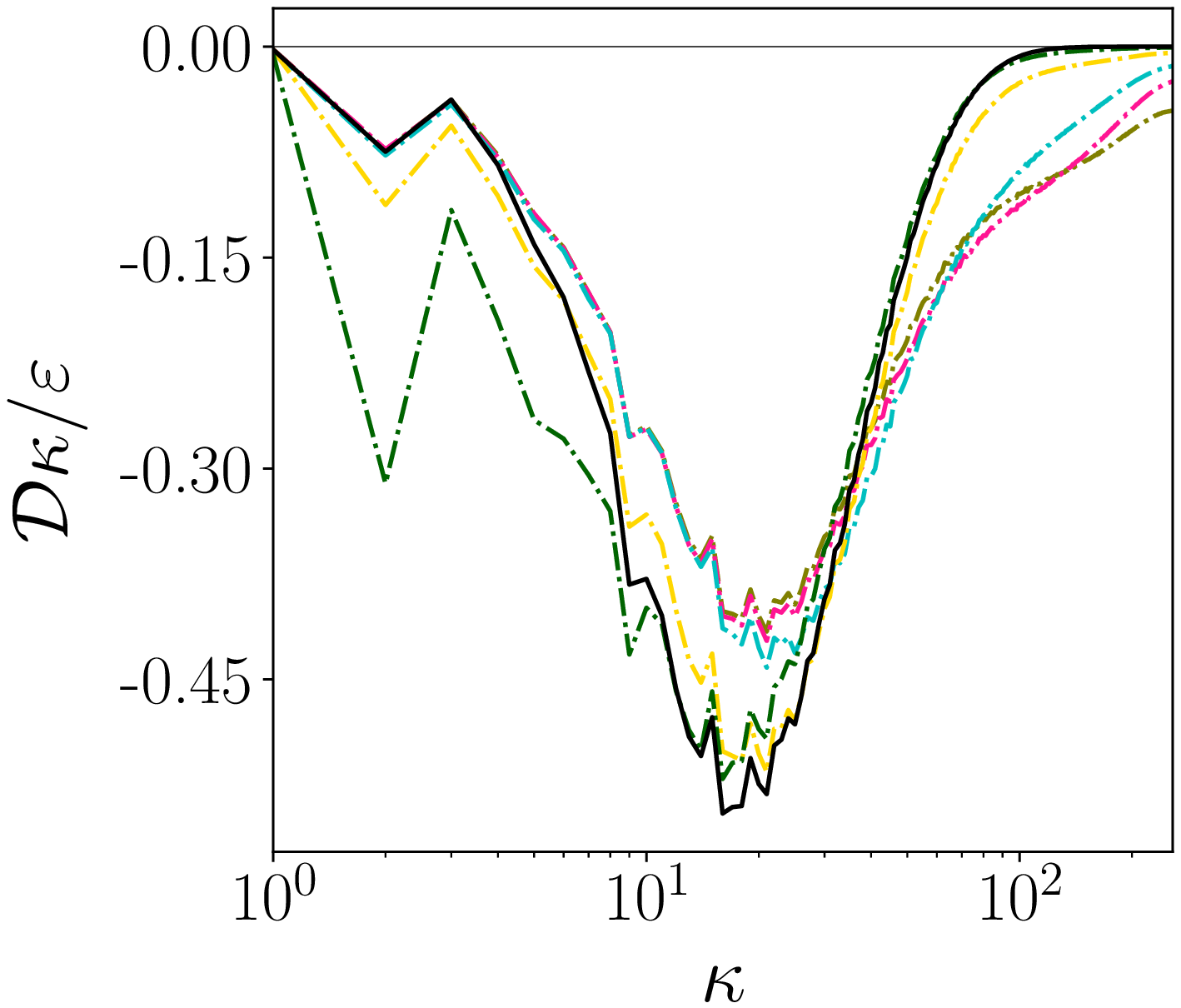}
	\put(-190,140){(\textit{d})}
	\caption{Scale-by-scale energy budget for different viscosity ratios $\mu_d/\mu_c$ at $\alpha=0.1$. (\textit{a}) displays the complete energy balance for case V22 with $\mu_d/\mu_c=0.1$; panels (\textit{b-d}) show the non-linear energy transfer $T$, the term $\mathcal{S}_\sigma$ associated with the surface tension and the energy dissipation transfer function $\mathcal{D}$.
	}
	\label{fig:balance_mu10}
\end{figure}

We present the SBS energy budget for $\alpha=0.1$ in \Cref{fig:balance_mu10}. The results  for $\alpha=0.03$ show similar trends, see \Cref{app:visc003} for the details.
%
Following the same scheme as in the previous section, we depict in panel (\textit{a}) the energy balance for case  V22, when $\mu_b/\mu_c=0.1$. 
Similarly to previous observations, the dispersed phase absorbs energy at large scales and redistributes it to small scales, that is  the presence of the interface provides an alternative path for energy transfer from small to large wavenumbers and no inverse cascade is observed. The non-linear energy transfer $T$, see panel (\textit{b}), displays a weak sensitivity to the viscosity ratio (almost negligible for $\alpha=0.03$ as shown in \Cref{fig:balance_mu} in  \Cref{app:visc003}). 
Thus, the differences in the $Re_\lambda$ and energy spectra discussed above are not associated with an extension of the inertial range.
For wavenumbers larger than that of the forcing,
the non-linear energy transfer is higher at large scales and lower at small scales than for the single-phase case. 

Panel (\textit{c}) in \Cref{fig:balance_mu10} show the  energy transport due to the surface tension term,  $\mathcal{S}_\sigma$.  
As $\mu_d/\mu_c$ increases, the wavelength where the positive energy transport is maximum shifts to larger scales. This behavior is possibly due to increased coalescence  for high $\mu_d/\mu_c$ 
( as  discussed later in this section). 
We also observe that with decreasing viscosity ratio, $\mu_d/\mu_c<1$, the curves tend to collapse, as the data for $\mu_d/\mu_c=0.1$ and $\mu_d/\mu_c=0.01$ are approximately overlapping.
At the injection scale, $\kappa=2$, almost all cases behave similarly. At intermediate wavelengths, the lower the viscosity ratio, the higher the energy absorbed by the surface tension forces. 
 As previously observed, the Hinze  scale represents, to a good approximation, the point where the energy transfer towards small scales by the surface tension term $\mathcal{S}_\sigma$ is maximum. All these observations apply to the two values of $\alpha$ considered, see also \Cref{app:visc003}.

A note should be made on the flow with the highest viscosity ratio: in this case, the energy is not transferred down to the dissipative range. 
A qualitative explanation is given by the following scenario. When the interface interacts with a sufficiently large vortex in the carrier phase, it tends to deform and, in doing so, absorbs energy through the work of the interfacial stresses. The deformation of the interface induces shear in the dispersed phase which is opposed by viscous forces. A higher viscosity in the dispersed phase will therefore dump larger and more energetic structures, reducing the energy available at small scales.  

Finally, panel (\textit{d}) of the same figures shows  the transfer function of the energy dissipation term $\mathcal{D}$. 
We observe that simulations with higher viscosity of the dispersed phase dissipate more energy at large scales, hence dumping turbulence in the inertial range, as expected for more viscous flows. 
This trend is maintained until the dissipative range, where, instead, a lower viscosity ratio produces higher dissipation. 
This causes the apparently paradoxical situation that, despite there is limited  energy transport  by the non-linear terms, dissipation is still active at smalls scales because of the energy brought by the interfacial stresses; this may suggest the need of a specific definition of dissipative range for multiphase flows.

\begin{figure}
	\centering
	\includegraphics[width=0.5\textwidth]{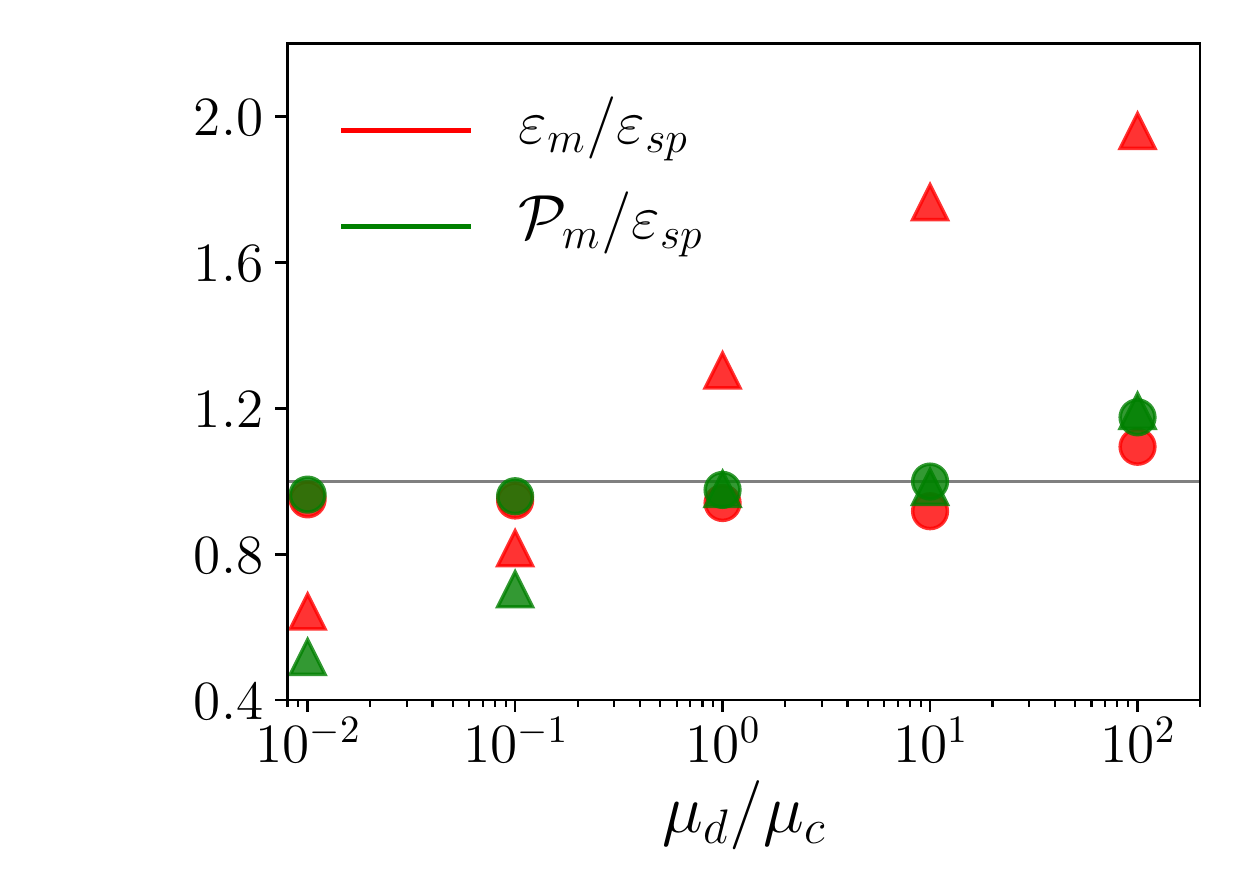}
	\put(-180,130){(\textit{a})}
	\includegraphics[width=0.5\textwidth]{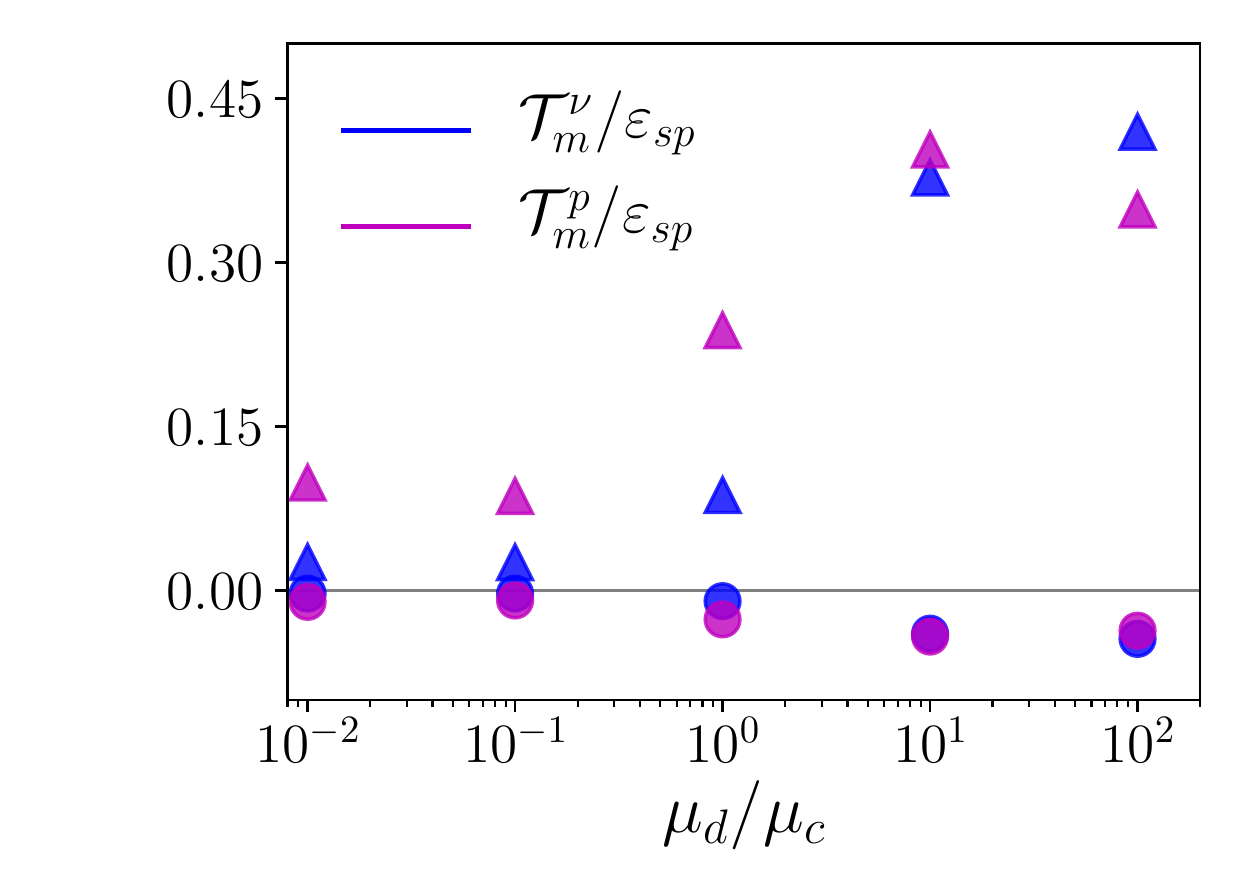}
	\put(-180,130){(\textit{b})}
	\caption{Phase-averaged energy balance versus the emulsion viscosity ratio, see definitions of each term in \Cref{eq:enBalMP}.
	Colored triangles (\mytriangle{black}) represent the dispersed phase ($m=d$) while circles (\mycircle{black}) are used for the carrier phase ($m=c$). Each term is normalized by the single phase energy dissipation $\varepsilon_{sp}$ computed for case SP2. The energy production $\mathcal{P}_m$ and energy dissipation $\varepsilon_m$  are reported in panel (\textit{a}), while viscous energy transport $\mathcal{T}^\nu_m$  and the pressure energy transport $\mathcal{T}^p_m$ in panel (\textit{b}).
	}
	\label{fig:balPhase_mu10}
\end{figure}

Next, we discuss the influence of the viscosity ratio on the phase-averaged energy budget, shown in  \Cref{fig:balPhase_mu10} for $\alpha=0.1$. As for the SBS balance, the same analysis for $\alpha=0.03$ can be found in \Cref{app:visc003}, as the variation of volume fraction does not significantly chance the underlying physical process.
The production density (green symbols in panel \textit{a}) shows only slight variations with viscosity ratios in the carrier phase, whereas it increases in the dispersed phase  when its viscosity increases; in particular, $\mathcal{P}_d<\mathcal{P}_c$ when $\mu_d<\mu_c$.
A similar trend is observed for the dissipation rate (red symbols in panel \textit{a}), when the differences between dispersed and carrier phases become more evident. In this case, the dissipation in the dispersed case increases with its viscosity until $\mu_d/\mu_c=100$, when it decreases because of the lower energy transferred to smaller scales inside the droplets.
The transport terms, $\mathcal{T}^\mu$ and $\mathcal{T}^p$ in panel (\textit{b}), indicate that energy is always transferred from the carrier to the dispersed phase. 
Both terms increase in magnitude when decreasing the viscosity ratio, 
indicating that energy needs to be supplied to the dispersed phase to sustain turbulence when viscous forces are increasing. 
The pressure transport is the preferential path for energy transfer from the carrier to the dispersed phase for low and moderate values of $\mu$. For the case with largest viscosity of  the dispersed phase, the transfer due to pressure forces becomes lower than that associated to viscous forces.

\begin{figure}
	\centering
	\includegraphics[width=0.33\textwidth]{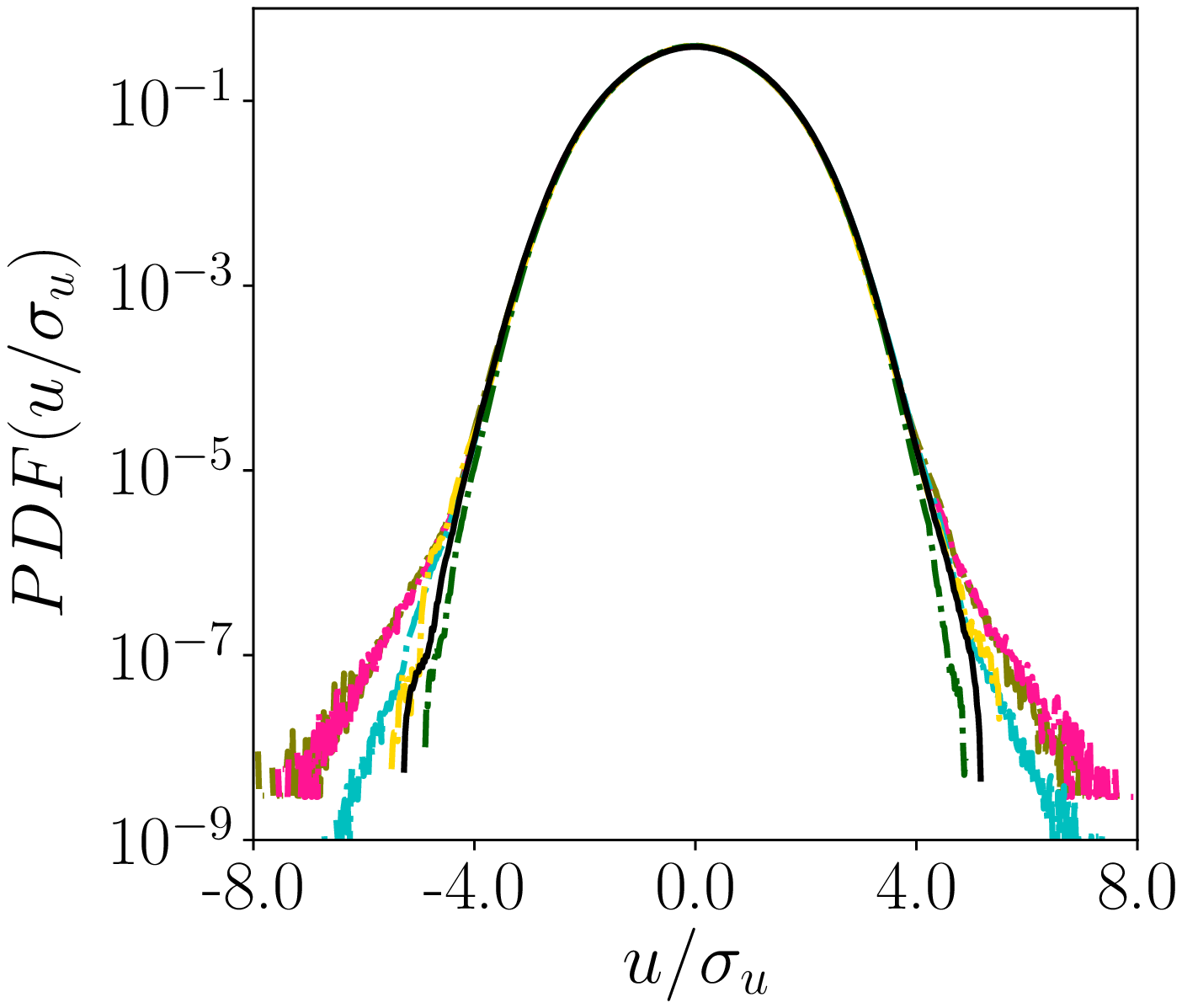}
	\put(-120,110){(\textit{a})}
	\includegraphics[width=0.33\textwidth]{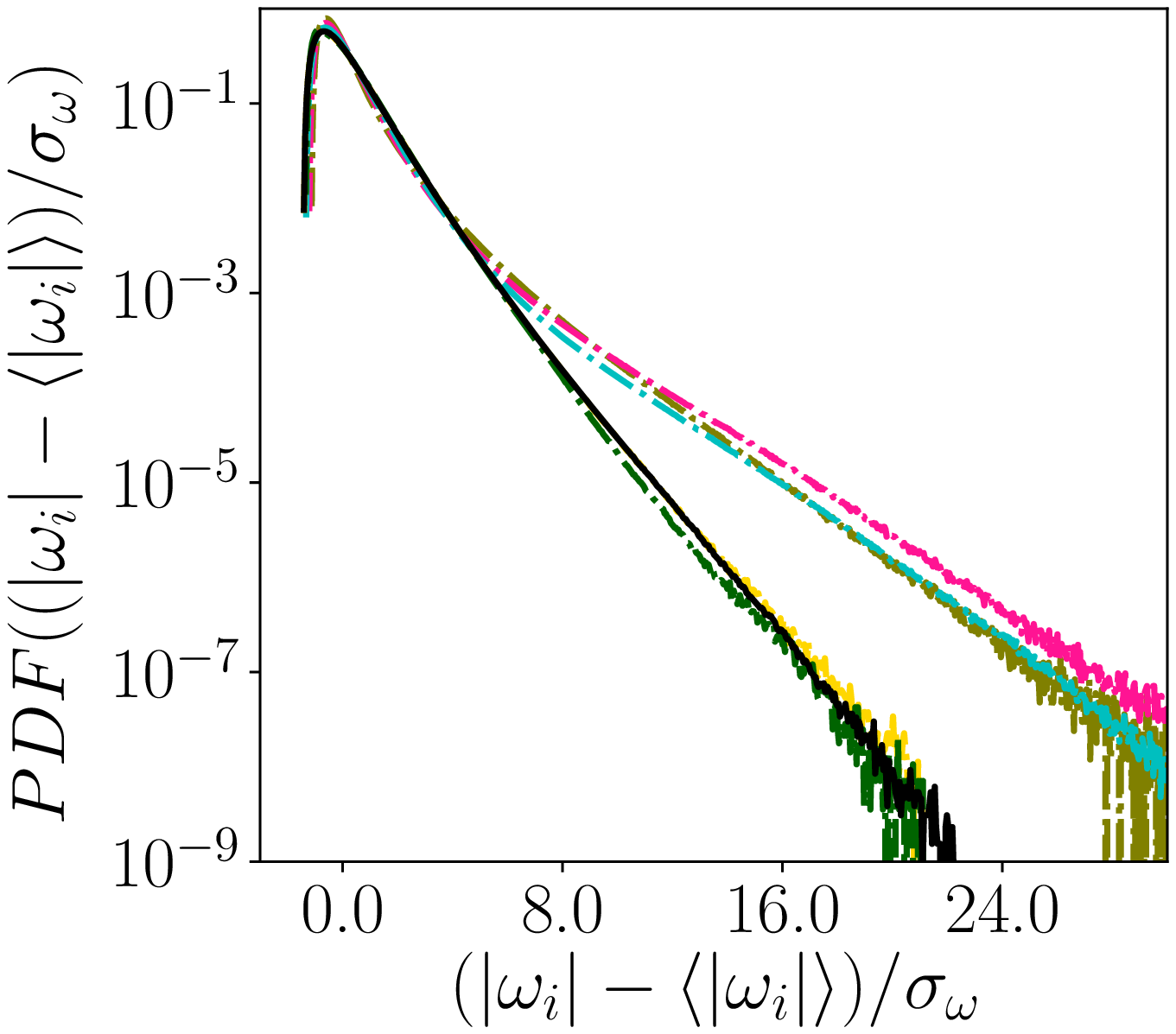}
	\put(-120,110){(\textit{b})}
	\includegraphics[width=0.33\textwidth]{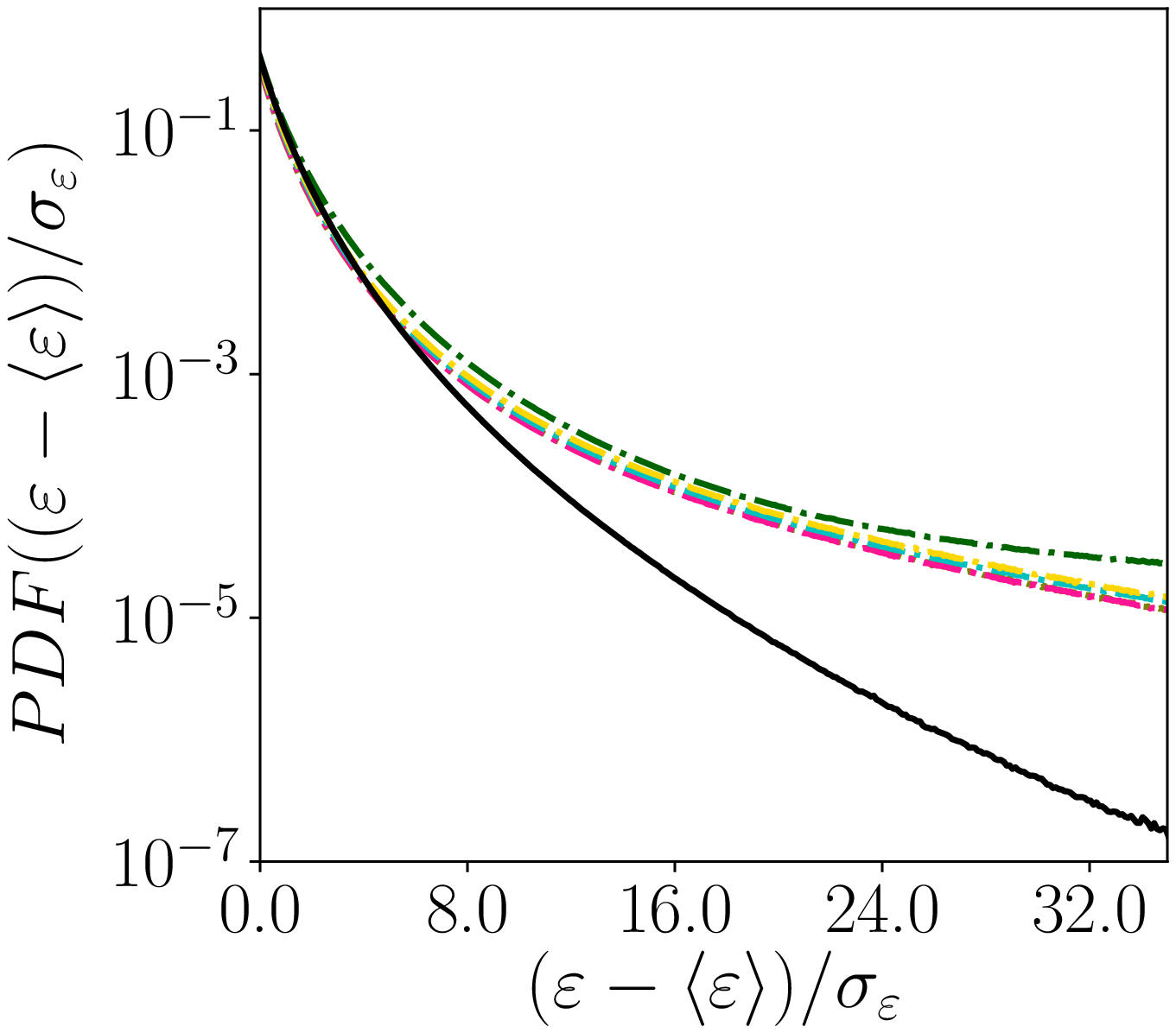}
	\put(-120,110){(\textit{c})}
	
	\includegraphics[width=0.8\textwidth]{figures/legend_mu.eps}
	\caption{PDF of velocity fluctuations $u$ (panel \textit{a}), vorticity $\omega$ (\textit{b}) and energy dissipation (\textit{c}). All quantities are normalized by their standard deviation. The data pertain cases V2x in \Cref{tab:testMat}, with $\alpha=0.1$.}
	\label{fig:PDF_mu10}
\end{figure}

Finally, we consider the effect of the viscosity ratio on the  PDFs of velocity, vorticity and dissipation rate, see \Cref{fig:PDF_mu10} for $\alpha=0.1$, while data for $\alpha=0.03$ can be found in \Cref{app:visc003}.
A low viscosity in the dispersed phase generates larger velocity fluctuations (see also \Cref{fig:enske_mu}), hence the tails of PDFs are more evident for small values of $\mu_d/\mu_c$
in panel (\textit{a}). 
Interestingly, when the viscosity ratio increases above unity, velocity fluctuations in the dispersed phase are quenched, the standard deviation decreases and the statistics are closer to those of the single-phase reference case. 
The distributions of the normalized vorticity are shown in panel (\textit{b}). As for the velocity fluctuations, a higher viscosity in the dispersed phase decreases the 
intermittency and the distributions  approach the single-phase values. 
For $\mu_d/\mu_c<1$,  intermittency increases and the tail of the distribution are more evident; nonetheless they can still be fitted  with decaying exponentials. 
As previously observed for varying $\alpha$, the pseudo-Gaussian part of the vorticity PDF collapse for all cases.

The PDF of the energy dissipation show almost no alteration between cases with different viscosity ratios, while intermittency is strongly increased with respect to the single-phase case. Due to the normalization with $\sigma_\varepsilon$,  the curve collapse indicates that 
variations induced by  $\mu_d/\mu_c$ of the small-scale dynamics are negligible.

To conclude, 
the turbulence is significantly affected by variations of the viscosity ratio already at small volume fractions. Higher viscosity of the dispersed phase
dampens the small-scale structures because of 
 higher viscous dissipation at all wavelengths. For emulsions with viscosity of the dispersed phase lower than that of the carrier phase,
 the activity at small scales increases and so does intermittency. The surface tension term $\mathcal{S}_\sigma$ significantly contributes to the transfer of energy to the smallest scales in this case.

\subsection{Influence of Weber number}
\label{sec:res:we}

\begin{figure}
	\centering
	\includegraphics[width=0.3\textwidth]{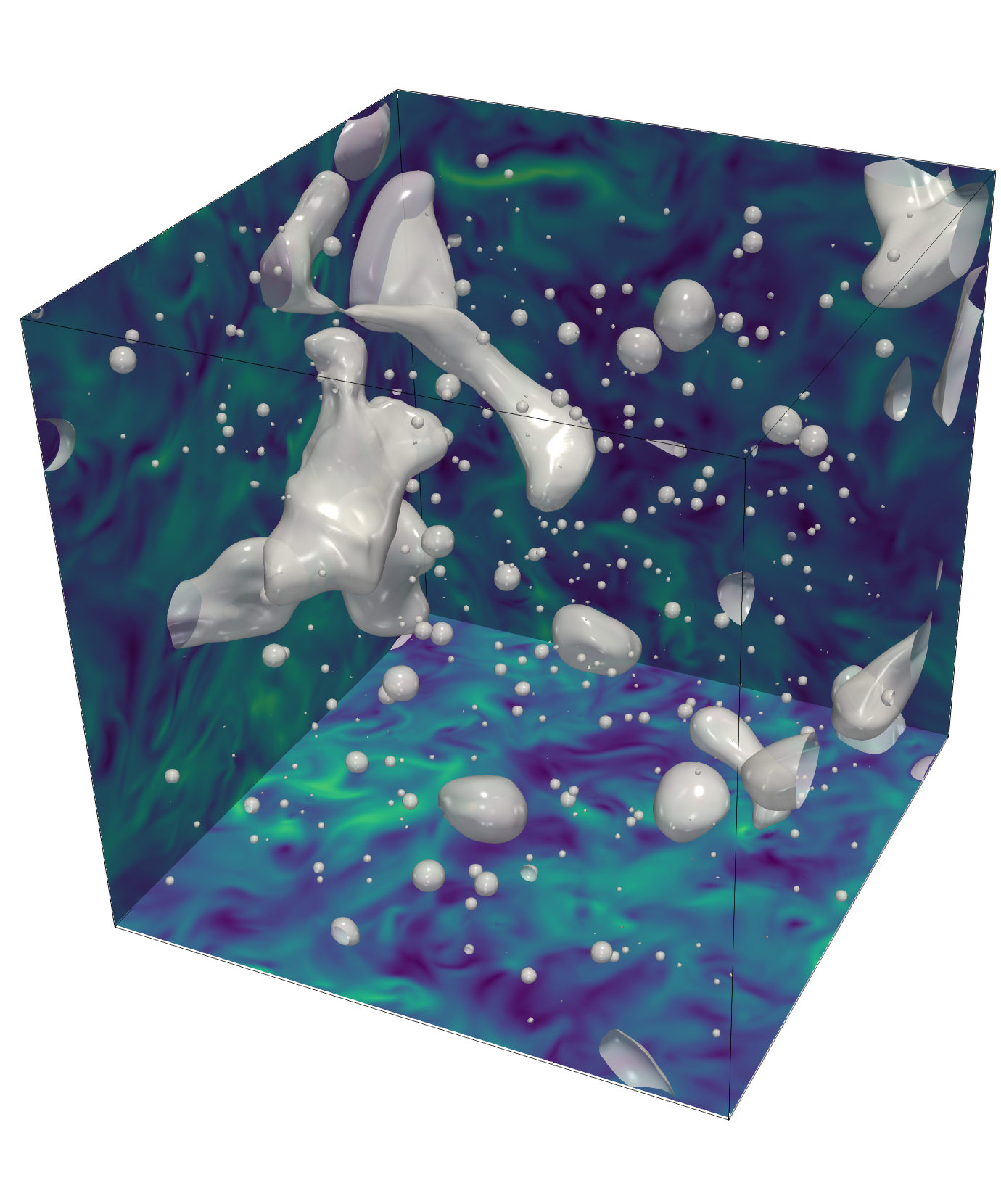}
	\put(-100,140){{$We_\mathcal{L}=10.6$}}
	\includegraphics[width=0.3\textwidth]{figures/tinf_c.png}
	\put(-100,140){{$We_\mathcal{L}=42.6$}}
	\includegraphics[width=0.3\textwidth]{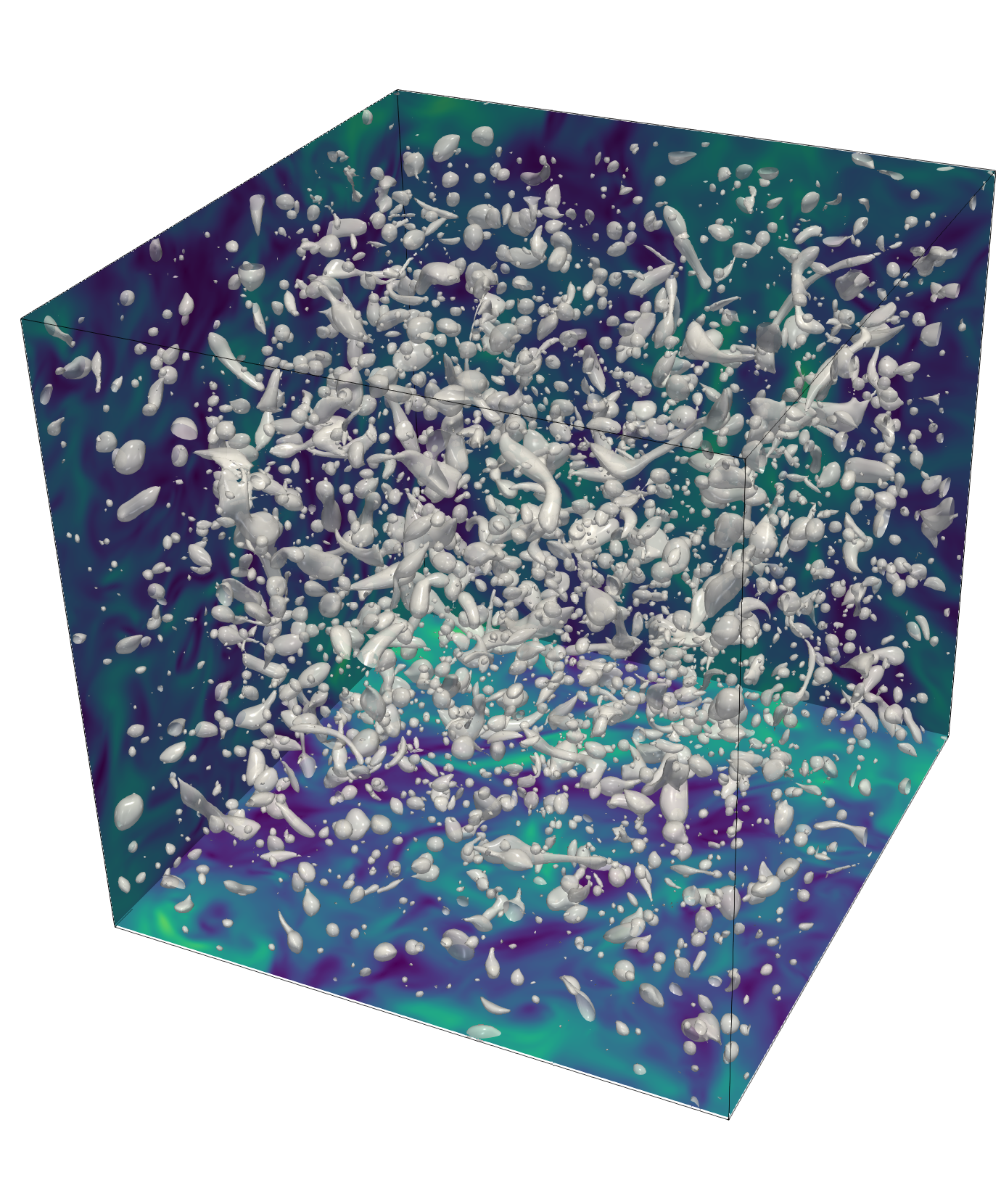}
	\put(-100,140){{$We_\mathcal{L}=106.5$}}
	\caption{Render of the two-fluid interface (corresponding to the value of the VOF function $\phi=0.5$) for different values of the Weber number $We_\mathcal{L}$ (left to right, 0.06, 0.2 and 0.5). 
	The vorticity fields are shown on the box faces on a planar view. 
	All simulations are performed at $\alpha=0.03$ and $\mu_d/\mu_c=1$.}
	\label{fig:weRender}
\end{figure}

The influence of the surface tension coefficient, expressed through the large scale $We_{\mathcal{L}}$ number, is examined in this section. As discussed in literature \citep{Komrakova2015, Roccon2017, Mukherjee2019}, the combination of volume fraction, surface tension coefficient and energy injection scale, $\mathcal{L}$, has to be accurately chosen because the HIT configuration is very sensitive to coalescence. 
Furthermore, high $We_{\mathcal{L}}$ may generate an excess of unresolved droplets, significantly affecting the results. 
Therefore, all the simulations discussed in this section are performed at $\alpha=0.03$, while the forcing is maintained at $\kappa_0=2$.
The cases discussed in this section are BE1 and the series W1x with reference to \Cref{tab:testMat}, covering a significant range of $We_{\mathcal{L}}$, from 10.6 to 106.5. In \Cref{fig:weRender} we show a render of the flow for different values of $We_\mathcal{L}$. 
As expected, at low $We_\mathcal{L}$ we observe the appearance of large liquid structures due to higher surface tension forces. At high $We_\mathcal{L}$, on the other hand, the dispersed phase undergoes severe fragmentation. 
The presence of small droplets resulting from fragmentation can be observed in all cases.

\begin{figure} 
	\centering
	\includegraphics[width=0.49\textwidth]{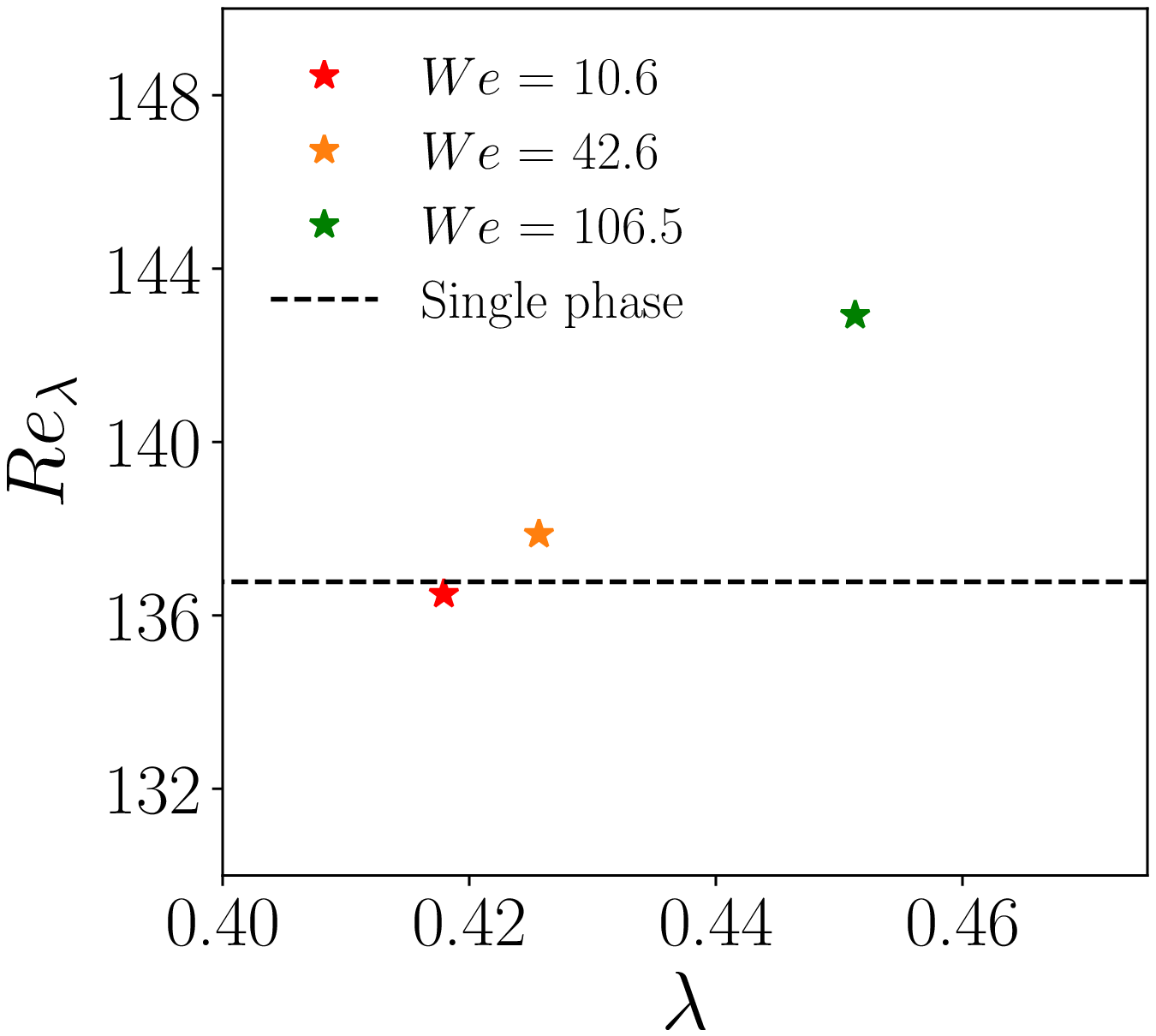}
	\put(-185,140){(\textit{a})}
	\put(-90,25){\includegraphics[height=1.75cm]{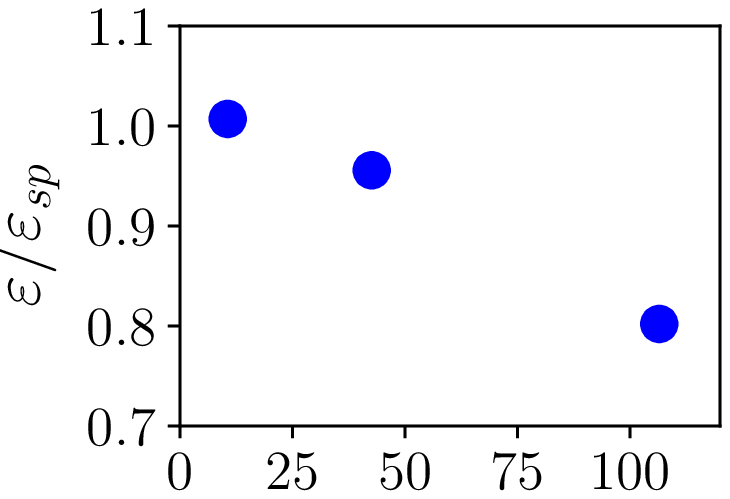}}
	\includegraphics[width=0.49\textwidth]{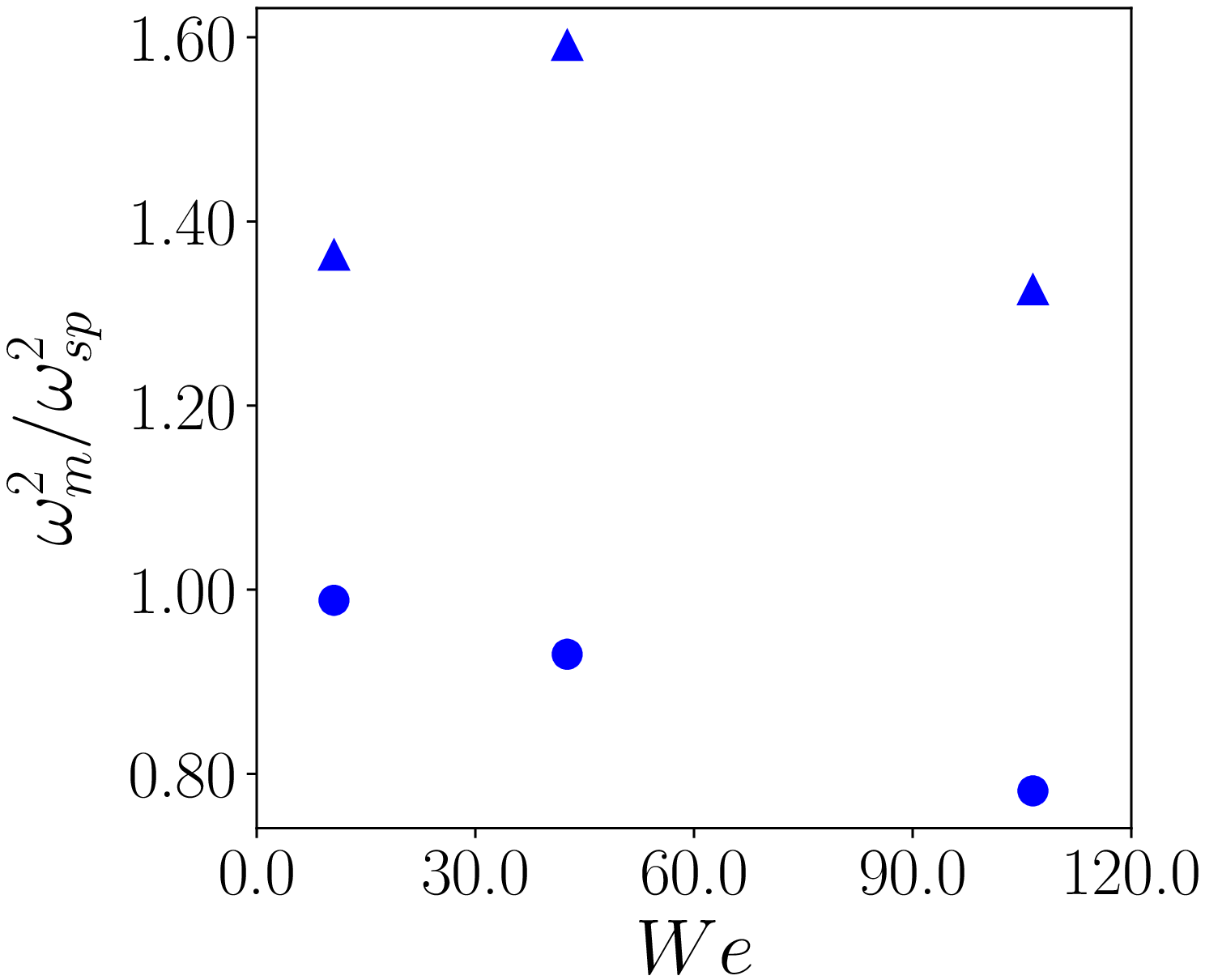}
	\put(-190,140){(\textit{b})}
	\caption{ (\textit{a}) $Re_\lambda$ versus the Taylor scale $\lambda$ and (\textit{b}) phase-averaged enstrophy versus the Weber number $We_{\mathcal{L}}$. 
	The inset of panel (\textit{a}) shows the global energy dissipation $\varepsilon$ (normalized by its single-phase value) as a function of $We_{\mathcal{L}}$. In panel (\textit{b}), colored triangles (\mytriangle{black}) represent the dispersed phase ($m=d$) while circles (\mycircle{black}) indicate data pertaining the carrier phase ($m=c$)}
	\label{fig:ReEns_We}
\end{figure}

We start by studying the global behavior of the flow through the integral quantities in \Cref{fig:ReEns_We}. $Re_\lambda$, reported in panel (\textit{a}), shows an almost linear increment both with the Taylor scale $\lambda$ and
 $We_{\mathcal{L}}$ (represented with colors). Unlike previous observations in \Cref{sec:res:alpha} where the increase of $Re_{\lambda}$ was mostly due to local variations of the ratio $k/\varepsilon$, decreasing surface tension also lowers the volume-averaged energy dissipation, as shown in the inset. These findings are in agreement with the results on turbulent emulsions in \cite{Rosti2020}.
As the viscosity ratio $\mu_d=\mu_c$ is constant, the decrease of the dissipation is caused by lower enstrophy levels, as it can be appreciated from the data for the carrier phase in panel (\textit{b}). The behavior of the enstrophy of the dispersed phase is less intuitive, exhibiting a non-monotonic behavior,  and will be addressed later when discussing the phase-averaged energy balance.

\begin{figure} 
\centering
	\includegraphics[width=0.49\textwidth]{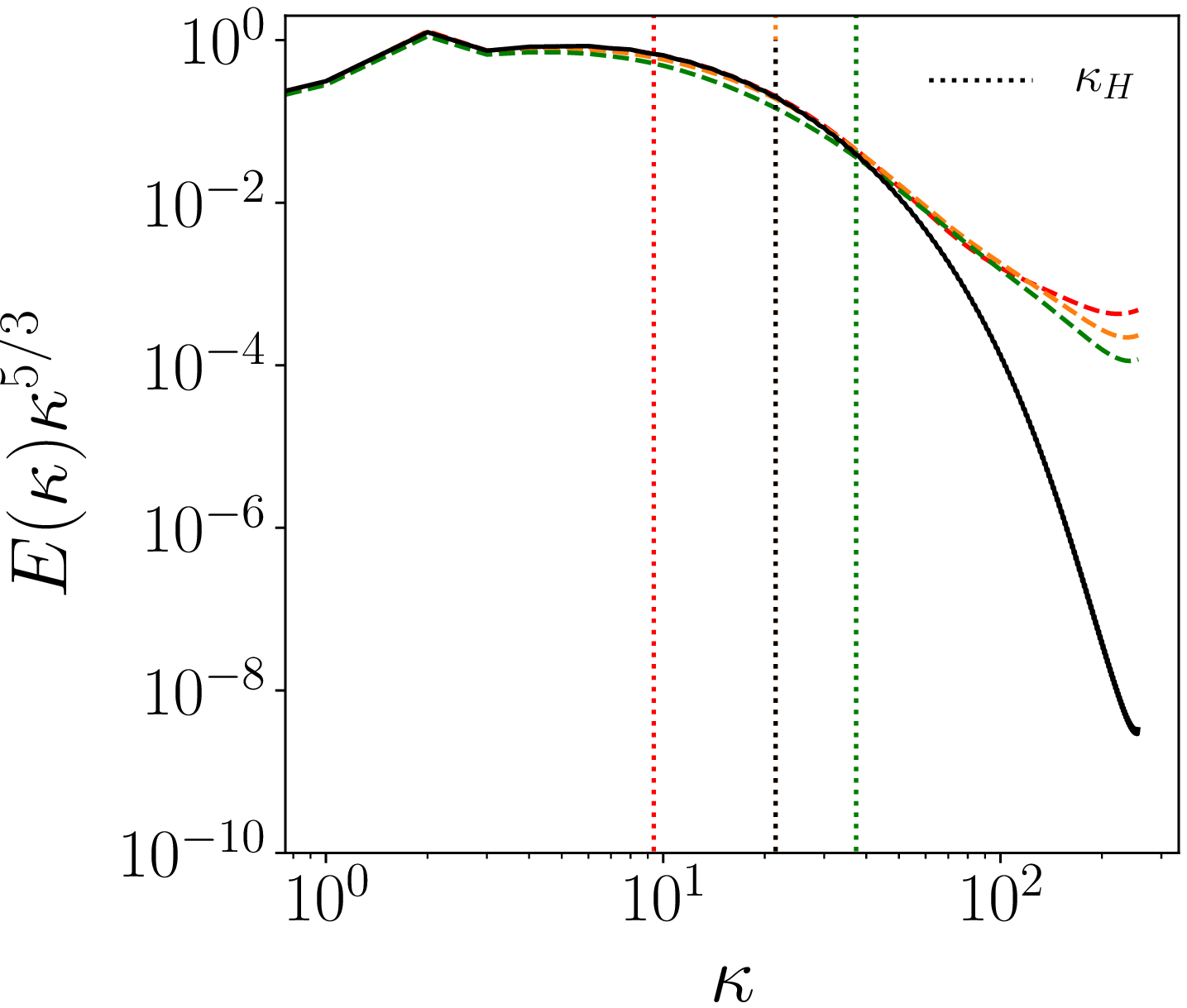}
	\put(-185,140){(\textit{a})}
	\includegraphics[width=0.49\textwidth]{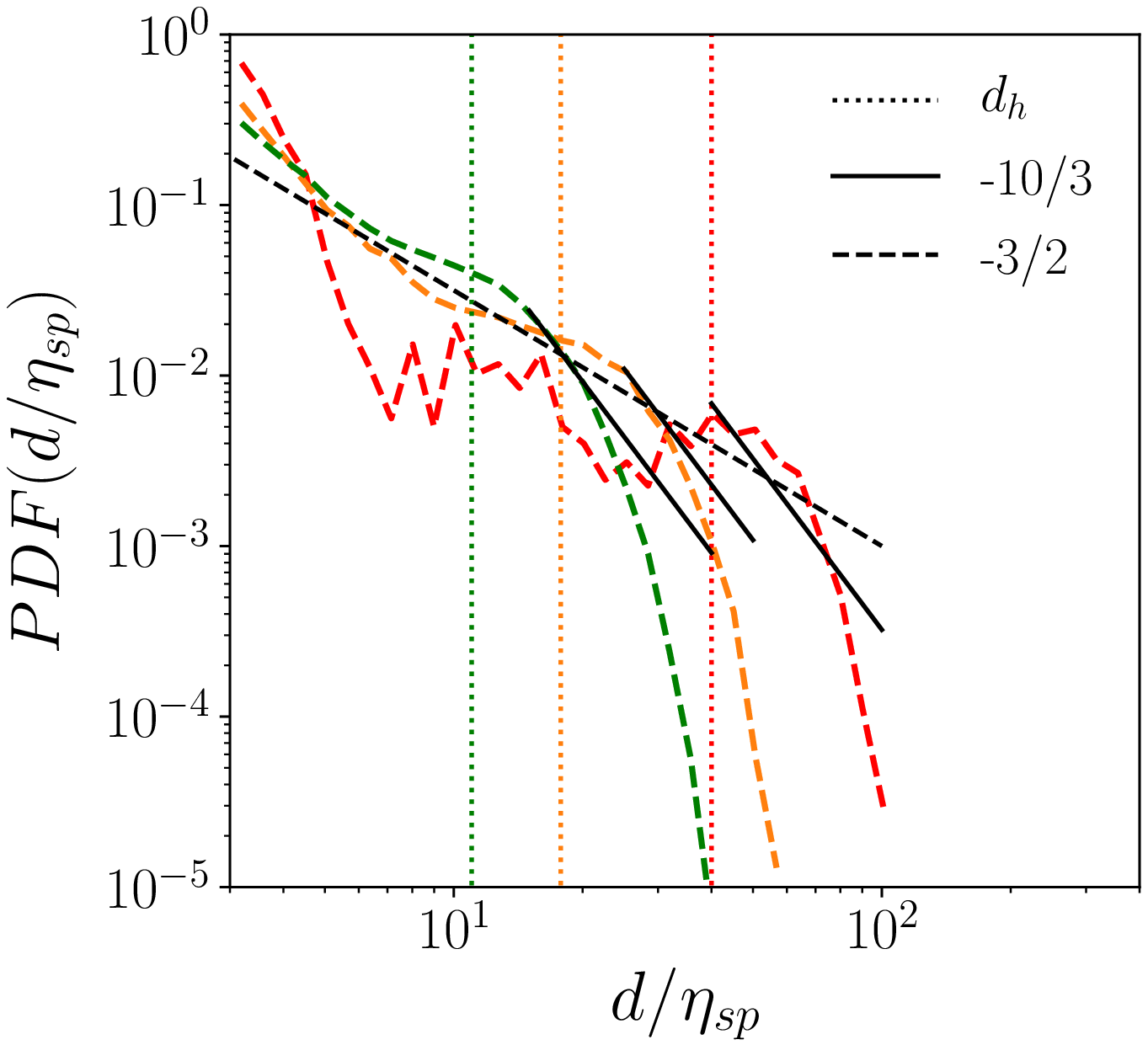}
    \put(-190,140){(\textit{b})}
    
    \includegraphics[width=0.8\textwidth]{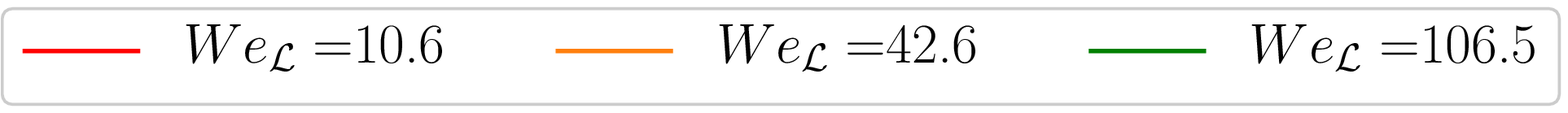}
\caption{(\textit{a}) One-dimensional compensated energy spectra for different large-scale Weber number $We_\mathcal{L}$; the wavelengths corresponding to the Hinze scale of each spectra are plotted with vertical dotted lines of corresponding colors. The inset highlights the differences with the single-phase spectrum. (\textit{b}) Droplet-size-distribution for different $We_\mathcal{L}$; the Hinze scale $d_H$  is reported with dotted lines of corresponding color. The continuous black line represents the region where the -10/3 law applies.}
\label{fig:spectra_We}
\end{figure}

\Cref{fig:spectra_We}(\textit{a}) shows the compensated energy spectra at different $We_\mathcal{L}$. As we are varying the surface tension, the Hinze scale varies in each case (see vertical dotted lines of corresponding color). 
As mentioned before, the Hinze scale defines with good approximation the spectra pivoting point. 
 As previously discussed, energy is reduced at larger scales in the inertial range and increases at smallest scales. 
 With increasing $We_\mathcal{L}$, higher energy is observed at high wavelengths.

\Cref{fig:spectra_We}(\textit{b}) shows the droplet-size-distribution for all the $We_\mathcal{L}$ under investigation. As for panel (\textit{a}) we show the Hinze scale for each case with vertical dotted lines. Again we observe that the -10/3 power-law from \cite{Garrett2000}
 provides a reasonable description for  the largest droplets, $d>d_H$; as we increase $\sigma$, i.e.\ low $We_{\mathcal{L}}$, larger droplets may appear, as expected by the increased cohesion forces. 
In this case, the energy required to breakup large droplets is only available in large eddies. As their turnover time is in the order of $\mathcal{T}$, large droplet breakup becomes a rare event and the distributions are more noisy, so that it is more difficult to identify a clear trend.
In addition, by reducing $We_\mathcal{L}$, the distribution becomes more irregular for $d<d_H$ as most of the dispersed phase is in large droplets. 

\begin{figure}
\centering
	\includegraphics[width=0.5\textwidth]{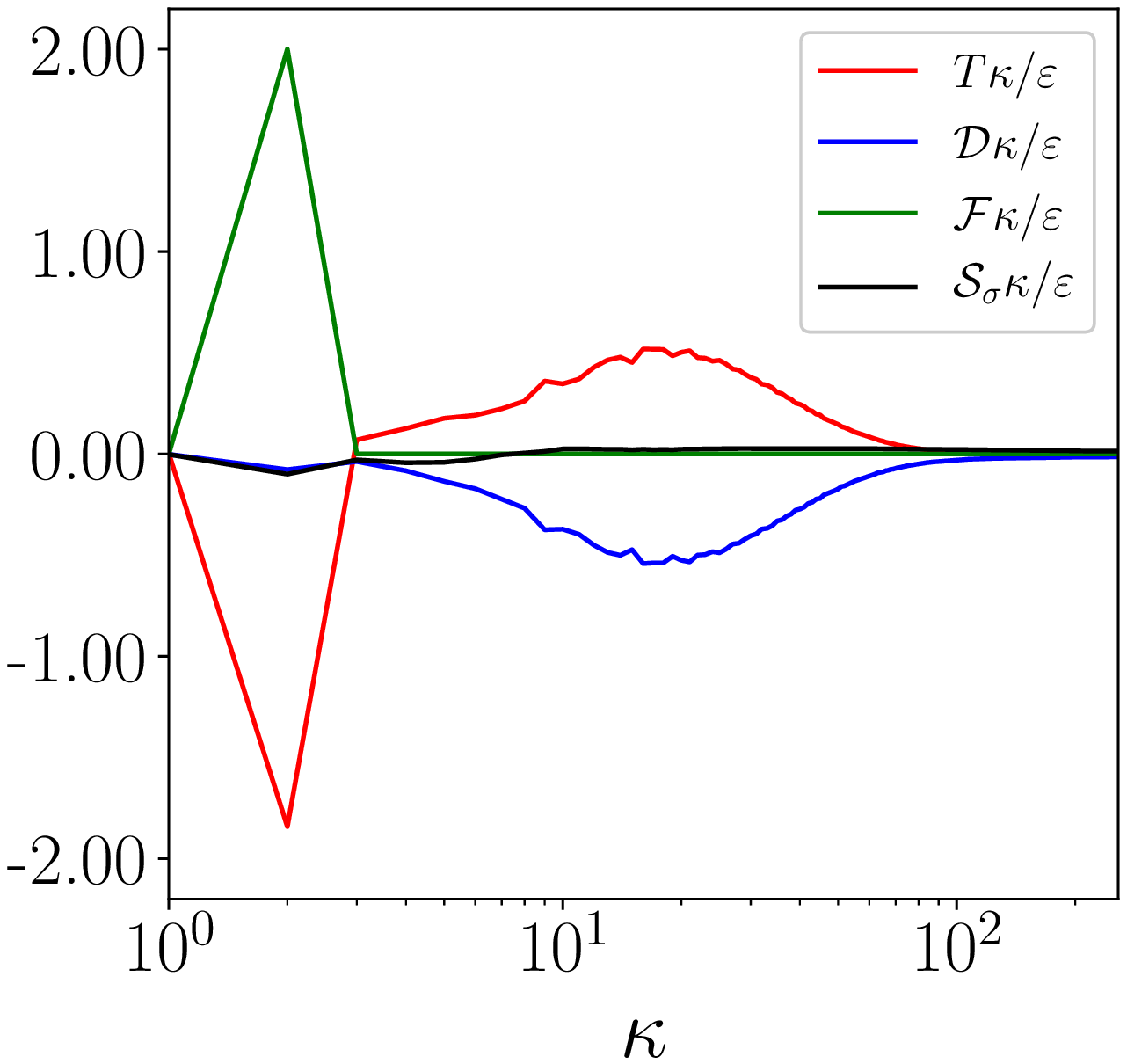}
	\put(-200,140){(\textit{a})}
	\includegraphics[width=0.5\textwidth]{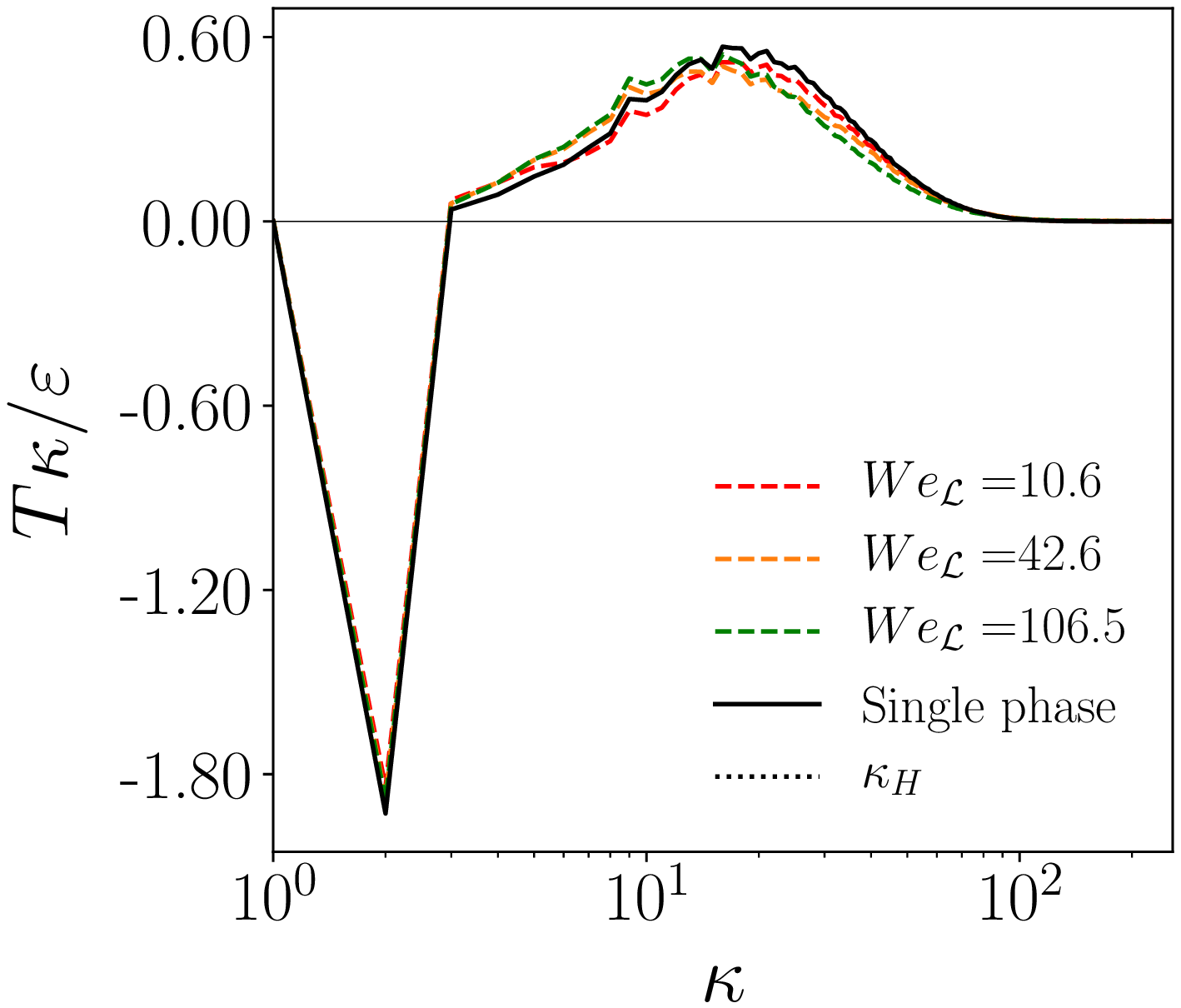}
	\put(-190,140){(\textit{b})}
	\begin{picture}(0,0)
	\end{picture}
	\includegraphics[width=0.5\textwidth]{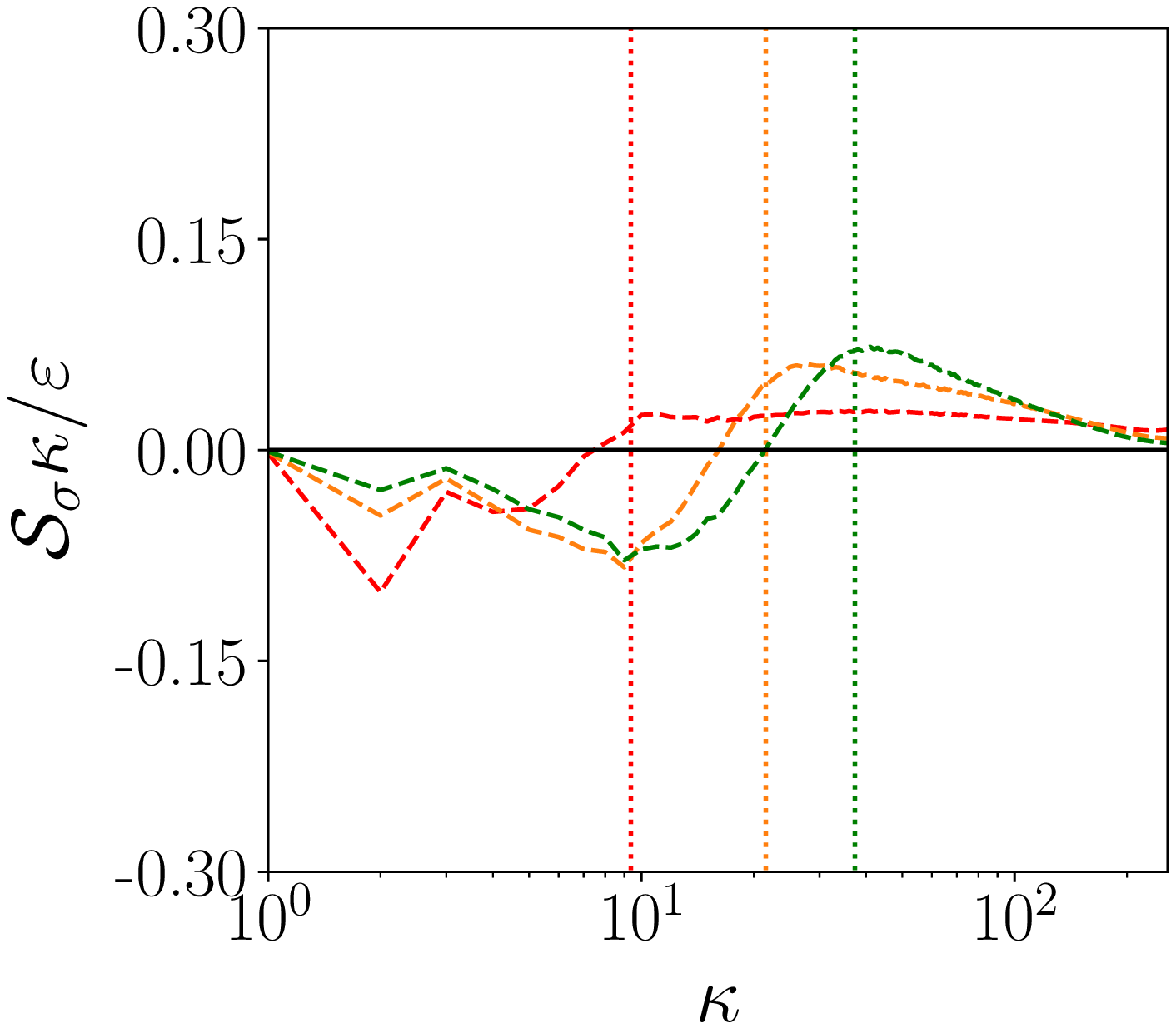}
	\put(-200,140){(\textit{c})}
	\includegraphics[width=0.5\textwidth]{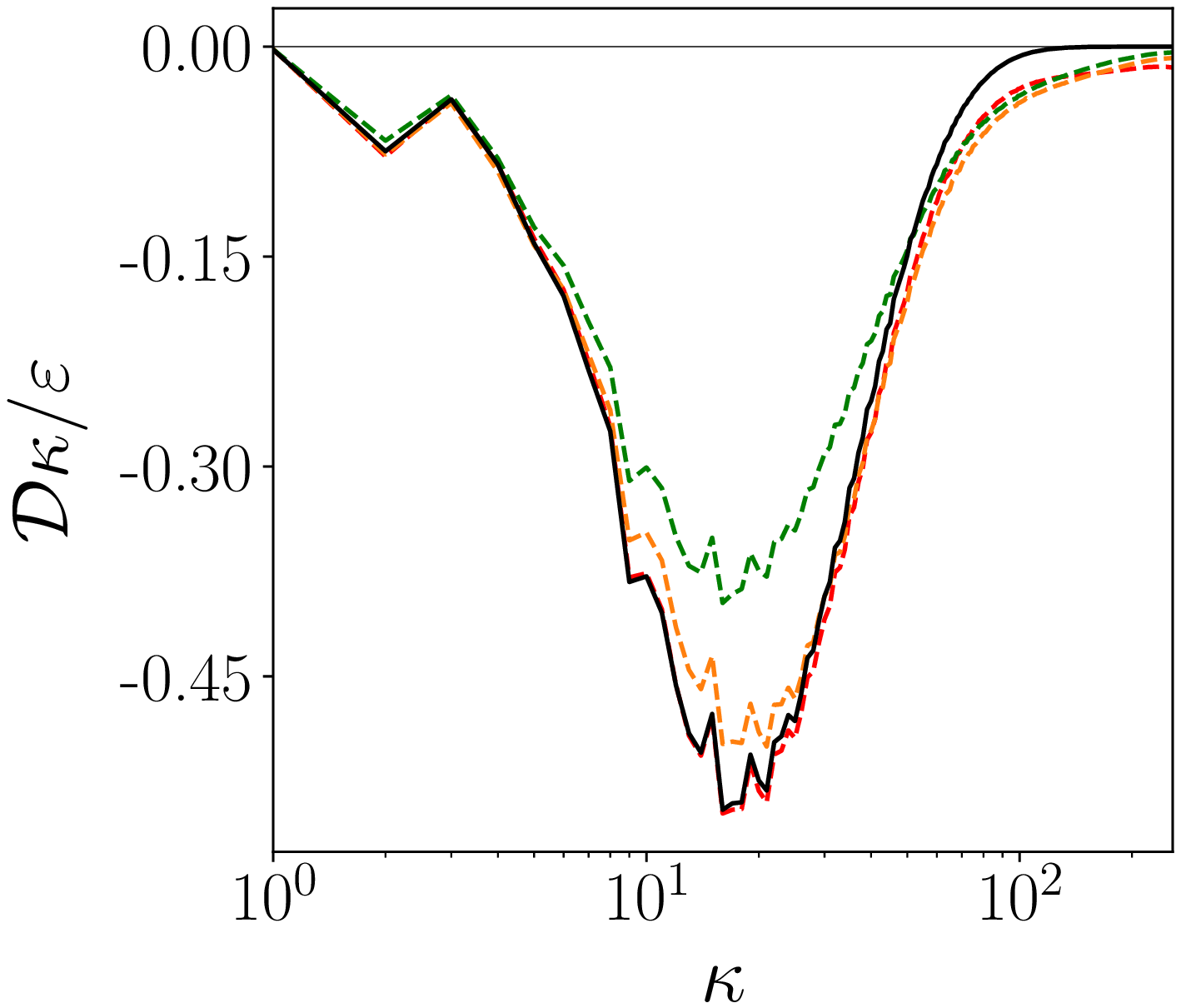}
	\put(-190,140){(\textit{d})}
\caption{Scale-by-scale energy budget for different large-scale Weber numbers, $We_\mathcal{L}$. Panel (\textit{a}) shows the full energy balance for case W11, with $We_\mathcal{L}=10.6$; (\textit{b}) the energy transfer $T$ due to the non-linear term; (\textit{c}) the energy flux $\mathcal{S}_\sigma$ associated with the surface tension term; (\textit{d}) the energy dissipation rate $\mathcal{D}$.}
\label{fig:balance_we}
\end{figure}

\begin{figure}
	\centering
	\includegraphics[width=0.5\textwidth]{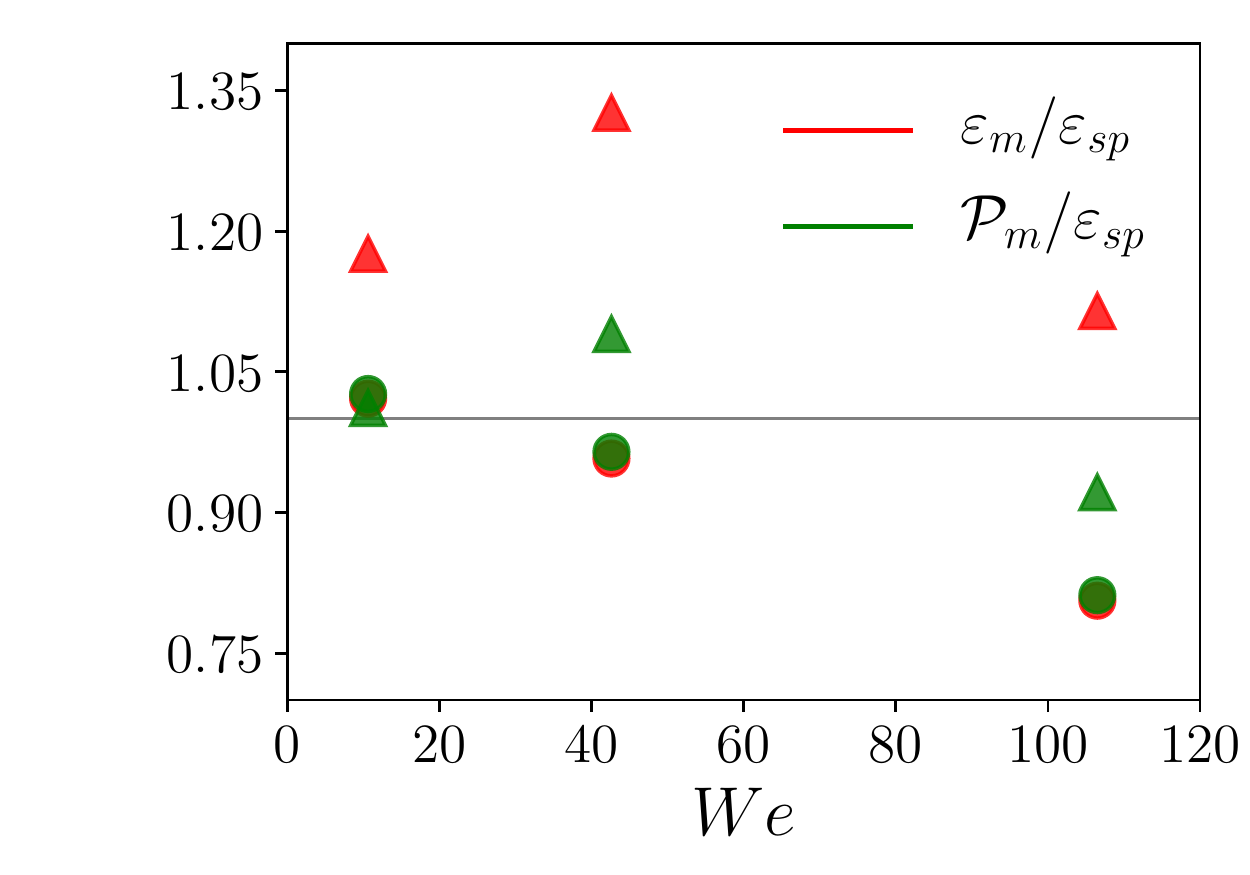}
	\put(-180,130){(\textit{a})}
	\includegraphics[width=0.5\textwidth]{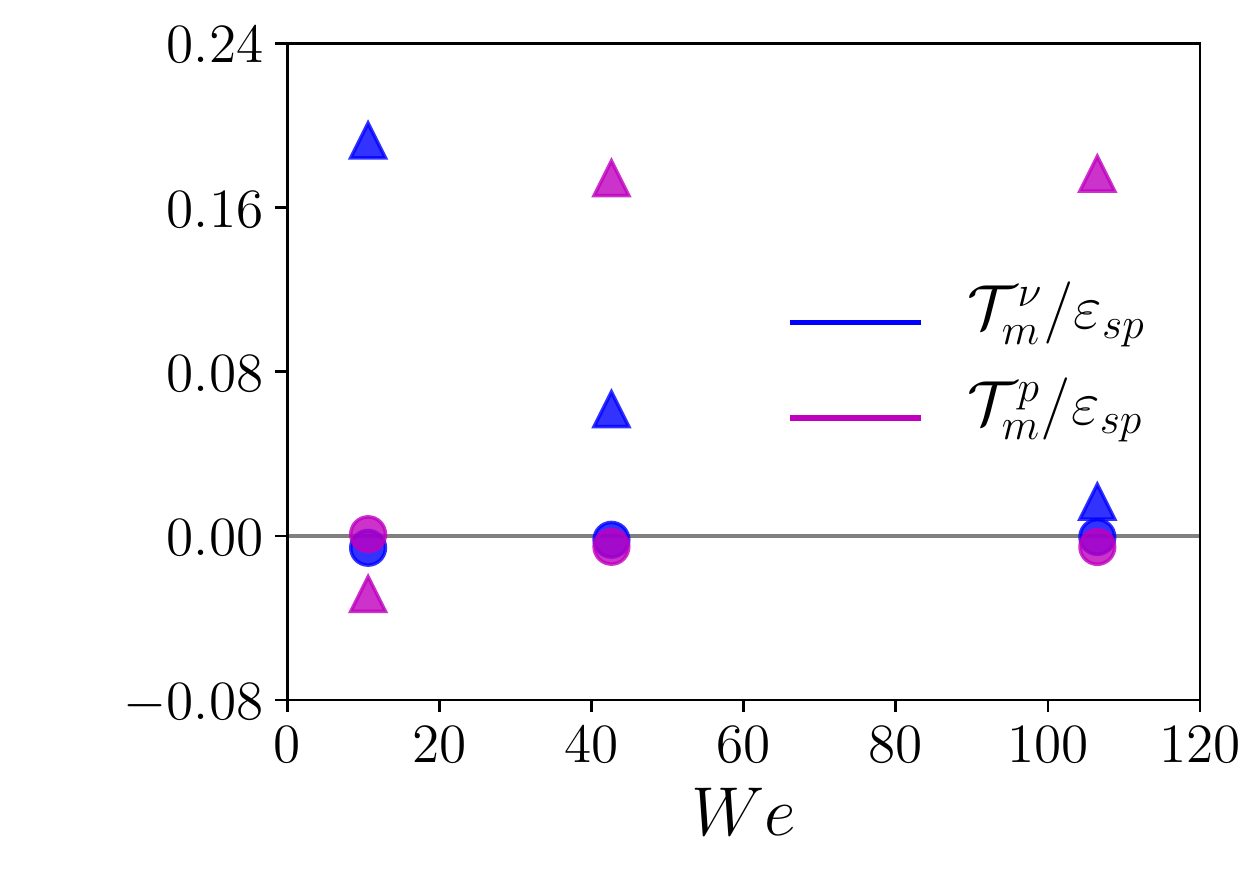}
	\put(-180,130){(\textit{b})}
	\caption{Phase-averaged energy balance versus the emulsion Weber number, see definitions of each term
	 in \Cref{eq:enBalMP}. Colored triangles (\mytriangle{black}) represent the dispersed phase ($m=d$) while circles (\mycircle{black}) indicate data pertaining the carrier phase ($m=c$). Each term is normalized by the single phase energy dissipation $\varepsilon_{sp}$ computed for case SP2. The energy production $\mathcal{P}_m$ and energy dissipation $\varepsilon_m$  are reported in panel (\textit{a}), while viscous energy transport $\mathcal{T}^\nu_m$  and the pressure energy transport $\mathcal{T}^p_m$ in panel (\textit{b})}
	\label{fig:balPhase_we}
\end{figure}

The effects of $We_{\mathcal{L}}$ can be better described by the scale-by-scale analysis, shown in \Cref{fig:balance_we}. 
The complete energy balance is shown for case W11 ($We_\mathcal{L}=10.6$) in panel (\textit{a}). Unlike the the cases shown previously for $We_\mathcal{L}=42.6$, the surface tension energy transfer $\mathcal{S}_\sigma$ is more uniform through the different scales and its effects are globally less evident. 
To deepen the analysis, we display the non-linear energy transfer function $T$ for each case at different $We_\mathcal{L}$ in figure \Cref{fig:balance_we}(\textit{b}). 
At the injection wavelength $\kappa=2$, no major differences are observed when varying the surface tension. At small wavelengths,  $\kappa>2$, we observe that 
the energy transfer by the non-linear term increases with
 $We_\mathcal{L}$,  compensating for the effect of the energy absorption from the surface tension. The energy transfer at smaller scales, after the peak, increases with $\sigma$, approaching the values of the single-phase flow. 

The energy transfer via the interfacial stresses,  $\mathcal{S}_{\sigma}$, is shown in panel (\textit{c}) of the same figure. The energy is again absorbed at large scales and distributed at small scales.  
Flows with small $We_\mathcal{L}$ absorb more energy at small wavenumbers and the transmission of energy (i.e. positive $\mathcal{S}_\sigma$) is smeared over a higher range of scales, hence the peak ($max(\mathcal{S}_\sigma)$) is also less evident. 
For all  $We_{\mathcal{L}}$ investigated, the surface tension term $\mathcal{S}_{\sigma}$ transfers energy also within the dissipative range at small scales, where the transport from non-linear terms has become negligible. 

The energy dissipation $\mathcal{D}$, panel (\textit{d}) of \Cref{fig:balance_we}, decreases with $We_{\mathcal{L}}$ at large and intermediate scales, as energy is partially absorbed by $\mathcal{S}_{\sigma}$. The amplitude of the dissipation rates becomes however almost independent of the Weber number at the smallest scales. Further, as previously observed, the presence of the dispersed phase delays the onset of the dissipative range.

The phase-averaged energy balance from simulations with different Weber number is shown in \Cref{fig:balPhase_we}. 
Both production  and dissipation (panel \textit{a}) are found to decrease in the carrier phase when increasing $We_\mathcal{L}$, while the former increases and then decreases in the carrier phase.
This can be possibly related to the droplet size distributions: decreasing the droplet size increases the internal dissipation, which may explain the behavior at the lower Weber examined. On the other hand, high deformability decreases the dissipation close to the interface, which may explain the decrease at the largest $We_\mathcal{L}$.
 For all values of $We_\mathcal{L}$ considered, the dispersed phase extracts kinetic energy from the carrier phase, as $\mathcal{T}_c>0$ (panel \textit{b}). The decrease of surface tension forces results in a monotonic decrease of the viscous transfer and an increase of the pressure transport for the dispersed phase.  Consistently, dissipation is always higher in the dispersed phase, while it decreases in the carrier phase when increasing $We_\mathcal{L}.$

\begin{figure}
	\centering
	\includegraphics[width=0.33\textwidth]{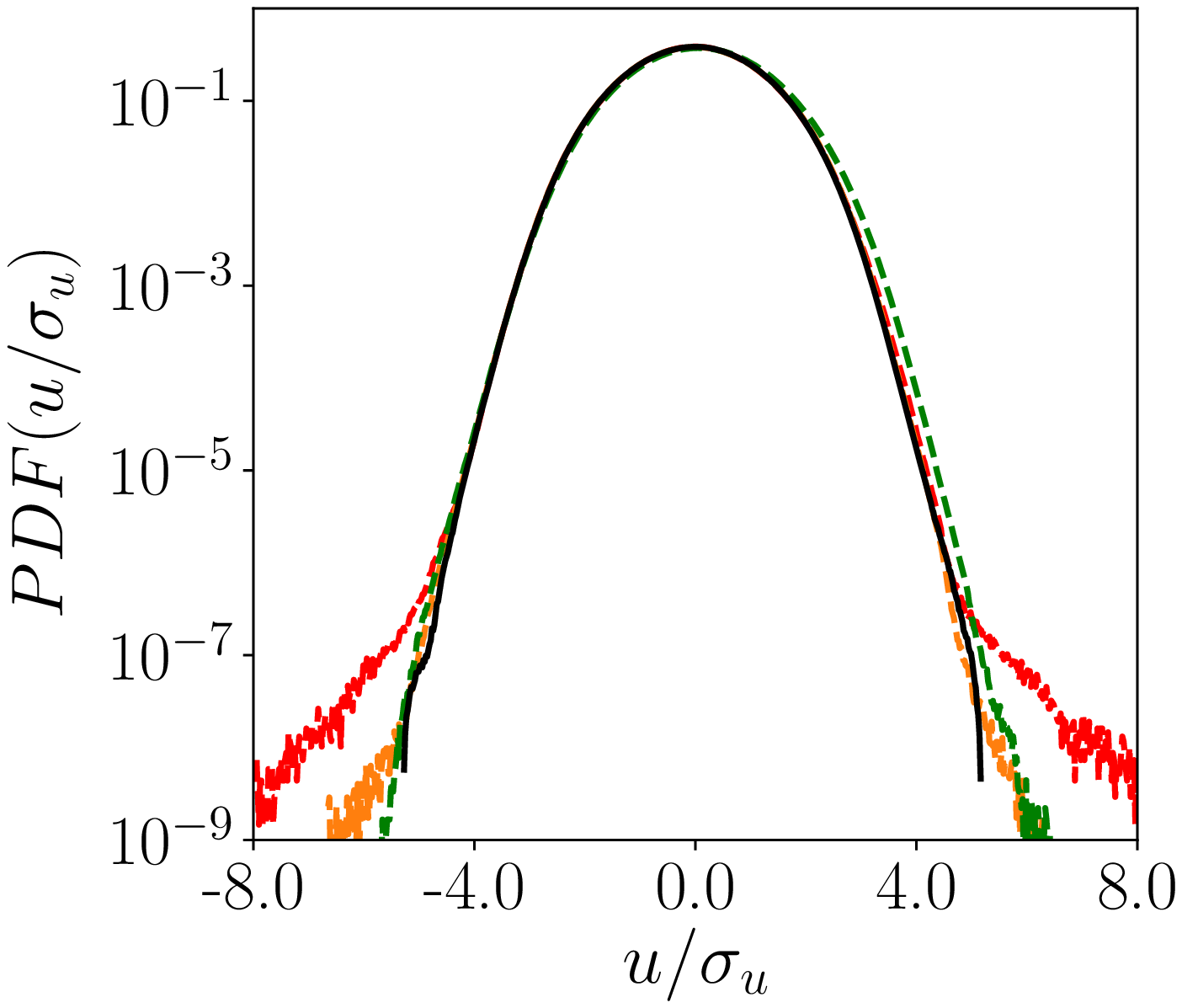}
	\put(-120,110){(\textit{a})}
	\includegraphics[width=0.33\textwidth]{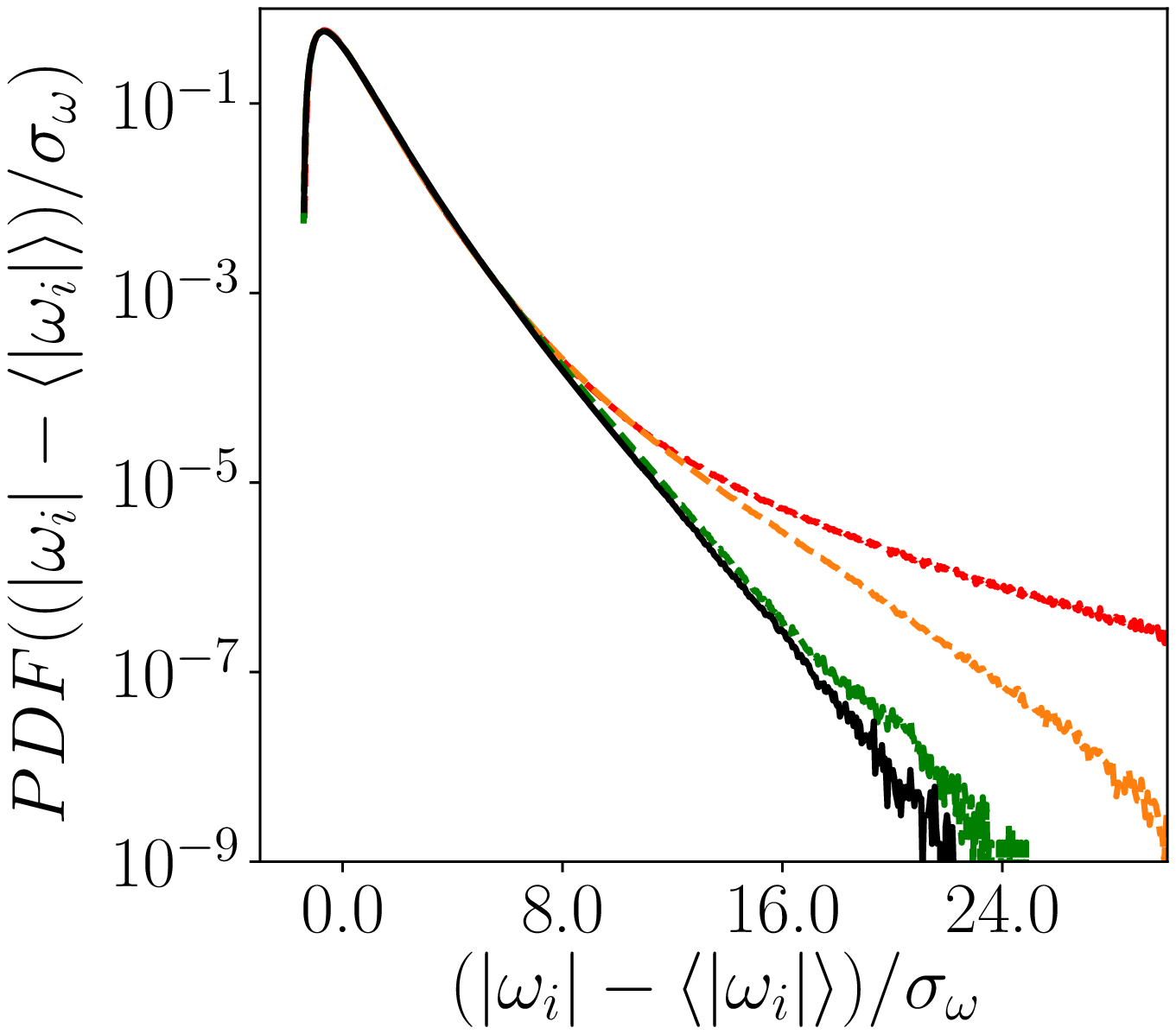}
	\put(-120,110){(\textit{b})}
	\includegraphics[width=0.33\textwidth]{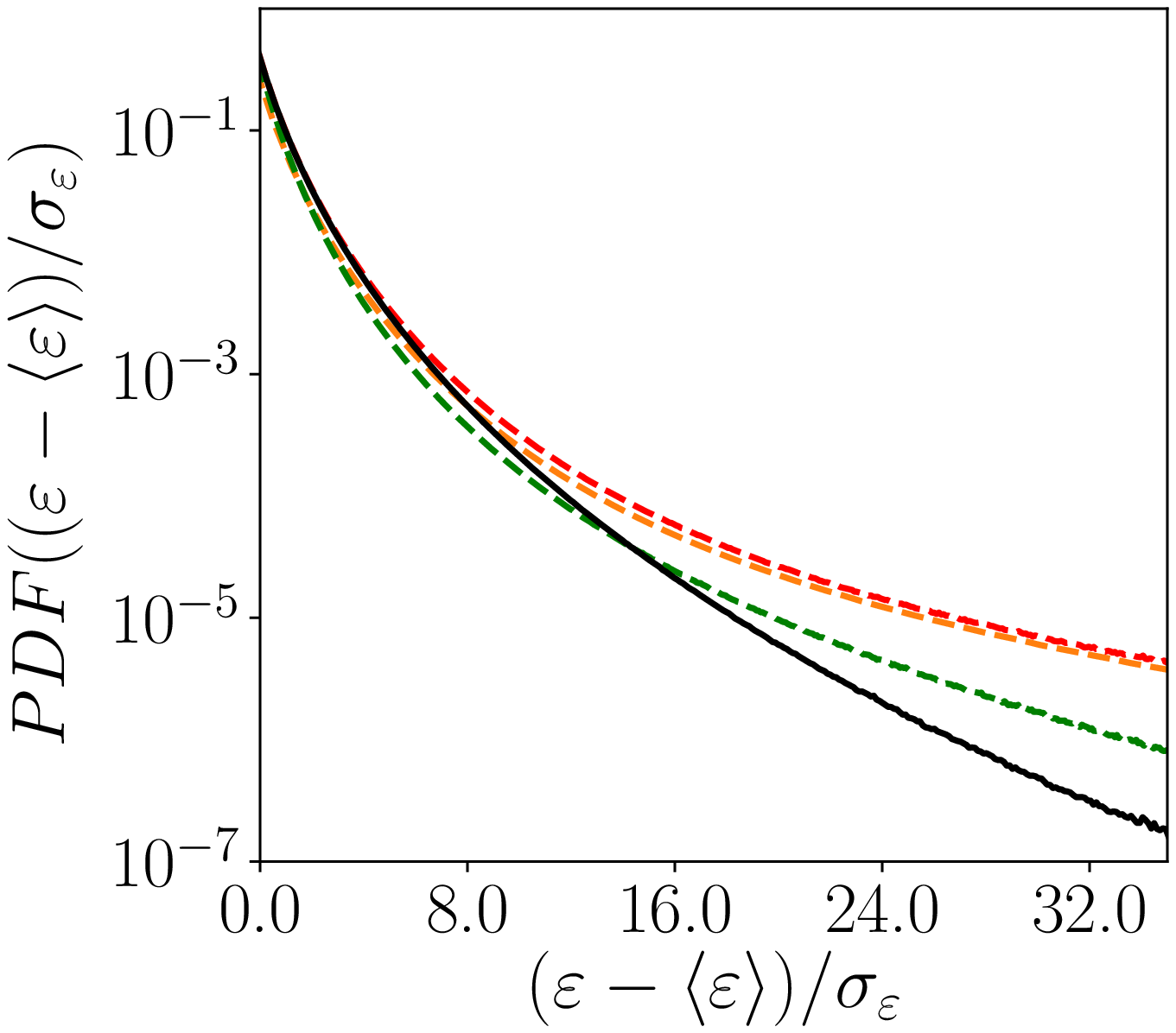}
	\put(-120,110){(\textit{c})}
	
	\includegraphics[width=0.95\textwidth]{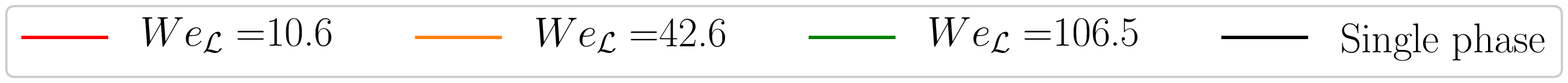}
	\caption{PDF of velocity fluctuations $u$ (panel \textit{a}),  vorticity $\omega$ (\textit{b}) and energy dissipation (\textit{c}). All quantities are normalized by their standard deviation. The data pertain  cases W1x, BE1 and SP1 in \Cref{tab:testMat}.}
	\label{fig:PDF_we}
\end{figure}

The analysis of the PDF for velocity, vorticity and dissipation are finally shown in panels (\textit{a}), (\textit{b}) and (\textit{c}) of \Cref{fig:PDF_we}. 
Strong variations are induced in all PDFs, showing that indeed a more rigid interface  favors the appearance of extreme events. 
Since a
 more deformable interface offers lower resistance to the propagation of velocity disturbances  from one phase to the other, less modifications of the PDFs with respect to the single-phase case  can be expected at higher  $We_{\mathcal{L}}$ \cite[see also][]{Rosti2020}. This is indeed observed in all PDFs, where the distributions are seen to approach the single-phase one when increasing $We_\mathcal{L}$. Nevertheless, rare events are still evident also at the largest Weber considered, especially for the energy dissipation. 
Vorticity shows, again, that the psuedo-Gaussian part of the distribution is identical for all  $We_\mathcal{L}$, while  the exponentially decaying tails display strong variations.

\section{Conclusions}
\label{sec:conclusions}

In this work we discuss how volume fraction, viscosity ratio and Weber number influence HIT in emulsions.
The analyses are performed at different levels of details, spanning from phase averaged balances to SBS energy transfer in spectral space. 
Some observations are common to all configurations and highlight some fundamental physical effects introduced by the dispersed phase. 
Here, we first consider these different aspects and then 
 discuss the modulation introduced by the variation of material properties.

\subsection{Spectra and SBS energy balance.}
In all simulations with a dispersed phase, the energy decreases  at large scales and increases at small scales,
corroborating previous findings~\citep{ten2004fully,Perlekar2014,Dodd2016,Mukherjee2019,Rosti2020,Olivieri2020a}.
Interestingly, this behaviour applies to both
solid and liquid dispersed phases in HIT.
Furthermore, the pivoting point of the energy spectra is found to be described, with a good approximation, by the Hinze scale. This has also been observed in binary mixtures \citep{Perlekar2014} and emulsions \citep{Mukherjee2019} and is here extended to several operating conditions.

In general, the mechanisms of energy transport  are modified as follows: the transfer by the non-linear advection terms decreases, 
as the surface tension forces absorb energy at large scales. 
In an emulsion, energy is transferred to small scales  also by the 
surface tension force,
well within the dissipative range of the corresponding single-phase flow, forcing the viscous dissipation to be active at even smaller scales. No inverse cascade has been observed in the present simulations.

The general idea, according to which coalescence and breakup are responsible for modifications  of the energy spectra
 seems to  only partially explain our observations. 
 In fact, according to this hypothesis, significant deviations should be observed when comparing spectra for different volume fractions.
 Here, instead, we observe the largest  deviations in the energy spectra, in particular at small scales. when varying the viscosity ratio.
  This issue may be further addressed in future studies if coalescence is inhibited, reduced or controlled numerically.

\subsection{Effects of the dispersed phase on the dissipative range}
The classical "far" dissipative range ($\kappa\sim\kappa_{max}$), where both non-linear energy transport and energy dissipation 
of the SBS budget are zero, is lost when a dispersed phase is introduced. 
In multiphase flows, despite the non-linear energy transfer vanishes at certain small scales, the energy dissipation does not 
because energy is brought to these smaller scales by the 
action of the surface tension.
 As discussed above, energy dissipation is thus forced to extend towards smaller scales, overall increasing the range of wavelengths where 
 there is activity.  In other words, this increased activity at small scale translates also into an extension of the dissipative range, with the non-linear transport substituted by the surface-tension transport. 

It is important to understand how the scaling in the inertial range might be affected by these modifications of the dissipation range. 
From a practical viewpoint, the results in \Cref{subsec:met:sp-case} shows that increasing the mesh resolution does not result in significant alterations of the inertial range, indicating that a relevant analysis of the inertial range dynamics is still possible even in simulations where the surface tension terms are slightly under-resolved at small scale. Nevertheless,
resolving the dissipative range is important for a complete discussion of the SBS budget and e.g.\ the DSD; understanding turbulence at small scales in multiphase flows remains therefore a relevant question also from a fundamental point of view.

\subsection{Flow intermittency}
We have observed that the presence of a dispersed phase increases intermittency, unless the dispersed phase is highly viscous. 
The probability of detecting rare events increases, mainly for energy dissipation and vorticity, as shown here by the PDF analysis.

In particular, at higher volume fractions and  constant $\mu_d/\mu_c$ and $We_\mathcal{L}$, the exponent describing the distribution tail exponential decay is independent of the volume fraction $\alpha$. 
The onset of the exponential tail (hence the probability of an extreme event) is, on the other hand, affected by $\alpha$, proving that these events are mostly occurring at the interface. 
This is a confirmation of the observations in \cite{Dodd2016} on the increased energy dissipation at the interface. The variation of the exponential tail for both energy dissipation rate and vorticity at different $\mu_d/\mu_c$ and $We_\mathcal{L}$  reveals that intermittency is significantly affected by the fluid properties. In cases with high $\mu_d$ and low surface tension, the vorticity intermittency is attenuated and similar to the single-phase cases. On the other hand, the dissipation seems to be 
always affected by the multiphase nature of the flow.

\subsection{Droplet statistics}
In all the conditions analyzed, the droplet-size distributions show both the -3/2 exponential scaling from \cite{Deane2002}  for the small droplets and the -10/3 from \cite{Garrett2000} for the larger ones, confirming and extending the previous findings of \cite{Mukherjee2019} to a significant number of different configurations. Moreover, employing a VOF approach, and its known mass conserving properties, allows to extend the -3/2 scaling to significantly small droplets. 

The power-law $d^{-10/3}$ well describes the distributions of larger droplets when the volume fraction is below $10\%$, 
with only a small loss in accuracy for higher values of $\alpha$, in agreement with the assumption of negligible coalescence in \cite{Garrett2000}. 
Although this power law was obtained under the assumption of a dilute dispersed phase, recent works based on a diffuse-interface approach report the same scaling in the 
presence of coalescence \citep{Soligo2019,Mukherjee2019}. However,
\cite{Deike:2016ir} estimate through accurate sharp-interface simulations
a similar exponent, $-3$, so that it might be difficult to have a clear distinction on the different effects. 
 Finally, we show
that the estimate of the Hinze scale as transition point between the two power laws  is less accurate for $\alpha>0.1$, suggesting 
different model coefficients may be needed when coalescence is relevant.

\subsection{Role of the fluid properties}

Our analyses demonstrate that the volume fraction $\alpha$ is the parameter that mostly modifies 
the energy fluxes in the flow; yet, increasing the volume of the dispersed phase
does not 
change the underlying physics. This is notably documented in \Cref{sec:res:alpha} where we show that the amount of total interface area determines 
 the energy transport across scales. 
 Moreover, the simulation data reveal that the energy transfer via surface tension forces is enhanced at low viscosity ratios, while high viscosity in the carrier phase inhibits the propagation of vortices through the interface, hence 
 reducing the overall energy transport. 
 Changing the Weber number amounts to modulating the pivoting frequency below which energy transfer through surface tension is directed towards smaller scales. 
 In particular,
 as the dispersed phase is less deformable, the energy absorption from the dispersed phase occurs at larger scales, and turbulence is progressively reduced. In fact, as the surface tension increases, more energy is required to deform the droplets, an energy which can only be found in large-scale eddies.

To study the role of the viscosity ratio, see \Cref{sec:res:visc}, we consider values ranging  from $10^{-2}$ (a value typical of bubbles) to $10^{2}$ (typical of droplets). 

The analysis reveals that for $\mu_d/\mu_c\leq 1$,  $Re_\lambda$ increase significantly, due to the lower viscosity in the dispersed phase. The scale-by-scale energy budget
 shows that the interfacial and non-linear transport terms are not strongly affected  at these low viscosity ratios. 
 For $\mu_d/\mu_c> 1$, on the other hand, the turbulence in the  dispersed phase is reduced, which implies a significantly smaller $Re_\lambda$, below the value of the single-phase case.
   In these cases, the energy transfer induced by the interfacial stresses is  significantly reduced, suggesting that large differences may be found
   in liquid-gas and gas-liquid emulsions. The droplet-size distribution does not show strong differences, although larger droplets are more likely to be generated by a more viscous dispersed phase. Note, as discussed above, that the viscosity ratio has a significant impact on the flow intermittency.

Finally, we have examined the role of the large-scale Weber number $We_{\mathcal{L}}$. 
At low $We_{\mathcal{L}}$, coalescence is more likely to occur, hence there is a higher probability to find large droplets. Nevertheless, the Hinze scale proves to be an accurate estimation of the transition between the -3/2 and -10/3 for all the cases analyzed. Changing $We_\mathcal{L}$ and thus
 the droplet size distribution also affects the energy transport across scales by the surface tension forces. Specifically,  when decreasing $We_\mathcal{L}$ the energy injection from interfacial tension moves to larger scales.

\section*{Acknowledgments}
This work was supported by the Swedish Research Council via the multidisciplinary research environment INTERFACE, Hybrid multiscale modelling of transport phenomena for energy efficient processes, Grant No. 2016-06119. The authors acknowledge computer time provided by the National Infrastructure for High Performance Computing and Data Storage in Norway, (Sigma2, project no. NN9561K) and by SNIC (Swedish National Infrastructure for Computing). 
M.E.R. was supported by the JSPS KAKENHI Grant No. JP20K22402 and acknowledges computer time provided by the Scientific Computing section of Research Support Division at OIST.

\section*{Declaration of Interests}
The authors report no conflict of interest.

\bibliographystyle{jfm}
\bibliography{references}

\begin{thebibliography}{62}
\expandafter\ifx\csname natexlab\endcsname\relax\def\natexlab#1{#1}\fi
\def\au#1{#1} \def\ed#1{#1} \def\yr#1{#1}\def\at#1{#1}\def\jt#1{\textit{#1}}
  \def\bt#1{#1}\def\bvol#1{\textbf{#1}} \def\vol#1{#1} \def\pg#1{#1}
  \def\publ#1{#1}\def\arxiv#1{#1}\def\org#1{#1}\def\st#1{\textit{#1}}

\bibitem[Alexakis \& Biferale(2018)]{Alexakis2018}
{\sc \au{Alexakis, A.} \& \au{Biferale, L.}} \yr{2018}  \at{{Cascades and
  transitions in turbulent flows}}.  \jt{Physics Reports}  \bvol{767-769},
  \pg{1--101}.

\bibitem[Bassenne {\em et~al.\/}(2016)Bassenne, Urzay, Park \&
  Moin]{Bassenne2016}
{\sc \au{Bassenne, Maxime}, \au{Urzay, Javier}, \au{Park, George~I.} \&
  \au{Moin, Parviz}} \yr{2016}  \at{{Constant-energetics physical-space forcing
  methods for improved convergence to homogeneous-isotropic turbulence with
  application to particle-laden flows}}.  \jt{Physics of Fluids}
  \bvol{28}~(3).

\bibitem[Biferale {\em et~al.\/}(2011)Biferale, Perlekar, Sbragaglia,
  Srivastava \& Toschi]{Biferale2011}
{\sc \au{Biferale, Luca}, \au{Perlekar, Prasad}, \au{Sbragaglia, Mauro},
  \au{Srivastava, Sudhir} \& \au{Toschi, Federico}} \yr{2011}  \at{{A lattice
  Boltzmann method for turbulent emulsions}}.  \jt{Journal of Physics:
  Conference Series}  \bvol{318}~(SECTION 5).

\bibitem[Brackbill {\em et~al.\/}(1992)Brackbill, Kothe \&
  Zemach]{Brackbill1992}
{\sc \au{Brackbill, J.~U.}, \au{Kothe, D.~B.} \& \au{Zemach, C.}} \yr{1992}
  \at{{A Continuum Method for Modeling Surface Tension}}.  \jt{J. Comput.
  Phys.}  \bvol{100}~(2),  \pg{335--354}.

\bibitem[Chan {\em et~al.\/}(2021)Chan, Johnson, Moin \& Urzay]{Chan2020a}
{\sc \au{Chan, Wai Hong~Ronald}, \au{Johnson, Perry~L.}, \au{Moin, Parviz} \&
  \au{Urzay, Javier}} \yr{2021}  \at{{The turbulent bubble break-up cascade.
  Part 2. Numerical simulations of breaking waves}}.  \jt{Journal of Fluid
  Mechanics}  \bvol{912},  \pg{A43},  \arxiv{arXiv: 2009.04804}.

\bibitem[Costa(2018)]{Costa2018}
{\sc \au{Costa, Pedro}} \yr{2018}  \at{{A FFT-based finite-difference solver
  for massively-parallel direct numerical simulations of turbulent flows}}.
  \jt{Computers and Mathematics with Applications}  \bvol{76}~(8),
  \pg{1853--1862},  \arxiv{arXiv: arXiv:1802.10323v3}.

\bibitem[{De Vita} {\em et~al.\/}(2019){De Vita}, Rosti, Caserta \&
  Brandt]{DeVita2019}
{\sc \au{{De Vita}, Francesco}, \au{Rosti, Marco~Edoardo}, \au{Caserta, Sergio}
  \& \au{Brandt, Luca}} \yr{2019}  \at{{On the effect of coalescence on the
  rheology of emulsions}}.  \jt{Journal of Fluid Mechanics}  \pg{pp. 969--991},
   \arxiv{arXiv: 1908.08383}.

\bibitem[Deane \& Stokes(2002)]{Deane2002}
{\sc \au{Deane, Grant~B.} \& \au{Stokes, M.~Dale}} \yr{2002}  \at{{Scale
  dependence of bubble creation mechanisms in breaking waves}}.  \jt{Nature}
  \bvol{418}~(6900),  \pg{839--844}.

\bibitem[Debue {\em et~al.\/}(2018)Debue, Shukla, Kuzzay, Faranda, Saw, Daviaud
  \& Dubrulle]{Debue2018}
{\sc \au{Debue, P.}, \au{Shukla, V.}, \au{Kuzzay, D.}, \au{Faranda, D.},
  \au{Saw, E.~W.}, \au{Daviaud, F.} \& \au{Dubrulle, B.}} \yr{2018}
  \at{{Dissipation, intermittency, and singularities in incompressible
  turbulent flows}}.  \jt{Physical Review E}  \bvol{97}~(5),  \pg{1--21}.

\bibitem[Deike {\em et~al.\/}(2016)Deike, Melville \& Popinet]{Deike:2016ir}
{\sc \au{Deike, L}, \au{Melville, W~K} \& \au{Popinet, S}} \yr{2016}  \at{{Air
  entrainment and bubble statistics in breaking waves}}.  \jt{J. Fluid Mech.}
  \bvol{801},  \pg{91--129}.

\bibitem[Dodd \& Ferrante(2014)]{Dodd2014}
{\sc \au{Dodd, Michael~S.} \& \au{Ferrante, Antonino}} \yr{2014}  \at{{A fast
  pressure-correction method for incompressible two-fluid flows}}.  \jt{Journal
  of Computational Physics}  \bvol{273},  \pg{416--434}.

\bibitem[Dodd \& Ferrante(2016)]{Dodd2016}
{\sc \au{Dodd, Michael~S.} \& \au{Ferrante, Antonino}} \yr{2016}  \at{{On the
  interaction of Taylor length scale size droplets and isotropic turbulence}}.
  \jt{Journal of Fluid Mechanics}  \bvol{806},  \pg{356--412}.

\bibitem[Dubrulle(2019)]{Dubrulle2019}
{\sc \au{Dubrulle, B{\'{e}}reng{\`{e}}re}} \yr{2019}  \at{{Beyond Kolmogorov
  cascades}}.  \jt{Journal of Fluid Mechanics}  \bvol{867},  \pg{P1}.

\bibitem[Einstein(1906)]{einstein1906neue}
{\sc \au{Einstein, A}} \yr{1906}  \at{{Eine neue Bestimmung der
  Molek{\"{u}}ldimensionen. Section 2: Berechnung des Reibungskoeffizienten
  einer Fl{\"{u}}ssigkeit, in welcher sehr viele kleine Kugeln in regelloser
  Verteilung suspendiert sind}}.  \jt{Annalen der Physik, IV}  \pg{pp.
  297--306}.

\bibitem[Einstein(1911)]{einstein1911berichtigung}
{\sc \au{Einstein, Albert}} \yr{1911}  \at{{Berichtigung zu meiner arbeit: Eine
  neue bestimmung der molek{\"{u}}ldimensionen}}.  \jt{Annalen der Physik}
  \bvol{339}~(3),  \pg{591--592}.

\bibitem[Eswaran \& Pope(1988)]{Eswaran1988}
{\sc \au{Eswaran, V.} \& \au{Pope, S.~B.}} \yr{1988} {An examination of forcing
  in direct numerical simulations of turbulence}.

\bibitem[French-McCay(2004)]{french2004oil}
{\sc \au{French-McCay, Deborah~P}} \yr{2004}  \at{Oil spill impact modeling:
  development and validation}.  \jt{Environmental Toxicology and Chemistry: An
  International Journal}  \bvol{23}~(10),  \pg{2441--2456}.

\bibitem[Frisch(1995)]{Frisch1995}
{\sc \au{Frisch, Uriel}} \yr{1995} {\em Turbulence: the legacy of AN
  Kolmogorov\/}.  \publ{Cambridge university press}.

\bibitem[Garrett {\em et~al.\/}(2000)Garrett, Li \& Farmer]{Garrett2000}
{\sc \au{Garrett, C.}, \au{Li, M.} \& \au{Farmer, D.}} \yr{2000}  \at{{The
  connection between bubble size spectra and energy dissipation rates in the
  upper ocean}}.  \jt{Journal of Physical Oceanography}  \bvol{30}~(9),
  \pg{2163--2171}.

\bibitem[Gopalan \& Katz(2010)]{Gopalan2010}
{\sc \au{Gopalan, Balaji} \& \au{Katz, Joseph}} \yr{2010}  \at{{Turbulent
  shearing of crude oil mixed with dispersants generates long microthreads and
  microdroplets}}.  \jt{Physical Review Letters}  \bvol{104}~(5),  \pg{1--4}.

\bibitem[Hinze(1955)]{Hinze1955}
{\sc \au{Hinze, J.~O.}} \yr{1955}  \at{{Fundamentals of the hydrodynamic
  mechanism of splitting in dispersion processes}}.  \jt{AIChE Journal}
  \bvol{1}~(3),  \pg{289--295}.

\bibitem[Ii {\em et~al.\/}(2012)Ii, Sugiyama, Takeuchi, Takagi, Matsumoto \&
  Xiao]{Ii2012}
{\sc \au{Ii, Satoshi}, \au{Sugiyama, Kazuyasu}, \au{Takeuchi, Shintaro},
  \au{Takagi, Shu}, \au{Matsumoto, Yoichiro} \& \au{Xiao, Feng}} \yr{2012}
  \at{{An interface capturing method with a continuous function: The THINC
  method with multi-dimensional reconstruction}}.  \jt{Journal of Computational
  Physics}  \bvol{231}~(5),  \pg{2328--2358}.

\bibitem[Ishihara {\em et~al.\/}(2009)Ishihara, Gotoh \& Kaneda]{Ishihara2009}
{\sc \au{Ishihara, Takashi}, \au{Gotoh, Toshiyuki} \& \au{Kaneda, Yukio}}
  \yr{2009}  \at{{Study of High–Reynolds Number Isotropic Turbulence by
  Direct Numerical Simulation}}.  \jt{Annual Review of Fluid Mechanics}
  \bvol{41}~(1),  \pg{165--180}.

\bibitem[Jansen {\em et~al.\/}(2001)Jansen, Agterof \& Mellema]{Jansen2001}
{\sc \au{Jansen, K. M.~B.}, \au{Agterof, W. G.~M.} \& \au{Mellema, J.}}
  \yr{2001}  \at{{Droplet breakup in concentrated emulsions}}.  \jt{Journal of
  Rheology}  \bvol{45}~(1),  \pg{227--236}.

\bibitem[Jimenez(2000)]{Jimenez2000a}
{\sc \au{Jimenez, Javier}} \yr{2000}  \at{{Turbulent velocity fluctuations need
  not be Gaussian}}.  \jt{Journal of Fluid Mechanics}  \bvol{376}~(-1),
  \pg{139--147}.

\bibitem[Kilpatrick(2012)]{kilpatrick2012water}
{\sc \au{Kilpatrick, Peter~K}} \yr{2012}  \at{Water-in-crude oil emulsion
  stabilization: review and unanswered questions}.  \jt{Energy \& Fuels}
  \bvol{26}~(7),  \pg{4017--4026}.

\bibitem[Knutsen {\em et~al.\/}(2020)Knutsen, Baj, Lawson, Bodenschatz, Dawson
  \& Worth]{Knutsen2020}
{\sc \au{Knutsen, Anna~N}, \au{Baj, Pawel}, \au{Lawson, John~M},
  \au{Bodenschatz, Eberhard}, \au{Dawson, James~R} \& \au{Worth, Nicholas~A}}
  \yr{2020}  \at{{The inter-scale energy budget in a von K{\'{a}}rm{\'{a}}n
  mixing flow}}.  \jt{Journal of Fluid Mechanics}  \bvol{895},  \pg{1--40}.

\bibitem[Kokal \& Others(2005)]{kokal2005crude}
{\sc \au{Kokal, Sunil~Lalchand} \& \au{Others}} \yr{2005}  \at{{Crude oil
  emulsions: A state-of-the-art review}}.  \jt{SPE Production {\&} facilities}
  \bvol{20}~(01),  \pg{5--13}.

\bibitem[Kolmogorov(1949)]{Kolmogorov1949}
{\sc \au{Kolmogorov, A.}} \yr{1949}  \at{On the breakage of drops in a
  turbulent flow.}  \jt{Dokl. Akad. Navk SSSR 66,}  \bvol{66},  \pg{825--828}.

\bibitem[Komrakova {\em et~al.\/}(2015)Komrakova, Eskin \&
  Derksen]{Komrakova2015}
{\sc \au{Komrakova, Alexandra~E.}, \au{Eskin, Dmitry} \& \au{Derksen, J.~J.}}
  \yr{2015}  \at{{Numerical study of turbulent liquid-liquid dispersions}}.
  \jt{AIChE Journal}  \bvol{61}~(8),  \pg{2618--2633},  \arxiv{arXiv:
  0201037v1}.

\bibitem[Lemenand {\em et~al.\/}(2017)Lemenand, Valle, Dupont \&
  Peerhossaini]{Lemenand2017}
{\sc \au{Lemenand, Thierry}, \au{Valle, Dominique~Della}, \au{Dupont, Pascal}
  \& \au{Peerhossaini, Hassan}} \yr{2017}  \at{{Turbulent spectrum model for
  drop-breakup mechanisms in an inhomogeneous turbulent flow}}.  \jt{Chemical
  Engineering Science}  \bvol{158}~(September 2016),  \pg{41--49}.

\bibitem[Li \& Garrett(1998)]{Li1998}
{\sc \au{Li, Ming} \& \au{Garrett, Chris}} \yr{1998}  \at{{The relationship
  between oil droplet size and upper ocean turbulence}}.  \jt{Marine Pollution
  Bulletin}  \bvol{36}~(12),  \pg{961--970}.

\bibitem[Mallouppas {\em et~al.\/}(2013)Mallouppas, George \& van
  Wachem]{Mallouppas2013}
{\sc \au{Mallouppas, G.}, \au{George, W.~K.} \& \au{van Wachem, B.~G.M.}}
  \yr{2013}  \at{{New forcing scheme to sustain particle-laden homogeneous and
  isotropic turbulence}}.  \jt{Physics of Fluids}  \bvol{25}~(8).

\bibitem[Mandal {\em et~al.\/}(2010)Mandal, Samanta, Bera \&
  Ojha]{mandal2010characterization}
{\sc \au{Mandal, Ajay}, \au{Samanta, Abhijit}, \au{Bera, Achinta} \& \au{Ojha,
  Keka}} \yr{2010}  \at{Characterization of oil- water emulsion and its use in
  enhanced oil recovery}.  \jt{Industrial \& Engineering Chemistry Research}
  \bvol{49}~(24),  \pg{12756--12761}.

\bibitem[Masuk {\em et~al.\/}(2021)Masuk, Salibindla \& Ni]{Masuk2021}
{\sc \au{Masuk, Ashik Ullah~Mohammad}, \au{Salibindla, Ashwanth~K.R.} \&
  \au{Ni, Rui}} \yr{2021}  \at{{Simultaneous measurements of deforming
  Hinze-scale bubbles with surrounding turbulence}}.  \jt{Journal of Fluid
  Mechanics} ,  \arxiv{arXiv: 2101.07349}.

\bibitem[McClements(2015)]{mcclements2015food}
{\sc \au{McClements, David~Julian}} \yr{2015} {\em {Food emulsions: principles,
  practices, and techniques}\/}.  \publ{CRC press}.

\bibitem[Mininni {\em et~al.\/}(2006)Mininni, Alexakis \& Pouquet]{Mininni2006}
{\sc \au{Mininni, P.~D.}, \au{Alexakis, A.} \& \au{Pouquet, A.}} \yr{2006}
  \at{{Large-scale flow effects, energy transfer, and self-similarity on
  turbulence}}.  \jt{Physical Review E - Statistical, Nonlinear, and Soft
  Matter Physics}  \bvol{74}~(1),  \pg{1--13}.

\bibitem[Mukherjee {\em et~al.\/}(2019)Mukherjee, Safdari, Shardt, Kenjeres,
  den Akker, Kenjere{\v{s}} \& {Van Den Akker}]{Mukherjee2019}
{\sc \au{Mukherjee, Siddhartha}, \au{Safdari, Arman}, \au{Shardt, Orest},
  \au{Kenjeres, Sasa}, \au{den Akker, Harry E. A.~Van}, \au{Kenjere{\v{s}},
  Sa{\v{s}}a} \& \au{{Van Den Akker}, Harry~E.A.}} \yr{2019}
  \at{{Droplet-Turbulence interactions and quasi-equilibrium dynamics in
  turbulent emulsions}}.  \jt{Journal of Fluid Mechanics}  \bvol{878},
  \pg{221--276},  \arxiv{arXiv: 1902.09929}.

\bibitem[Nielloud(2000)]{nielloud2000pharmaceutical}
{\sc \au{Nielloud, Fran{\c{c}}oise}} \yr{2000} {\em Pharmaceutical emulsions
  and suspensions: revised and expanded\/}.  \publ{CRC Press}.

\bibitem[Olivieri {\em et~al.\/}(2020{\natexlab{{\em a\/}}})Olivieri, Akoush,
  Brandt, Rosti \& Mazzino]{Olivieri2020a}
{\sc \au{Olivieri, Stefano}, \au{Akoush, Assad}, \au{Brandt, Luca}, \au{Rosti,
  Marco~E.} \& \au{Mazzino, Andrea}} \yr{2020{\natexlab{{\em a\/}}}}
  \at{Turbulence in a network of rigid fibers}.  \jt{Phys. Rev. Fluids}
  \bvol{5},  \pg{074502}.

\bibitem[Olivieri {\em et~al.\/}(2020{\natexlab{{\em b\/}}})Olivieri, Brandt,
  Rosti \& Mazzino]{Olivieri2020}
{\sc \au{Olivieri, Stefano}, \au{Brandt, Luca}, \au{Rosti, Marco~E.} \&
  \au{Mazzino, Andrea}} \yr{2020{\natexlab{{\em b\/}}}}  \at{Dispersed fibers
  change the classical energy budget of turbulence via nonlocal transfer}.
  \jt{Phys. Rev. Lett.}  \bvol{125},  \pg{114501}.

\bibitem[Pacek {\em et~al.\/}(1998)Pacek, Man \& Nienow]{Pacek1998}
{\sc \au{Pacek, A.~W.}, \au{Man, C.~C.} \& \au{Nienow, A.~W.}} \yr{1998}
  \at{{On the Sauter mean diameter and size distributions in turbulent
  liquid/liquid dispersions in a stirred vessel}}.  \jt{Chemical Engineering
  Science}  \bvol{53}~(11),  \pg{2005--2011}.

\bibitem[Pal(2000)]{Pal2000}
{\sc \au{Pal, Rajinder}} \yr{2000}  \at{{Shear viscosity behavior of emulsions
  of two immiscible liquids}}.  \jt{Journal of Colloid and Interface Science}
  \bvol{225}~(2),  \pg{359--366}.

\bibitem[Pal(2001)]{Pal2001}
{\sc \au{Pal, Rajinder}} \yr{2001}  \at{{Novel viscosity equations for
  emulsions of two immiscible liquids}}.  \jt{Journal of Rheology}
  \bvol{45}~(2),  \pg{509--520}.

\bibitem[Perlekar(2019)]{Perlekar2019}
{\sc \au{Perlekar, Prasad}} \yr{2019}  \at{{Kinetic energy spectra and flux in
  turbulent phase-separating symmetric binary-fluid mixtures}}.  \jt{Journal of
  Fluid Mechanics}  \bvol{873},  \pg{459--474}.

\bibitem[Perlekar {\em et~al.\/}(2014)Perlekar, Benzi, Clercx, Nelson \&
  Toschi]{Perlekar2014}
{\sc \au{Perlekar, Prasad}, \au{Benzi, Roberto}, \au{Clercx, Herman~J.H.},
  \au{Nelson, David~R.} \& \au{Toschi, Federico}} \yr{2014}  \at{{Spinodal
  decomposition in homogeneous and isotropic turbulence}}.  \jt{Physical Review
  Letters}  \bvol{112}~(1),  \pg{1--5}.

\bibitem[Perlekar {\em et~al.\/}(2012)Perlekar, Biferale, Sbragaglia,
  Srivastava \& Toschi]{Perlekar2012}
{\sc \au{Perlekar, Prasad}, \au{Biferale, Luca}, \au{Sbragaglia, Mauro},
  \au{Srivastava, Sudhir} \& \au{Toschi, Federico}} \yr{2012}  \at{{Droplet
  size distribution in homogeneous isotropic turbulence}}.  \jt{Physics of
  Fluids}  \bvol{24}~(6),  \pg{065101},  \arxiv{arXiv: 1112.6041}.

\bibitem[Perlekar {\em et~al.\/}(2017)Perlekar, Pal \& Pandit]{Perlekar2017}
{\sc \au{Perlekar, Prasad}, \au{Pal, Nairita} \& \au{Pandit, Rahul}} \yr{2017}
  \at{{Two-dimensional Turbulence in Symmetric Binary-Fluid Mixtures:
  Coarsening Arrest by the Inverse Cascade}}.  \jt{Scientific Reports}
  \bvol{7}~(February),  \pg{1--7},  \arxiv{arXiv: 1506.08524}.

\bibitem[Podvigina \& Pouquet(1994)]{Podvigina1994}
{\sc \au{Podvigina, O.} \& \au{Pouquet, A.}} \yr{1994}  \at{{On the non-linear
  stability of the 1:1:1 ABC flow}}.  \jt{Physica D: Nonlinear Phenomena}
  \bvol{75}~(4),  \pg{471--508}.

\bibitem[Qi {\em et~al.\/}(2020)Qi, {Mohammad Masuk} \& Ni]{Qi2020}
{\sc \au{Qi, Yinghe}, \au{{Mohammad Masuk}, Ashik~Ullah} \& \au{Ni, Rui}}
  \yr{2020}  \at{{Towards a model of bubble breakup in turbulence through
  experimental constraints}}.  \jt{International Journal of Multiphase Flow}
  \bvol{132},  \pg{103397}.

\bibitem[Rivi{\`{e}}re {\em et~al.\/}(2021)Rivi{\`{e}}re, Mostert, Perrard \&
  Deike]{Riviere2021}
{\sc \au{Rivi{\`{e}}re, Ali{\'{e}}nor}, \au{Mostert, Wouter}, \au{Perrard,
  St{\'{e}}phane} \& \au{Deike, Luc}} \yr{2021}  \at{{Sub-Hinze scale bubble
  production in turbulent bubble break-up}}.  \jt{Journal of Fluid Mechanics}
  \bvol{917},  \pg{A40}.

\bibitem[Roccon {\em et~al.\/}(2017)Roccon, {De Paoli}, Zonta \&
  Soldati]{Roccon2017}
{\sc \au{Roccon, Alessio}, \au{{De Paoli}, Marco}, \au{Zonta, Francesco} \&
  \au{Soldati, Alfredo}} \yr{2017}  \at{{Viscosity-modulated breakup and
  coalescence of large drops in bounded turbulence}}.  \jt{Physical Review
  Fluids}  \bvol{2}~(8),  \pg{1--15}.

\bibitem[Rosales \& Meneveau(2005)]{Rosales2005}
{\sc \au{Rosales, Carlos} \& \au{Meneveau, Charles}} \yr{2005}  \at{{Linear
  forcing in numerical simulations of isotropic turbulence: Physical space
  Implementations and convergence properties}}.  \jt{Physics of Fluids}
  \bvol{17}~(9),  \pg{1--8}.

\bibitem[Rosti {\em et~al.\/}(2019)Rosti, {De Vita} \& Brandt]{Rosti2019}
{\sc \au{Rosti, Marco~E.}, \au{{De Vita}, Francesco} \& \au{Brandt, Luca}}
  \yr{2019}  \at{{Numerical simulations of emulsions in shear flows}}.
  \jt{Acta Mechanica}  \bvol{230}~(2),  \pg{667--682}.

\bibitem[Rosti {\em et~al.\/}(2020)Rosti, Ge, Jain, Dodd \& Brandt]{Rosti2020}
{\sc \au{Rosti, Marco~E}, \au{Ge, Zhouyang}, \au{Jain, Suhas~S}, \au{Dodd,
  Michael~S} \& \au{Brandt, Luca}} \yr{2020}  \at{{Droplets in homogeneous
  shear turbulence}}.  \jt{J. Fluid Mech}  \bvol{876},  \pg{962--984}.

\bibitem[Skartlien {\em et~al.\/}(2013)Skartlien, Sollum \&
  Schumann]{Skartlien2013}
{\sc \au{Skartlien, R.}, \au{Sollum, E.} \& \au{Schumann, H.}} \yr{2013}
  \at{{Droplet size distributions in turbulent emulsions: Breakup criteria and
  surfactant effects from direct numerical simulations}}.  \jt{Journal of
  Chemical Physics}  \bvol{139}~(17).

\bibitem[Soligo {\em et~al.\/}(2019)Soligo, Roccon \& Soldati]{Soligo2019}
{\sc \au{Soligo, Giovanni}, \au{Roccon, Alessio} \& \au{Soldati, Alfredo}}
  \yr{2019}  \at{{Breakage, coalescence and size distribution of
  surfactant-laden droplets in turbulent flow}}.  \jt{Journal of Fluid
  Mechanics}  \bvol{881},  \pg{244--282}.

\bibitem[Spernath \& Aserin(2006)]{spernath2006microemulsions}
{\sc \au{Spernath, Aviram} \& \au{Aserin, Abraham}} \yr{2006}
  \at{Microemulsions as carriers for drugs and nutraceuticals}.  \jt{Advances
  in colloid and interface science}  \bvol{128},  \pg{47--64}.

\bibitem[Sreenivasan \& Antonia(1997)]{Sreenivasan1997}
{\sc \au{Sreenivasan, K.~R.} \& \au{Antonia, R~A}} \yr{1997}  \at{{THE
  PHENOMENOLOGY OF SMALL-SCALE TURBULENCE}}.  \jt{Annual Review of Fluid
  Mechanics}  \bvol{29}~(1),  \pg{435--472}.

\bibitem[Ten~Cate {\em et~al.\/}(2004)Ten~Cate, Derksen, Portela \& Van
  Den~Akker]{ten2004fully}
{\sc \au{Ten~Cate, Andreas}, \au{Derksen, Jos~J}, \au{Portela, Luis~M} \&
  \au{Van Den~Akker, Harry~EA}} \yr{2004}  \at{Fully resolved simulations of
  colliding monodisperse spheres in forced isotropic turbulence}.  \jt{Journal
  of Fluid Mechanics}  \bvol{519},  \pg{233}.

\bibitem[Tryggvason {\em et~al.\/}(2011)Tryggvason, Scardovelli \&
  Zaleski]{tryggvason2011direct}
{\sc \au{Tryggvason, Gr{\'{e}}tar}, \au{Scardovelli, Ruben} \& \au{Zaleski,
  St{\'{e}}phane}} \yr{2011} {\em {Direct numerical simulations of gas--liquid
  multiphase flows}\/}.  \publ{Cambridge University Press}.

\bibitem[Yi {\em et~al.\/}(2021)Yi, Toschi \& Sun]{Yi2021}
{\sc \au{Yi, Lei}, \au{Toschi, Federico} \& \au{Sun, Chao}} \yr{2021}
  \at{{Global and local statistics in turbulent emulsions}}.  \jt{Journal of
  Fluid Mechanics}  \bvol{912},  \pg{1--17},  \arxiv{arXiv: 2011.00963}.

\end{thebibliography}

\appendix
\section{Effects of viscosity ratio at $\alpha=0.03$}
\label{app:visc003}
We report here the results for different values of $\mu_d/\mu_c$ at $We_\mathcal{L}=42.6$ and $\alpha=0.03$ (cases V1x and BE1), for completeness. The main discussions on the physical effects given by different viscosity ratios are provided in \Cref{sec:res:visc}, while here only main differences due to the lower volume fraction will be highlighted.  

\begin{figure}
\centering
	\includegraphics[width=0.5\textwidth]{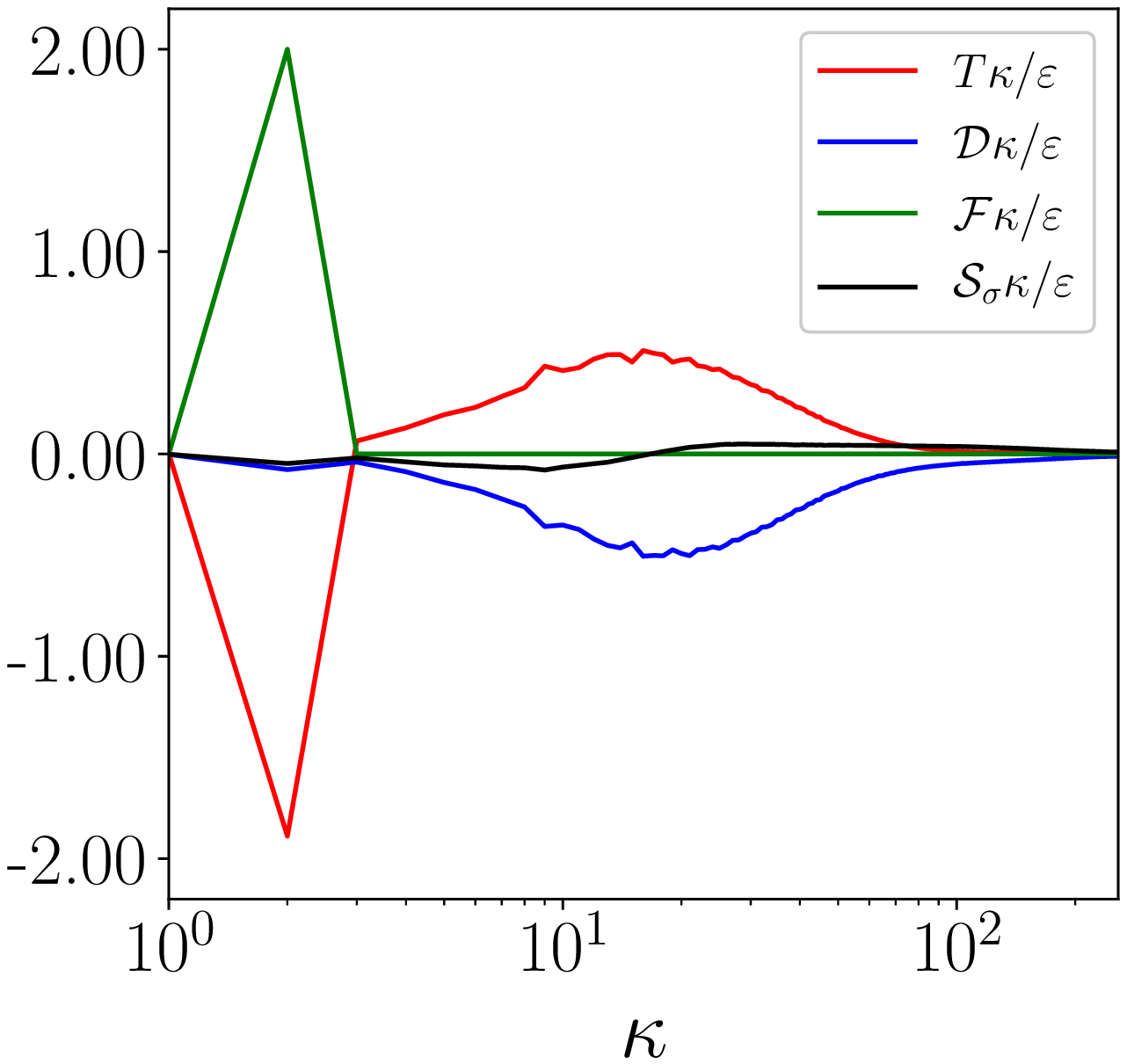}
	\put(-200,140){(\textit{a})}
	\includegraphics[width=0.5\textwidth]{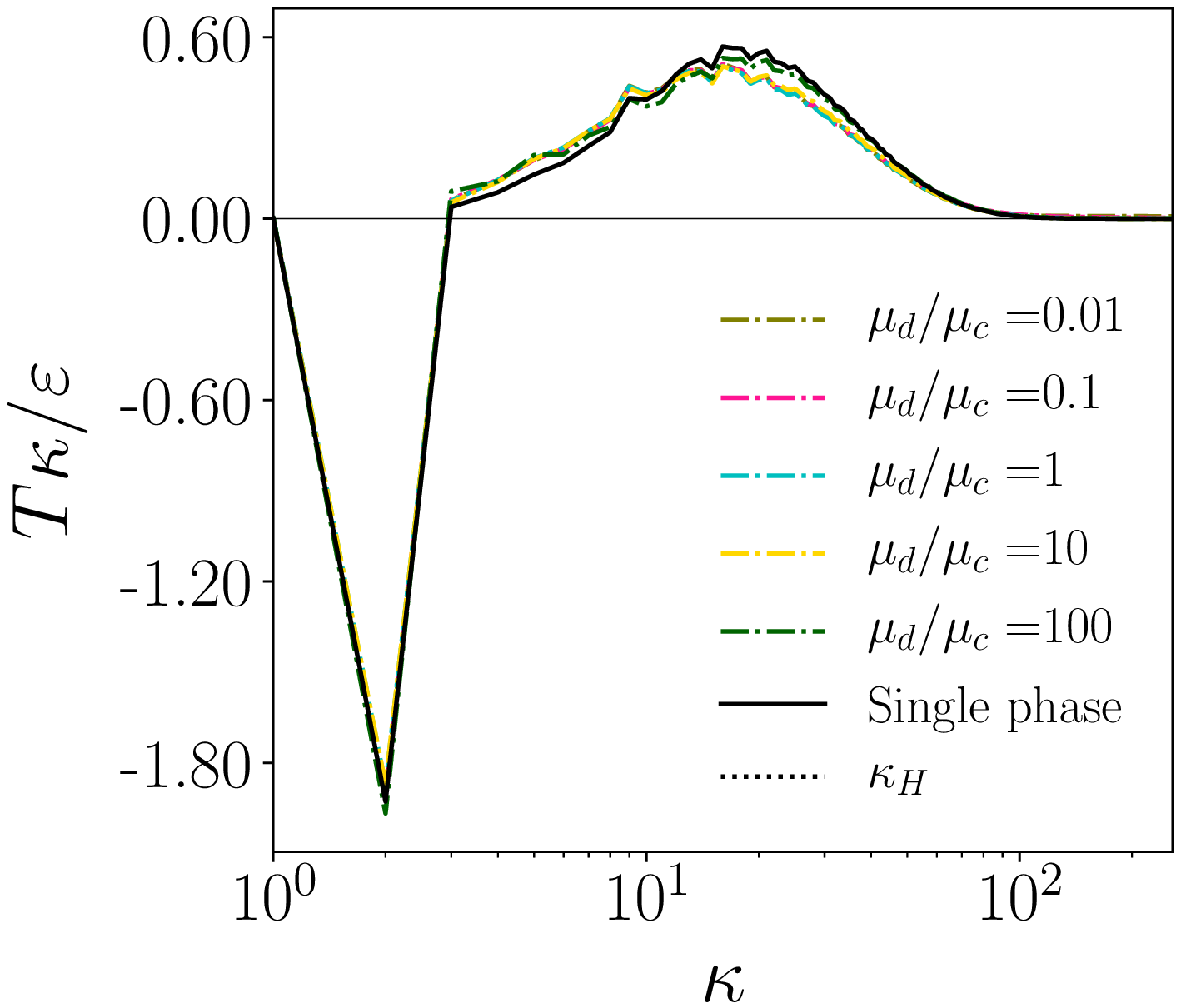}
	\put(-190,140){(\textit{b})}

	\includegraphics[width=0.5\textwidth]{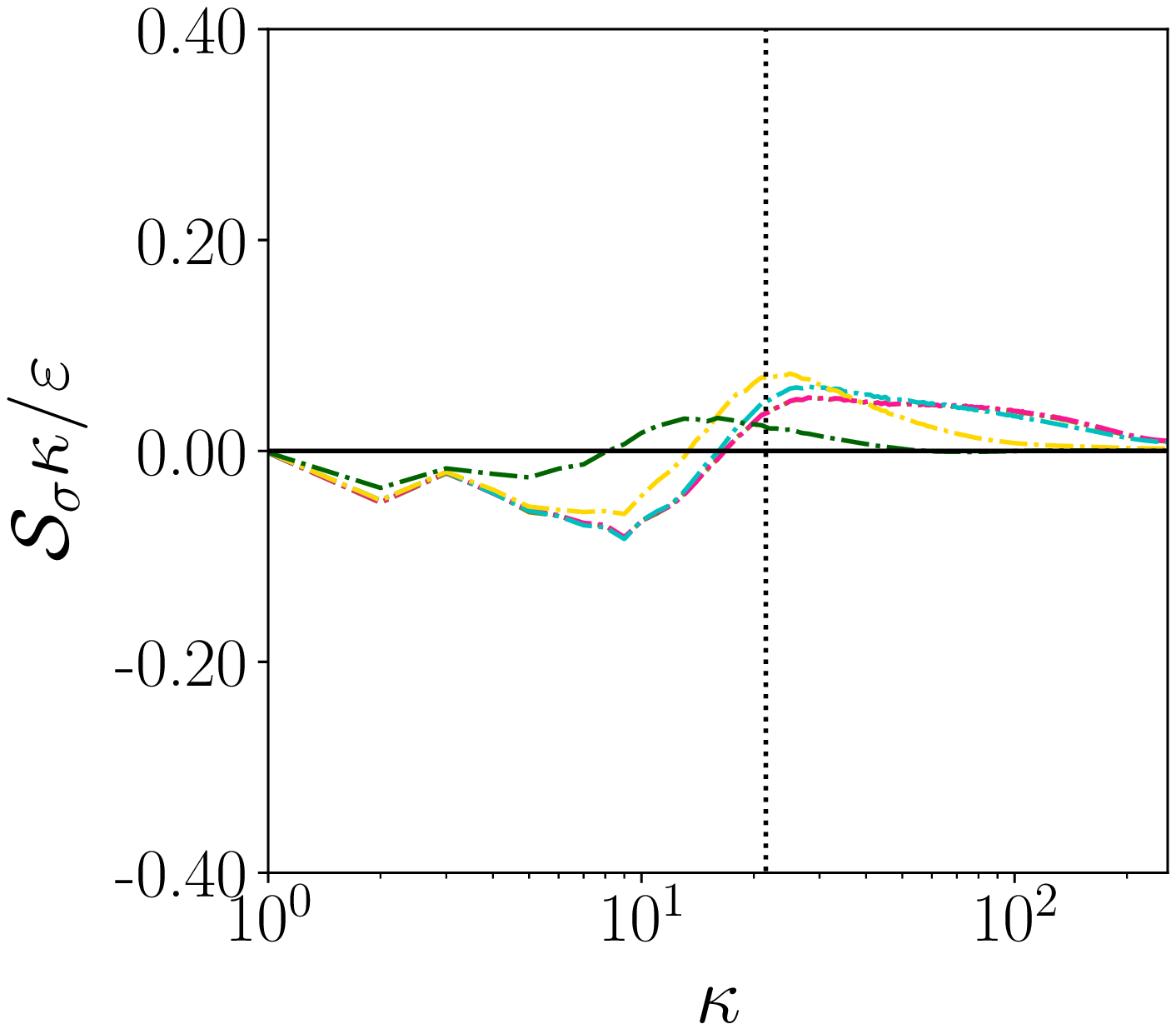}
	\put(-200,140){(\textit{c})}
	\includegraphics[width=0.5\textwidth]{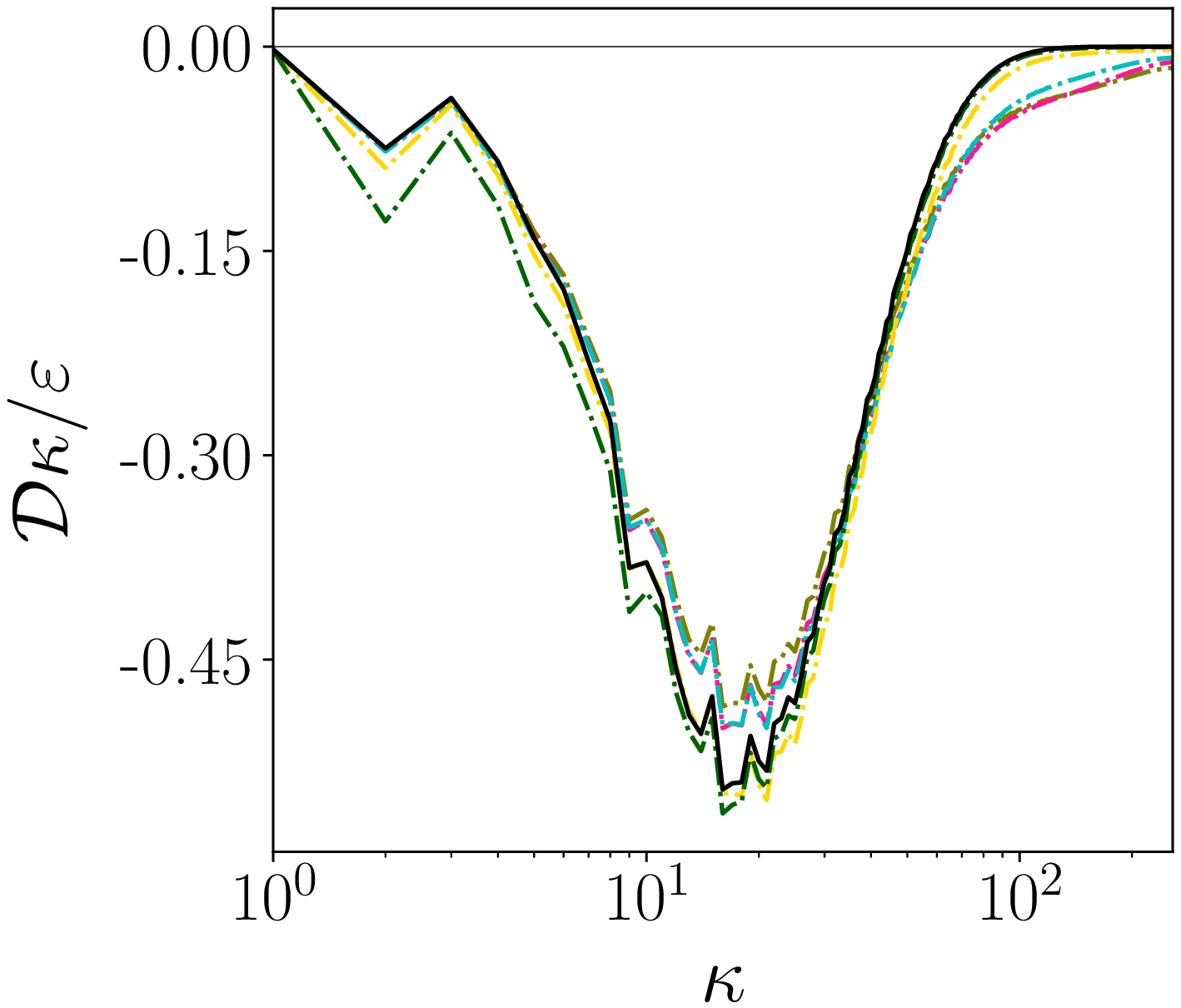}
	\put(-190,140){(\textit{d})}
\caption{Scale-by-scale energy budget for different viscosity ratios $\mu_d/\mu_c$ at $\alpha=0.03$. (\textit{a}) displays the complete energy balance for case V12 with $\mu_d/\mu_c=0.1$; panels (\textit{b-d}) show the non-linear energy transfer $T$, the term $\mathcal{S}_\sigma$ associated with the surface tension and the energy dissipation transfer function $\mathcal{D}$.
}
\label{fig:balance_mu}
\end{figure}

\Cref{fig:balance_mu}(\textit{a}) shows the full SBS energy balance for case V12 (see \Cref{tab:testMat}). The low volume fraction reduces significantly the effect of energy transport due to surface tension $\mathcal{S}_\sigma$. Consequently, the modifications of the non-linear transport with respect to the single-phase case are small,
see panel (\textit{a}) of the figure. The energy transport due to surface tension (panel \textit{c}) is attenuated at high viscosity ratios and shifts towards small wavelengths due to increased coalescence (see \Cref{sec:res:visc}). Finally, energy dissipation, panel (\textit{d}), shows again limited variations due to reduced volume fraction, although, it can be observed again that the small scale energy transfer is unaffected at high viscosity ratios. 

\begin{figure}
	\centering
	\includegraphics[width=0.5\textwidth]{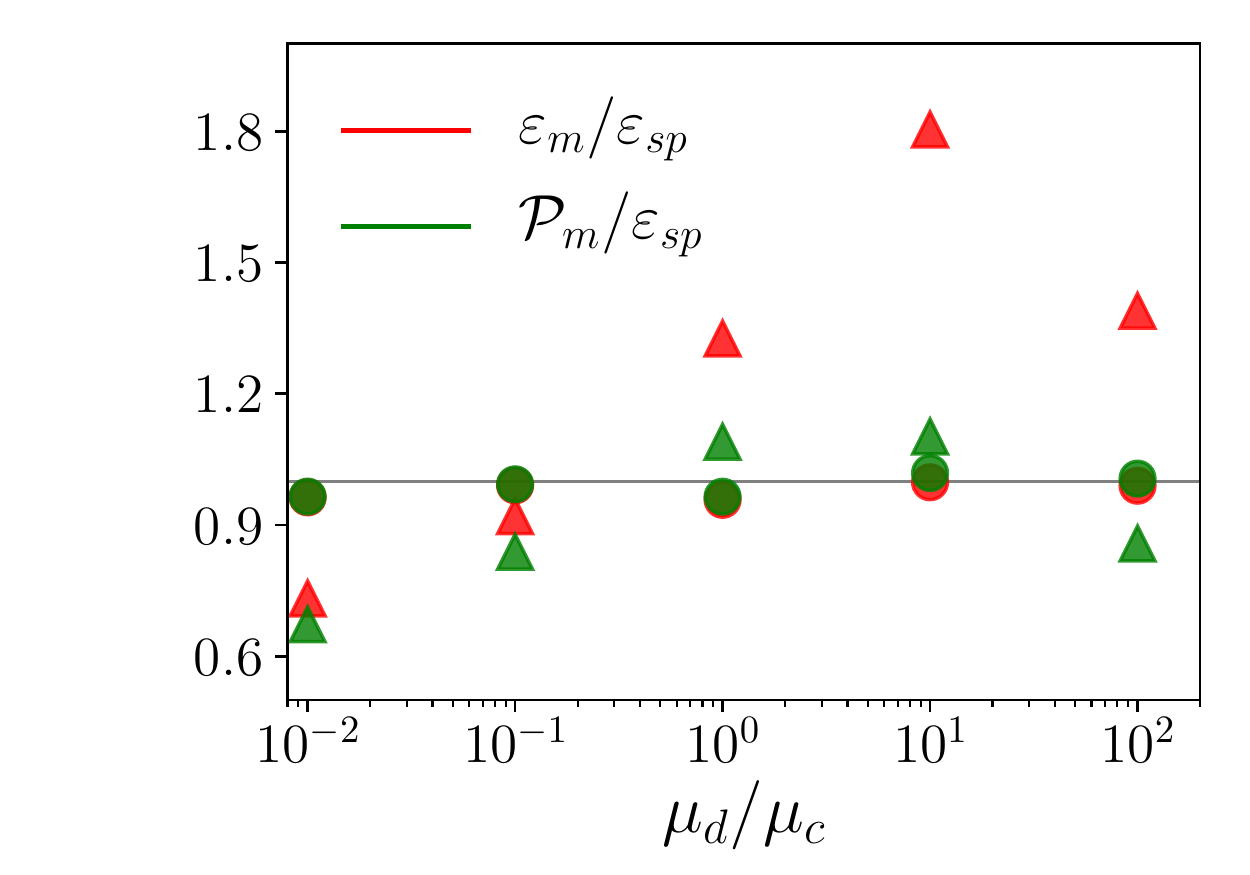}
	\put(-180,130){(\textit{a})}
	\includegraphics[width=0.5\textwidth]{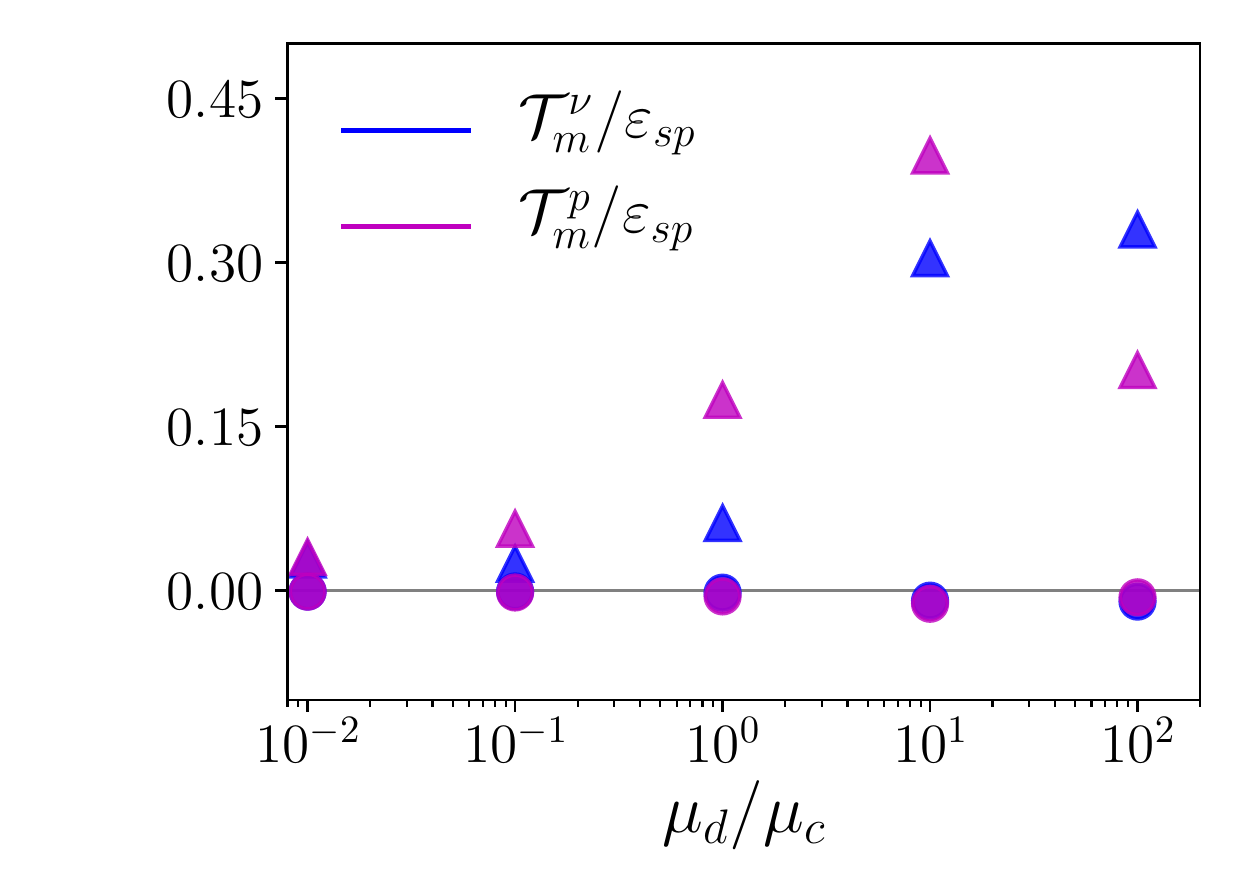}
	\put(-180,130){(\textit{b})}
	%
	\caption{Phase-averaged energy balance versus the emulsion viscosity ratio, see definitions of each term in \Cref{eq:enBalMP}. 
		Colored triangles (\mytriangle{black}) represent the dispersed phase ($m=d$) while circles (\mycircle{black}) are used for the carrier phase ($m=c$). Each term is normalized by the single phase energy dissipation $\varepsilon_{sp}$, computed for case SP2. The energy production $\mathcal{P}_m$ and energy dissipation $\varepsilon_m$  are reported in panel (\textit{a}), while viscous energy transport $\mathcal{T}^\nu_m$  and the pressure energy transport $\mathcal{T}^p_m$ in panel (\textit{b}).
	}
	\label{fig:balPhase_mu}
\end{figure}

The phase-averaged energy balance in \Cref{fig:balPhase_mu} shows only weak variations with respect to the cases at $\alpha=0.1$ in \Cref{fig:balPhase_mu10}.
 Again, we notice that energy dissipation in the dispersed phase increases at higher $\mu_d$, while energy is always transferred from the carrier to the dispersed phase, as for $\alpha=0.1$.

\begin{figure}
	\centering
	\includegraphics[width=0.33\textwidth]{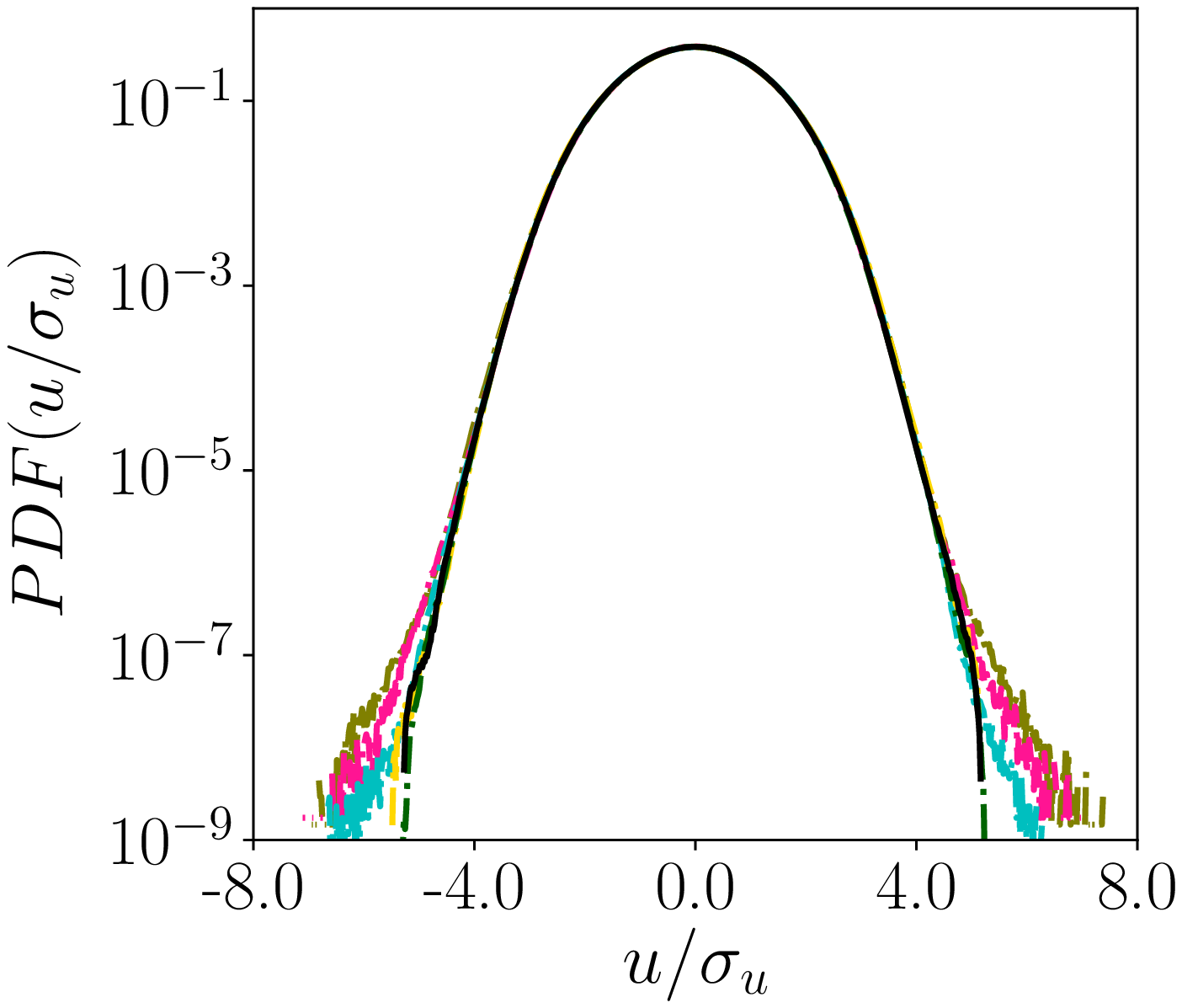}
	\put(-120,110){(\textit{a})}
	\includegraphics[width=0.33\textwidth]{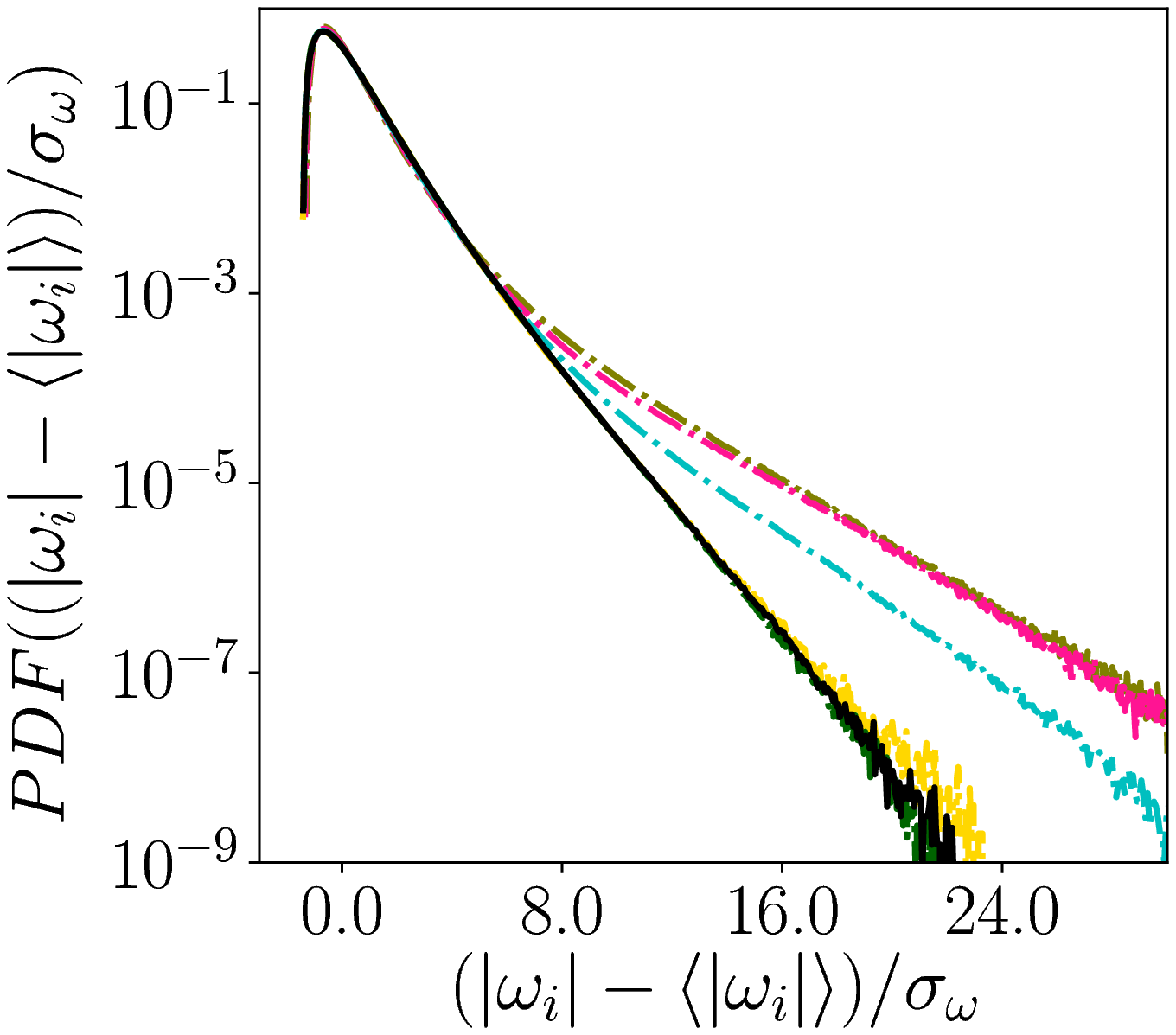}
	\put(-120,110){(\textit{b})}
	\includegraphics[width=0.33\textwidth]{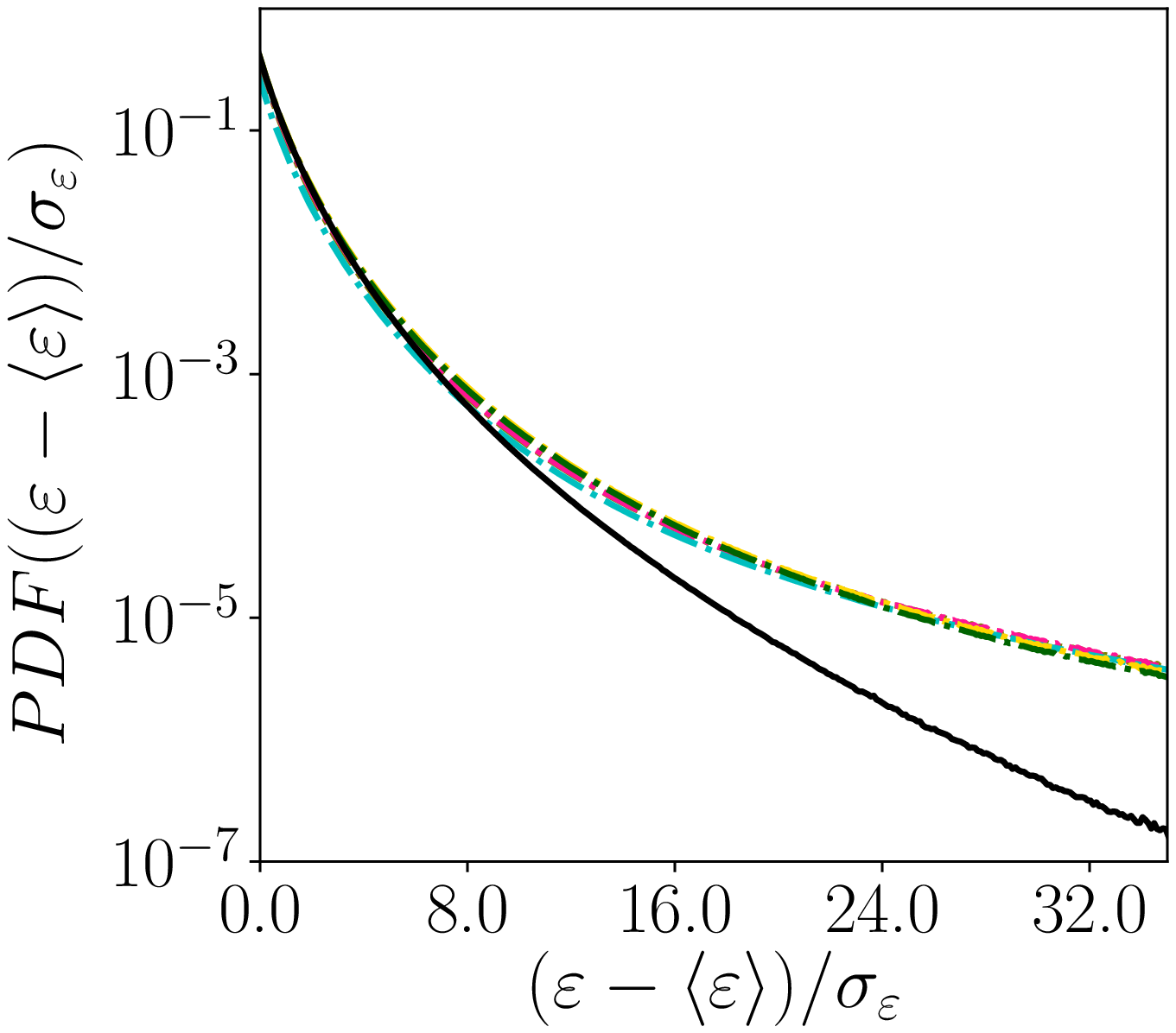}
	\put(-120,110){(\textit{c})}
	
	\includegraphics[width=0.8\textwidth]{figures/legend_mu.eps}
	\caption{PDF of velocity fluctuations $u$ (panel \textit{a}), vorticity $\omega$ (\textit{b}) and energy dissipation (\textit{c}). All quantities are normalized by their standard deviation. The data pertain cases V2x, with $\alpha=0.03$.}
	\label{fig:PDF_mu}
\end{figure}

We finally present the PDFs of velocity, vorticity and energy dissipation in \Cref{fig:PDF_mu}(\textit{a,b,c}). Again, small variations can be observed with respect to cases at $\alpha=0.1$ (\Cref{fig:PDF_mu10}). For vorticity and energy dissipation, we report
lower probability to observe rare events at lower volume fraction, as discussed in \Cref{sec:res:alpha}.

\end{document}